\documentclass[%
 reprint,
superscriptaddress,
nofootinbib,
 amsmath,amssymb,
 aps,
prb,
]{revtex4-2}

\usepackage[hidelinks,colorlinks=true,linkcolor=blue,citecolor=blue]{hyperref}
\usepackage{graphicx}
\usepackage{dcolumn}
\usepackage{bm}
\usepackage{bbold}
\usepackage{soul}
\usepackage{physics}
\usepackage{mdframed}
\usepackage{ulem}
\usepackage[skins,theorems]{tcolorbox}

\usepackage{amsthm}
\makeatletter
\def\l@subsubsection#1#2{}
\makeatother
\usepackage{tikz}

\newcommand{\lgen}{\langle\!\langle}
\newcommand{\rgen}{\rangle\!\rangle}
\newcommand{\defn}{\equiv}

\newcommand{\tlt}{\widetilde{\leftarrow}}
\newcommand{\trt}{\widetilde{\rightarrow}}

\begin{document}

\preprint{APS/123-QED}
\title{Dynamics in the presence of local symmetry-breaking impurities}
\author{Yahui Li}
\thanks{These authors contributed equally.}
\affiliation{
Technical University of Munich, 
TUM School of Natural Sciences, 
Physics Department,
James-Franck-Str. 1,
85748 Garching,
Germany}%
\affiliation{Munich Center for Quantum Science and Technology (MCQST), Schellingstr. 4, 80799 M{\"u}nchen, Germany}
\author{Pablo Sala}%
\thanks{These authors contributed equally.}
\affiliation{Department of Physics and Institute for Quantum Information and Matter, California Institute of Technology, Pasadena, CA 91125, USA}
\affiliation{Walter Burke Institute for Theoretical Physics, California Institute of Technology, Pasadena, CA 91125, USA}

\author{Frank Pollmann}
\affiliation{
Technical University of Munich, 
TUM School of Natural Sciences, 
Physics Department,
James-Franck-Str. 1,
85748 Garching,
Germany}%
\affiliation{Munich Center for Quantum Science and Technology (MCQST), Schellingstr. 4, 80799 M{\"u}nchen, Germany}

\author{Sanjay Moudgalya}
\affiliation{
Technical University of Munich, 
TUM School of Natural Sciences, 
Physics Department,
James-Franck-Str. 1,
85748 Garching,
Germany}%
\affiliation{Munich Center for Quantum Science and Technology (MCQST), Schellingstr. 4, 80799 M{\"u}nchen, Germany}

\author{Olexei Motrunich}
\affiliation{Department of Physics and Institute for Quantum Information and Matter, California Institute of Technology, Pasadena, CA 91125, USA}
\affiliation{Walter Burke Institute for Theoretical Physics, California Institute of Technology, Pasadena, CA 91125, USA}

\date{\today}

\begin{abstract}
Continuous symmetries lead to universal slow relaxation of correlation functions in quantum many-body systems.
In this work, we study how local symmetry-breaking impurities affect the dynamics of these correlation functions using Brownian quantum circuits, which we expect to apply to generic non-integrable systems with the same symmetries.
While explicitly breaking the symmetry is generally expected to lead to eventual restoration of full ergodicity, we find that approximately conserved quantities that survive under such circumstances can still induce slow relaxation.  
This can be understood using a super-Hamiltonian formulation, where low-lying excitations determine the late-time dynamics and exact ground states correspond to conserved quantities.
We show that in one dimension, symmetry-breaking impurities modify diffusive and subdiffusive behaviors associated with U$(1)$ and dipole conservation at late-times, e.g., by increasing power-law decay exponents of the decay of autocorrelation functions.
This stems from the fact that for these symmetries, impurities are relevant in the renormalization group sense, e.g., bulk impurities effectively disconnect the system, completely modifying both temporal and spatial correlations.
On the other hand, for an impurity that disrupts strong Hilbert space fragmentation, the super-Hamiltonian only acquires an exponentially small gap, leading to prethermal plateaus in autocorrelation functions which extend for times that scale exponentially with the distance to the impurity.
Overall, our approach systematically characterizes how symmetry-breaking impurities affect relaxation dynamics in symmetric systems. 

\end{abstract}

\maketitle

\tableofcontents

\section{Introduction}
Generic quantum many-body systems out of equilibrium eventually equilibrate in a sense that the behavior of extensive local observables can be understood in terms of standard thermodynamic ensembles (see e.g., Ref.~\cite{1991_Deutsch, Srednicki_1999, rigol_thermalization_2008,  2016_Alessio,2016_eisert_thermalization,Mori_2018}).
The approach to this equilibrium is particularly interesting in the presence of symmetries and constraints, which lead to a rich phenomenology of universal behaviors in the non-equilibrium dynamics of quantum many-body systems, which are captured by the theory of hydrodynamics~\cite{kadanoff1963hydro, crossley2017effective, glorioso2018lectures, landry2020thecoset}.
For example, generic short-range interacting systems with a U$(1)$ symmetry generically exhibit universal diffusive behavior at late times, which can be effectively described by classical hydrodynamics~\cite{Mukerjee06, Rosch13, Bohrdt16,leviatan2017quantum, Rakovszky18,Khemani18Hyd}.
More exotic behavior, such as anomalous diffusion, can be obtained in the presence of more exotic symmetries, such as dipole or higher moment conservation~\cite{guardadosanchez2020subdiffusion,Gromov_2020,  2020_Pablo_automato, Zhang_2020_subdiffusion,  Morningstar_2020, moudgalya2021spectral,Iaconis_2021, glorioso2023goldstone,ogunnaike2023unifying,morningstar2023hydrodynamics}, subsystem and general types of spatially-modulated symmetries~\cite{Iaconis_2019,spat_mod_2022}, or the presence of various types of constraints~\cite{singh2021subdiffusion,feldmeier2021critically}.
Even more exotic behavior that completely precludes thermalization and the relaxation of observables to their thermodynamic values can be achieved in systems exhibiting strong Hilbert space fragmentation~\cite{2020_sala_ergodicity-breaking, 2020_khemani_local, Moudgalya_2021_thermalization, 2020_Iadecola_HSF, 2022_Moudgalya_review_HSF_SCAR, papic2021review}, which can appear due to simple dynamical constraints. 

\begin{figure*}[bt]
\includegraphics[width=1\linewidth, scale=1]{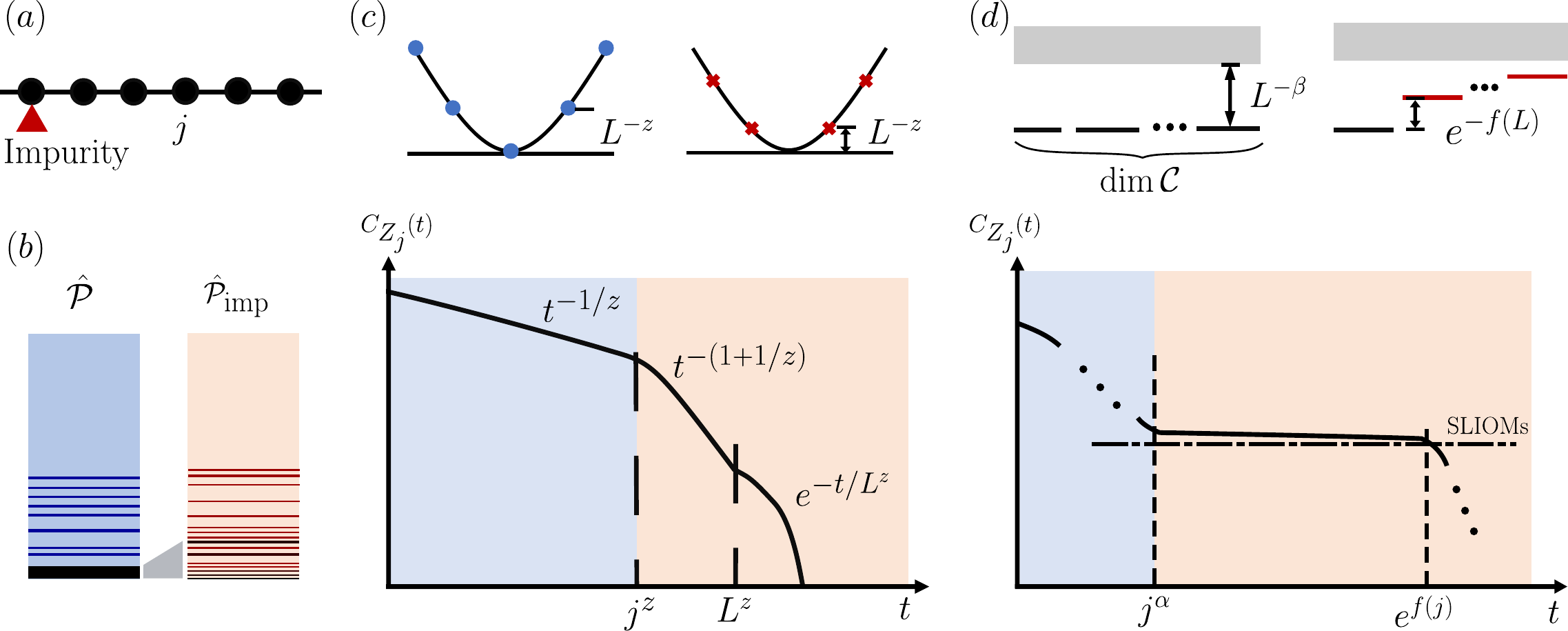}
\caption{ \textbf{Effect of symmetry-breaking impurities.}
(a) A one-dimensional chain with a local symmetry-breaking impurity (red triangle) at the boundary.
(b) Energy spectra of the super-Hamiltonians.
Without any impurity, the super-Hamiltonian $\hat{\mathcal{P}}$ has (possibly highly) degenerate zero-energy ground states corresponding to (exact) conserved quantities. 
With an impurity, the ground state degeneracy is lifted, modifying the low-lying spectrum of $\hat{\mathcal{P}}_{\mathrm{imp}}$, which in turn alters the long-time behaviors.
(c) Nature of energy spectra and autocorrelation functions $C_{Z_j}(t)$ at site $j$ for charge- or dipole-conserving systems with fully symmetry-breaking impurities. 
The energy spectrum of relevant eigenstates that contribute to correlation functions, which correspond to approximately conserved quantities (blue dots for unperturbed eigenstates and red crosses for eigenstates under the perturbation).
The energy gap scales as $\mathcal{O}(L^{-z})$ with or without impurity, but the structure of the approximately conserved quantities is modified, with $z$ the corresponding dynamical exponent.
The autocorrelation function $C_{Z_j}(t)$ 
 (shown in log-log scale) exhibits three regimes: 
(i) At $t \lesssim \mathcal{O}(j^z)$ before the impurity takes effect, it decays as $t^{-1/z}$.
(ii) In the time window $\mathcal{O}(j^z) \lesssim t \lesssim \mathcal{O}(L^z)$, the impurity has taken effect and it decays as $j^z t^{-(1+1/z)}$.
(iii) At timescales $t\gtrsim \mathcal{O}(L^z)$, the correlation decays exponentially as $j^z L^{-(z+1)} e^{-t/L^z}$ due to the finiteness of the system.
(d) Nature of the energy spectrum and autocorrelation functions of strongly fragmented systems with fully symmetry-breaking impurities.
The exponentially large unperturbed ground state degeneracy ($\dim \mathcal{C}~\sim \mathcal{O}(e^{L})$) is lifted in the presence of symmetry-breaking impurities, opening up an energy gap $\mathcal{O}(e^{-f(L)})$, where $f(x)$ depends on the specific model.
The gray area indicates a continuous spectrum with a polynomial gap $L^{-\beta}$ (distance between the lowest black solid line and the lower part of the gray area).
The autocorrelation function then exhibits the following regimes: 
(i) At $t\lesssim O(j^\alpha)$, it decays to a prethermal plateau whose value is mainly determined by the SLIOMs localized around the evaluation point $j$.
(ii) The prethermal plateau lasts exponentially long for a time determined by the distance $j$ to the impurity, i.e., for $O(j^\alpha) \lesssim t \lesssim \mathcal{O}(e^{f(j)})$.
This is exponentially long in system size for $j = x L$ keeping $x$ fixed making the flat plateau window sharply defined. 
(iii) 
Finally, at the longest times the autocorrelation function decays exponentially at times $t \gtrsim \mathcal{O}(e^{f(L)})$ when all approximately conserved quantities have decayed.
The possible emergent ``hydrodynamic" behaviors connecting these different regimes depends on the specifics of the strongly-fragmented models and lies beyond the scope of this work. 
}
\label{fig:schematic}
\end{figure*}

A relevant question is to understand how much of this phenomenology survives when the symmetries are explicitly broken.
Generically, the presence of an extensive symmetry-breaking perturbation is expected to lead to quick thermalization to non-symmetric thermodynamic ensembles.
What we mean more specifically is that even in thermodynamically large systems we expect a finite thermalization time.
This time can be large for small perturbation strengths, e.g., in generic systems this time is expected to grow polynomially with the inverse perturbation strength~\cite{Berges2004Prethermalization, Kollar2011Generalized, Langen2016Prethermalization, Krishnanand_2019_prethermalization, Krishnanand_2021_prethermalization, 2019_Krishnanand_floquet}.
In certain systems governed by the rigorous theory of prethermalization~\cite{abanin_rigorous_2017, Abanin_2018_floquet_prethermalization, Chao_prethermalization_2023, abdelshafy2025onsetquantumchaosergoditicy}, the thermalization time can diverge exponentially with the inverse of perturbation strength in isolated systems or in Floquet systems with the driving frequency~\cite{Abanin_2018_floquet_prethermalization, Francisco_2019_floquet_heating, Takashi_2016_floquet, kuwahara_floquetmagnus_2016}.
Moreover, for systems with Hilbert space fragmentation that possess higher-form symmetries, the thermalization timescale is exponentially long in the inverse of the perturbation strength for arbitrary $k$-body perturbations~\cite{2022_stephen_ergodicity_hsf, 24_Stahl_quantum_loop_frag, 2025_Khudorozhkov_group_valued_loop_model}.
Nevertheless, in all these cases with extensive perturbations, the thermalization time is independent of the system size.
However, the thermalization can also be unexpectedly slow with system size if the symmetry-breaking perturbation is spatially localized---even with $\mathcal{O}(1)$ perturbation strength---as illustrated by Fig.~\ref{fig:schematic}(a).
For example, certain systems where strong Hilbert space fragmentation is broken with a boundary impurity have been shown to exhibit thermalization that is exponentially slow in system size, which can be understood from the existence of various bottlenecks in the connectivity of states in the Hilbert space~\cite{han2024exponentially, wang2025exponentiallyslowthermalization1d}.

In this work, we provide a systematic method to study the dynamics of correlation functions under local symmetry-breaking impurities for noisy Brownian circuits~\cite{lashkari2013towards,bauer2017stochastic,  sunderhauf2019quantum, bauer2019equilibrium, xu2019locality, jian2021note, ogunnaike2023unifying, moudgalya2024symmetries}.
In the absence of impurity, a powerful method for quantitatively studying hydrodynamics for a wide range of symmetries and constraints is using a superoperator formalism. 
Namely, the averaged operator dynamics of the Brownian circuits can be described by a super-Hamiltonian on the double Hilbert space, which has the form of a Lindbladian~\cite{ogunnaike2023unifying, moudgalya2024symmetries}.
The conserved quantities correspond to the ground states of the super-Hamiltonian, while the late-time dynamics are governed by the low-lying excited states. 
As we discuss in this work, this formalism naturally extends to the case where local symmetry-breaking impurities are added to the Brownian circuit, which produces a perturbation of the super-Hamiltonian.
This results in a modification of the low-lying excited states of the super-Hamiltonian of the perturbed system, which can be interpreted as approximately conserved quantities that govern the late-time dynamics in the presence of impurities.
One possibility is that these low-energy states are modifications of the original hydrodynamic modes (low-energy excitations of the original super-Hamiltonian), as shown in the upper part of Fig.~\ref{fig:schematic}(c). 
Another possibility is that they can be related to unperturbed degenerate ground states, which correspond to conserved quantities of the unperturbed system, that become weakly lifted under the addition of the impurity, as schematically shown in the upper part of Fig.~\ref{fig:schematic}(d).
This systematic approach to identifying all relevant approximate conserved quantities allows us to provide a detailed characterization of the effects associated with a symmetry-breaking impurity, e.g., how the effects of the impurity are felt at varying distances from it as measured by local autocorrelation functions [setup shown schematically in Fig.~\ref{fig:schematic}(a)].
We expect the qualitative behaviors observed in this work to universally apply to generic non-integrable systems (in the absence of energy conservation or for finite energy-density initial states) with the same symmetries in the presence of local symmetry-breaking impurities, e.g., quantum random circuits and classical cellular automaton dynamics.
We apply this framework to a number of symmetries.
We first show that hydrodynamic modes in U($1$) conserving systems with a local impurity can be mapped to a single-particle problem with an absorbing boundary, where the standard diffusive behavior (autocorrelations decaying as $t^{-1/2}$) transits to that of a system with an absorbing boundary at late times (with autocorrelations decaying as $t^{-3/2}$ once the impurity effects are felt at the observation location).
Next, we consider systems with a combination of charge and dipole conservation.
We reproduce the subdiffusive behavior with dynamical exponent $z=4$ (power law decay $t^{-1/4}$) in the case without impurity~\cite{guardadosanchez2020subdiffusion,Morningstar_2020,Zhang_2020_subdiffusion, Gromov_2020,2020_Pablo_automato} by mapping to (an approximate) single-particle problem for the corresponding hydrodynamic mode. 
Then we show the emergent boundary conditions that arise in the presence of two kinds of local impurities: one that breaks only the dipole conservation, or one that breaks both charge and dipole moment conservation.
Surprisingly, we find that the former kind of impurity does not affect the dynamical exponent of the subdiffusive behavior, while the latter leads to decay of charge correlations with a power-law $t^{-5/4}$.
In the superoperator language, these results can all be attributed to the modification of the structure of the original hydrodynamic modes (low-energy excitations) as a consequence of the impurity perturbation.
Finally, we investigate locally breaking strong Hilbert space fragmentation using two canonical examples: the $t-J_z$ model and the dipole-conserving spin-$1$ model with $3$-local interactions~\cite{2020_SLIOMs}. 
For both models, an impurity at the boundary leads to prethermal plateaus of correlation functions, which is related to the statistically local integrals of motions (SLIOMs) that are exponentially (statistically) localized at various locations in the system~\cite{2020_SLIOMs}.
SLIOMs that are localized furthest away from the impurity set the exponentially long (in system size) time scale for the decay of the corresponding longest-lasting boundary autocorrelation plateau, which can be associated with the restoration of ergodicity across the entire system.
On the other hand, SLIOMs that are localized in the bulk of the system set the time scale for the restoration of ergodicity at their location, which takes an exponentially long time scaling with the distance to the impurity.
Different natures of the bulk SLIOMs in the $t-J_z$ and $3$-local dipole conserving spin-$1$ models lead to qualitative differences in such bulk ergodicity restoration between the two cases, even though both show a strong Hilbert space fragmentation in the absence of an impurity. 
In the super-Hamiltonian language, the exponentially large ground state degeneracy associated with fragmentation is lifted in the presence of a symmetry-breaking impurity, leading to an exponentially small (in system size) gap.
The corresponding exponentially slow modes (which we show are related to the SLIOMs) dominate the long-time dynamics, leading to prethermal plateaus in charge autocorrelation functions, as shown in Fig.~\ref{fig:schematic}(d).
The super-Hamiltonian language is rather general and allows for more detailed analysis in fragmented systems going beyond results in  Refs.~\cite{han2024exponentially, wang2025exponentiallyslowthermalization1d}, and in App.~\ref{app:rel_graph} we show that the graph theory interpretation there reduces to a particular perturbative calculation of the super-Hamiltonian.
In all of the above cases, we validate our analytic super-Hamiltonian approach predictions by performing extensive numerical simulations using classical cellular automata that exactly realize the relevant symmetries (see e.g., Refs.~\cite{Iaconis_2019,  2020_Pablo_automato, spat_mod_2022,2022_Lehmann_Pablo_Markov,Iaconis_2021, Morningstar_2020, Hart_2022}).
Since in this paper we deal with systems with Abelian symmetries (i.e., Abelian commutant algebras) that are diagonal in the computational basis, we expect that the corresponding cellular automata have the same qualitative long-time dynamics, providing unbiased checks on the analytical predictions.

The rest of this paper is organized as follows.
In Sec.~\ref{sec:review} we review the superoperator formalism to study the hydrodynamics of Brownian circuits.
In Sec.~\ref{sec:U1} we introduce our method with the breaking of U($1$) charge conservation with a local impurity.
Then n Sec.~\ref{sec:dipole} we study the effect of two kinds of impurities in the charge- and dipole-conserving system, which breaks only dipole conservation or both charge and dipole conservation. 
In Sec.~\ref{sec:fragmentation}, we investigate the relaxation of strongly fragmented systems under local impurities, and compare the results to those obtained for weakly fragmented systems.
We summarize our results and discuss open questions in Sec.~\ref{sec:conclude}.
We present many technical details in the appendices.

\section{Review: Hydrodynamics of symmetric Brownian circuits}\label{sec:review}
We consider the Brownian circuit dynamics with a time-dependent Hamiltonian of the form 
\begin{equation}\label{eq:Brownian_circuit}
    H = \sum_\alpha J_\alpha^{(t)} h_\alpha. 
\end{equation}
Here $\{J_\alpha^{(t)}\}$ are uncorrelated random variables chosen from a Gaussian distribution for each time step $\Delta t$, with zero mean $\overline{J_\alpha^{(t)}}=0$ and variance $\overline{J_{\alpha}^{(t)} J_{\alpha'}^{(t^\prime)}} = \sigma^2 \delta_{t,t^\prime}\delta_{\alpha,\alpha^\prime}$.
We consider $\sigma^2 = 2/\Delta t$. This distribution is referred to as shot noise in the limit of $\Delta t \rightarrow 0$~\cite{majumdar2005brownian}. 
As we review below, the symmetries and their effect on the evolution of correlation functions under the Brownian circuit can be efficiently studied in the superoperator picture, where the average operator dynamics is governed by certain \textit{superoperators} that act on two copies of the Hilbert space. 

\subsection{Symmetries}\label{subsec:symmetries}
The symmetries of the family of Brownian circuits $H$ in Eq.~\eqref{eq:Brownian_circuit} can be characterized by the bond and commutant algebras~\cite{2007_Read_Commutant, moudgalya2024symmetries, li2023hilbert}. 
This algebraic language is a generalized framework that captures many kinds of symmetries including conventional symmetries~\cite{2000_Zanardi_quantuminfo, 01_Paolo_virtual_quantum_subsystems, 2007_Bartlett_SU2_decompose_Hilbert_space, moudgalya2022from}, quantum many-body scars~\cite{moudgalya2022exhaustive, hartse2024stabilizerscars}, and Hilbert space fragmentation~\cite{2007_Read_Commutant, moudgalya_hilbert_2022, 2024_Parameswaran_SY_commutant, li2023hilbert}, into a single framework, and can also be used to understand their effect on closed and open quantum systems~\cite{2007_Bartlett_SU2_decompose_Hilbert_space, li2023hilbert,li2024highlyentangledstationarystatesstrong, 2023_Zanardi_scambling_algebra}.
Given the family of Brownian circuits $H$, we can define the bond algebra $\mathcal{A}$ as being generated by linear combinations of arbitrary products of local terms $h_\alpha$.
Symmetries are then given by the \textit{commutant} of the bond algebra, $\mathcal{C}$, which we refer to as the commutant algebra, and this consists of all operators that commute with every $h_\alpha$.
These algebras are given by
\begin{equation}
    \mathcal{A} = \lgen \{h_\alpha\}, \mathbb{1} \rgen,\,\,\, \mathcal{C} = \{O: [h_\alpha, O] = 0, \forall j\}.
\end{equation}
For example, for a spin-$1/2$ chain and $\{h_\alpha\}$ chosen to be generic terms with U$(1)$ conservation $S_{\mathrm{tot}}^z = \sum_j S_j^z$ (e.g., $\{h_\alpha\} = \{X_j X_{j+1} + Y_j Y_{j+1}, Z_{j}\}$ where $X_j$, $Y_j$, $Z_j$ are the Pauli operators on site $j$), the commutant algebra is given by $\mathcal{C} = \mathrm{span}\{\mathbb{1}, S_{\mathrm{tot}}^z, (S_{\mathrm{tot}}^z)^2, \ldots, (S_{\mathrm{tot}}^z)^{L}\}$, where $L$ is the system size.
Hence such a Brownian circuits is said to have $L+1$ linearly independent conserved quantities.
We now review how these symmetries in $\mathcal{C}$ can be understood as ground states of certain superoperators~\cite{moudgalya2023numerical, moudgalya2024symmetries}, which will be helpful in connecting them to Brownian circuits.
First, the operators acting on Hilbert space $\mathcal{H}$ can be mapped to the states in the double Hilbert space $\mathcal{H}_t \otimes \mathcal{H}_b$ using Choi's isomorphism~\cite{1975_Choi}
\begin{equation}
    O = \sum_{ij} o_{ij} \ket{i}\bra{j} \Longleftrightarrow |O) = \sum_{ij} o_{ij} \ket{i}_t \ket{j}_b,
\end{equation}
with an orthonormal basis $\{\ket{i}_t\} \in \mathcal{H}_t$ and $\{\ket{j}_b\} \in \mathcal{H}_b$.
Here we adopt the bilayer interpretation and label the two parts of the double Hilbert space with subscripts $t$ and $b$, which refer to the `top' and `bottom' layers, respectively.
The inner product of the double Hilbert space is defined as $ (A|B)\equiv \mathrm{Tr}(A^\dagger B)$.
The superoperators acting on the operators can be denoted as 
\begin{equation}
    O_{\alpha; t} := O_{\alpha} \otimes \mathbb{1},\;\; O_{\alpha; b} := \mathbb{1} \otimes O_{\alpha},
\end{equation}
with $O_{\alpha;t(b)}$ acting on $\mathcal{H}_{t(b)}$.
Therefore, the commutator of two operators interpreted as a state on the doubled Hilbert space, is given by
\begin{equation}
|[h_\alpha, O]) \Longleftrightarrow  \hat{\mathcal{L}}_\alpha |O) := (h_{\alpha;t} - h_{\alpha;b}^T) |O),
\end{equation}
with $T$ being the transposition with respect to a fixed orthonormal basis (in this case the computational basis).
Therefore, elements of the commutant algebra $O\in\mathcal{C}$ satisfy $\hat{\mathcal{P}}_\alpha|O) = \hat{\mathcal{L}}_\alpha^\dagger \hat{\mathcal{L}}_\alpha |O) = 0$. 
Consider the super-Hamiltonian corresponding to the local Hermitian terms $h_{\alpha} \in\mathcal{A}$, 
\begin{equation}
    \hat{\mathcal{P}} = \sum_{\alpha} \hat{\mathcal{P}}_{\alpha} = \sum_\alpha (h_{\alpha;t}^2 + (h_{\alpha;b}^T)^2 - 2h_{\alpha; t} h_{\alpha;b}^T).
\label{eq:superhamiltonianexpr}
\end{equation}
note that its action $\hat{\mathcal{P}}|O)$ on a state $|O)$ in $\mathcal{H}_t \otimes \mathcal{H}_b$ maps to $ -\sum_{\alpha} (2h_\alpha O h_\alpha - \{h_\alpha^2, O\})$, which also corresponds to the dissipation term in a Lindblad evolution (with a minus sign)~\cite{ogunnaike2023unifying}.
Since the super-Hamiltonian $\hat{\mathcal{P}}$ is a sum of positive semi-definite Hermitian terms $\hat{\mathcal{P}}_\alpha$, it is positive semi-definite as a whole, and has a real energy spectrum with energies $E\geq 0$. 
This property also guarantees that elements of the commutant algebra map to the ground states of the super-Hamiltonian at zero energy,
\begin{equation}
    \hat{\mathcal{P}}|O) = \sum_\alpha \hat{\mathcal{P}}_\alpha |O)= 0 ~~\Leftrightarrow~~ \hat{\mathcal{P}}_\alpha|O) = 0\;\; \forall \alpha.
\end{equation}
\subsection{Two-point correlation functions}

In addition to the ground states of $\hat{\mathcal{P}}$ having an interpretation as symmetries, its low-energy eigenstates are the hydrodynamic modes that govern the long-time operator dynamics, as we review below.
We consider the infinite-temperature correlation function averaging over random variables $\{J_\alpha^{(t)}\}$, 
\begin{equation} \label{eq:Inn_prod}
    \overline{C_{B,A}(t)} = \frac{1}{\dim (\mathcal{H})} \overline{\mathrm{Tr}\big(B^\dagger A(t)\big)} = \frac{1}{\dim (\mathcal{H})} \overline{(B|A(t))},
\end{equation}
where $\overline{\cdots}$ denotes the ensemble average over different circuit realizations.
The average operator evolution is given by $\overline{|O(t+\Delta t))} = \overline{e^{i\Delta t\sum_\alpha J_\alpha^{(t)}\mathcal{L}_\alpha }}|O)$, leading to a continuous-time ($\Delta t \rightarrow 0$) evolution of the form~\cite{lashkari2013towards, sunderhauf2019quantum, bauer2019equilibrium, xu2019locality, jian2021note, ogunnaike2023unifying, moudgalya2024symmetries}
\begin{equation} \label{eq:O_t}
    \overline{|O(t))} = e^{- \hat{\mathcal{P}}t} \overline{|O(0))},
\end{equation}
where $\hat{\mathcal{P}}$ is precisely the super-Hamiltonian of Eq.~(\ref{eq:superhamiltonianexpr}).
Using the eigenspectrum of $\hat{\mathcal{P}}$, the average correlation function can be expressed as
\begin{equation}
\label{eq:CBAt}
\begin{aligned}
    \overline{C_{B,A}(t)} &= \frac{1}{\dim (\mathcal{H})} (B|e^{- \hat{\mathcal{P}}t}|A),\\
    &= \frac{1}{\dim (\mathcal{H})} \sum_{\mu} e^{- E_\mu t}(B|E_\mu)(E_\mu|A),
\end{aligned}
\end{equation}
where $|E_{\mu})$ are the orthonormal eigenstates of $\hat{\mathcal{P}}$ with energy $E_\mu$.
In the long time limit, the correlation function saturates to 
\begin{equation}\label{eq:Brownian_AC_Mazur}
    \lim_{t\rightarrow\infty} \overline{C_{B,A}(t)} = \frac{1}{\dim (\mathcal{H})} \sum_{\mu} \delta_{E_\mu,0} (B|E_\mu)(E_\mu|A),
\end{equation}
which is given by the overlap of operators $A$ and $B$ with the zero energy eigenstates, i.e., the symmetry operators discussed in Sec.~\ref{subsec:symmetries}, which are elements of the commutant. 
Equation~\eqref{eq:Brownian_AC_Mazur} reduces to the well-known Mazur bound of the autocorrelation function~\cite{MAZUR1969, Mazur_2021}, which is the long-time average of $C_{A,A}(t)$.

Low-lying excited eigenstates of the super-Hamiltonian correspond instead to \textit{approximate} conserved quantities, which give rise to long-time hydrodynamic tails in correlation functions of operators that have non-zero overlap with them. 
For example, as we will discuss for the case of U$(1)$ symmetric Brownian circuits, the low-lying excitations have the form of spin-waves and lead to a gap that vanishes with system size as $\sim 1/L^2$,  signifying diffusive behavior~\cite{ogunnaike2023unifying, moudgalya2024symmetries}.
Such analysis of low-energy excitations has also been used to find universal subdiffusive behavior in the presence of dipole moment conservation~\cite{ogunnaike2023unifying, moudgalya2021spectral}, as well as other types of symmetries and fragmentation~\cite{moudgalya2024symmetries}.
\subsection{Stochastic cellular automaton dynamics}\label{subsec:CA}
Before we proceed to the analysis of super-Hamiltonians in the presence of impurities, we present an overview of the complementary non-perturbative numerical methods we use in this work.
Since we focus on examples of symmetries which admit an eigenbasis of product states, such as U($1$), dipole conservation, and classical fragmentation, i.e., those that can also exist in classical Markov processes, we utilize stochastic cellular automata to numerically study the dynamics of correlation functions similarly to Refs.~\cite{Iaconis_2019,  2020_Pablo_automato, spat_mod_2022,2022_Lehmann_Pablo_Markov,Iaconis_2021, Morningstar_2020, Hart_2022} with and without the impurity.
A stochastic cellular automaton employs the update rules that satisfy the symmetries and constraints of the quantum evolution, and allows for large-scale simulations as follows.
Consider a string of classical spins $\{s\} = \{s_1, s_2, \ldots, s_L\}$, with $s_j \in \{1, 2, \ldots m\}$ for local Hilbert space dimension $m$ and $j = 1,\ldots, L$. 
At each discrete time step, $\{s\}$ is updated to $\{s^\prime\}$, where $\{s^\prime\}$ is randomly chosen from those configurations that are allowed by symmetry constraints determined by $\{h_{\alpha}\}$. 
To be more specific, consider two-site gates $G_{i,i+1}$ for a U($1$) charge-conserving system with spin-$1/2$ or two-level system per site.
Such a two-site gate $G_{i,i+1}$ maps between configurations $\{s\}$ and $\{s^\prime\}$, where $\{s_i, s_{i+1}\} =\{0, 1\}_{i,i+1}$ and $\{s_i^\prime, s_{i+1}^\prime\} =\{1, 0\}_{i,i+1}$, leaving other configurations unchanged. 
For each local gate $G_{i,i+1}$, and if the configuration $\{s_i, s_{i+1}\}$ is given by $\{0, 1\}_{i,i+1}$, it is flipped to $\{1, 0\}_{i,i+1}$ with probability $1/2$, and remains unchanged with probability $1/2$; and similarly for $\{s_i, s_{i+1}\} = \{1, 0\}_{i,i+1}$.
At each discrete time step, we randomly select a layer of fully-packed local gates, either $\{G_{i\in\text{odd}}\}$ or $\{G_{i\in\text{even}}\}$, with equal probability $1/2$.
Adding a local symmetry-breaking impurity can be treated similarly.
For example, in charge-conserving systems with a charge-breaking impurity at $j_s$, we consider the impurity gate $G^{\text{imp}}_{j_s}$ that maps $\{s_{j_s}\} = \{0\}_{j_s}$ to $\{1\}_{j_s}$ with probability $1/2$ and keeps the configuration unchanged with probability $1/2$; and similarly for ${\{s_{j_s}\} =} \{1\}_{j_s}$.
At each time step, we randomly select either a layer of gates $\{G_{i \in \text{odd}}\}$, $\{G_{i \in \text{even}}\}$, or the impurity gate $G_{j_s}^{\text{imp}}$ on a single site, each with equal probability $1/3$.
Note that this implementation of the cellular automaton dynamics guarantees detailed balance with respect to the infinite-temperature (i.e., uniform) probability distribution over the spin states.
The correlation functions are then calculated by averaging over random realizations of initial configurations.

\section{Locally breaking U$(1)$ symmetry}\label{sec:U1}
As a starting example, we first recover standard diffusion due to a U$(1)$ symmetry using the superoperator formalism, including in the presence of a charge-conserving boundary.
We then investigate the effect of a local symmetry-breaking impurity, showing how the impurity modifies the long-time hydrodynamic behavior.
Throughout, we will consider a spin-$1/2$ XY chain of length $L$ with bond algebra
\begin{equation}\label{eq:bond_alg_U1}
    \mathcal{A}_{\mathrm{U}(1)} = \lgen \mathbb{1}, \{X_j X_{j+1} + Y_{j}Y_{j+1}\}, \{Z_j\} \rgen,
\end{equation}
whose commutant is given by
\begin{equation}\label{eq:comm_alg_U1}
    \mathcal{C}_{\mathrm{U}(1)} = \lgen Z_{\mathrm{tot}} \rgen = \mathrm{span}\{\mathbb{1}, Z_{\mathrm{tot}}, (Z_{\mathrm{tot}})^2,\ldots, (Z_{\mathrm{tot}})^L\},
\end{equation}
namely the algebra generated by the total magnetization $Z_{\mathrm{tot}} = \sum_{j=1}^L Z_j$, and spanned by $L+1$ linearly independent conserved quantities~\cite{moudgalya_hilbert_2022, moudgalya2022from}.
\subsection{Diffusion in the Superoperator Formalism}
In this subsection, we review the superoperator derivation of late-time diffusion due to a U($1$) conservation law as discussed in Refs.~\cite{ogunnaike2023unifying, moudgalya2024symmetries}, including new results which did not appear in any of those references.
We first illustrate it on a lattice, and then discuss taking the continuum limit; the latter will be useful when working with impurities in subsequent sections.

\subsubsection{Super-Hamiltonian on a Lattice}
Consider the super-Hamiltonian $\hat{\mathcal{P}}_{\mathrm{U}(1)}$ corresponding to $\mathcal{A}_{\mathrm{U}(1)}$, and denote by $\ket{\sigma}_j$ with $\sigma=\uparrow,\downarrow$ the basis states in the local $Z$-basis.
Then, ground states $|O)$ of $\hat{\mathcal{P}}_{\mathrm{U}(1)}$ satisfy $\hat{\mathcal{P}}_{Z_j} |O) = (Z_{j;t}-Z_{j;b})^2 |O) = 0$ for all $j$, which enforces $|\sigma\rangle_{j;t} = |\sigma\rangle_{j;b}$ on all lattice sites.
Thus, all ground states lie in the composite spin subspace spanned by $\otimes_j \ket{\widetilde{\sigma}}_j$, with the following composite spins defined on the rungs of the ladder
\begin{equation}
    |\widetilde{\uparrow}\rangle \defn \begin{pmatrix} \uparrow \\ \uparrow \end{pmatrix}, \quad |\widetilde{\downarrow}\rangle \defn \begin{pmatrix} \downarrow \\ \downarrow \end{pmatrix}.
\label{eq:compositespins}
\end{equation}
In this composite subspace, the super-Hamiltonian takes the form~\cite{moudgalya2024symmetries}
\begin{equation}\label{eq:Pcomp_U1}
    \mathcal{\hat{P}}_{\mathrm{U}(1)|\mathrm{comp}} = 4\sum_j \left(1-\widetilde{X}_{j}\widetilde{X}_{j+1}-\widetilde{Y}_{j}\widetilde{Y}_{j+1}-\widetilde{Z}_{j}\widetilde{Z}_{j+1} \right).
\end{equation}
Hence, for the bond algebra in Eq.~\eqref{eq:bond_alg_U1}, the super-Hamiltonian is the SU$(2)$-symmetric ferromagnetic Heisenberg chain, which is exactly solvable.
Notice nonetheless that a different set of generators of $\mathcal{A}_{\mathrm{U}(1)}$ (where we can also allow more local terms than required to generate the algebra) will lead to a different super-Hamiltonian, and hence it does not need to be either solvable or SU$(2)$ symmetric. 
However, since the ground states will be common to these different super-Hamiltonian instances, we expect to obtain the same qualitative results--- such as properties of the low lying spectrum\footnote{This can be seen by explicitly constructing trial wavefunctions for the low-energy excitations~\cite{ogunnaike2023unifying} or a continuum field theory for the low-energy spectrum~\cite{moudgalya2021spectral}.}---as long as the generators are local and their bond algebra faithfully captures the U$(1)$ symmetry.
Exploiting the solvability of the Heisenberg chain, one then indeed finds that the ground state manifold---the fully-polarized (ferromagnetic) multiplet---is $L+1$ degenerate, hence corresponding to the dimension of $\mathcal{C}_{\mathrm{U}(1)}$.
Moreover, for large system sizes $L$, low-lying excitations show an energy gap $\Delta \sim 1/L^2$, residing within the composite spin subspace.
In particular, only single-particle-like excitations (corresponding to a magnon excitation over a specific polarized state representing the identity operator) govern the behavior of spin-spin correlation functions of the form $C_Z(j,j_0;t)$.
Indeed, 
in the double space language and for a spin-$1/2$ degree of freedom,
\begin{equation}\label{eq:S_spins}
\begin{aligned}
    &Z = \ket{\uparrow}\bra{\uparrow} - \ket{\downarrow}\bra{\downarrow} \Longleftrightarrow |Z) = |\widetilde{\uparrow}\rangle - |\widetilde{\downarrow}\rangle,\\
    &\mathbb{1} = \ket{\uparrow}\bra{\uparrow} + \ket{\downarrow}\bra{\downarrow} \Longleftrightarrow |\mathbb{1}) = |\widetilde{\uparrow}\rangle + |\widetilde{\downarrow}\rangle.
\end{aligned}
\end{equation}
Hence, in the local $\widetilde{X}$-basis defined via
\begin{equation}\label{eq:X_spins}
\begin{aligned}
 \ket{\widetilde{\rightarrow}} := \frac{1}{\sqrt{2}}(|\widetilde{\uparrow}\rangle + |\widetilde{\downarrow}\rangle),\quad
 \ket{\widetilde{\leftarrow}} := \frac{1}{\sqrt{2}}(|\widetilde{\uparrow}\rangle - |\widetilde{\downarrow}\rangle),
\end{aligned}
\end{equation}
we have full Hilbert space operators mapping to states as $|\mathbb{1}) = 2^{L/2} \otimes_{j=1}^L |\widetilde{\rightarrow} \rangle_j$ and
\begin{equation}
\label{eq:Zj}
    |Z_j) = 2^{L/2} |\widetilde{\rightarrow} \ldots \widetilde{\rightarrow}_{j-1} \widetilde{\leftarrow}_j \widetilde{\rightarrow}_{j+1} \ldots \widetilde{\rightarrow}_L\rangle.
\end{equation}
Furthermore, $\mathcal{\hat{P}}_{\mathrm{U}(1)|\mathrm{comp}}$ takes the simple form
\begin{equation*}
\mathcal{\hat{P}}_{\mathrm{U}(1)|\mathrm{comp}} = 8 \sum_{j=1}^{L_{\text{max}}} (\ket{\trt \tlt} - \ket{\tlt \trt})(\bra{\trt \tlt} - \bra{\tlt \trt})_{[j, j+1]}, 
\end{equation*}
from which we see that it preserves the subspace spanned by $|Z_j)$.
Moreover, using that
\begin{equation} \label{eq:CZZ_formal}
    C_Z(j,j_0;t) = \frac{1}{\dim (\mathcal{H})} \sum_\mu e^{- E_\mu t} (Z_j|E_\mu)(E_{\mu}|Z_{j_0})
\end{equation}
with $\dim (\mathcal{H})=2^L$ the dimension of the Hilbert space, we also see that the only eigenstates of $\mathcal{\hat{P}}_{\mathrm{U}(1)|\mathrm{comp}}$ that contribute to the correlation $C_Z(j,j_0;t)$ are those that are spanned by $|Z_j)$.
Within this subspace, the super-Hamiltonian maps to the single-particle problem
\begin{align} \label{eq:H_U1}
\mathbb{H}_{\text{U(1)}} = 8 \sum_{j=1}^{L_{\text{max}}} &\left[ |j)(j| + |j+1)(j+1|\right. \nonumber \\
&\left.- |j)(j+1| - |j+1)(j| \right],
\end{align}
where for ease of notation we have denoted $|j) \defn |\widetilde{\rightarrow} \ldots \widetilde{\rightarrow}_{j-1} \widetilde{\leftarrow}_j \widetilde{\rightarrow}_{j+1} \ldots \widetilde{\rightarrow}_L\rangle$. 
In periodic boundary conditions (PBC, $L_\text{max}=L$), this is a uniform ``hopping'' with uniform ``on-site potential'', while in open boundary conditions (OBC, $L_\text{max}=L-1$), besides the absence of the hopping between sites $L$ and $1$, the on-site potentials on these sites is half of those on all the other sites.
The specific on-site potentials in both the PBC and OBC cases are such that $\sum_{j=1}^L |j)$ is an exact zero-energy eigenstate, corresponding to $\sum_{j=1}^L Z_j$ belonging to the commutant.
\subsubsection{Continuum Limits}\label{subsubsec:U1continuum}
Reference~\cite{moudgalya2024symmetries} discussed orbitals $\phi_k(j) \equiv (j|E_k)$ for the corresponding hopping problem and showed that the correlation function $C_Z(j,j_0;t)$ has the standard diffusion form in the thermodynamic limit.
Here we directly observe that because of the above simplification in terms of single-particle eigenstates, we can write a closed-form expression for the evolution equation of $C_Z(j,j_0;t)$, from which we will recover the expected long-time behavior.
Indeed, taking its time derivative we find that~\footnote{For OBC, we instead find $\frac{d}{dt} C_Z(j,j_0;t) = 8\Delta_xC_Z(j,j_0;t)$  close to the boundaries with $j=1,L$. Here, $\Delta_x$ is a finite difference operator defined by $\Delta_{x} C_Z(j,j_0;t) \equiv \pm (C_Z(j,j_0;t)-C_Z(j+1,j_0;t))$.}
\begin{equation}\label{eq:U1_CZZ}
\begin{aligned}
    \frac{d}{dt} C_Z(j,j_0;t) &= -\sum_{j_1=1}^L( j|\mathbb{H}_{\mathrm{U}(1)}|j_1) {C_Z(j_1,j_0;t)}  \\
    &=D^{\textrm{latt.}}\Delta_x^2 C_Z(j,j_0;t),
\end{aligned}
\end{equation}
with initial condition $C_Z(j,j_0;t=0)=\delta_{j,j_0}$ and the numerical factor $D^{\textrm{latt.}}=8$.
Here, $\Delta_x^2$ is the centered second finite-difference defined as $\Delta_x^2 C_Z(j,j_0;t) \defn C_Z(j-1,j_0;t)+C_Z(j+1,j_0;t)-2C_Z(j,j_0;t)$. 
Hence, the evolution of $C_Z(j,j_0;t)$ is governed by an unbiased random walk, which in the continuum limit recovers the diffusion equation 
\begin{equation}\label{eq:U1_cont}
\begin{aligned}
    \partial_tC_Z(x,x_0;t) &= D \partial_x^2 C_Z(x,x_0;t),
\end{aligned}
\end{equation}
where the diffusion constant $D$ captures microscopic details.
We take the initial condition to be $C_Z(x,x_0;t=0) = \delta(x-x_0)$, normalized so that the integrated ``initial density'' is $1$.
Hence, we have microscopically (rather than phenomenologically) derived the diffusion equation governing the behavior of two-point correlation functions.\footnote{\label{ft:charge_correlation}Note that the fact that the evolution of correlation functions is governed by the diffusion equation, which is also the evolution equation for the charge density in a U$(1)$ conserving system, is not accidental. 
In fact, the $C_Z(j,j_0,t)$ corresponds to a charge density-density correlation, where the ``total correlation function" is related to the global charge $\sum_j{Z_j}$, i.e., we have $\sum_{j}C_Z(j,j_0,t)=\langle Z_{\textrm{tot}}(t) Z_{j_0}\rangle$, which is conserved if the total charge is conserved. 
We will hence sometimes interpret the evolution of the correlation function starting from an initial $\delta$-function as an initial point charge spreading throughout the system.
}

For an infinite system, we can directly solve Eq.~(\ref{eq:U1_cont}) as
\begin{equation} \label{eq:CZ_diff_inf}
    C_Z(x,x_0;t) = \frac{1}{2\pi} \int_{-\infty}^\infty dk\, e^{-Dk^2t} \cos(k(x-x_0)),
\end{equation}
where we have used the structure of the eigenfunctions of the R.H.S. in Eq.~(\ref{eq:U1_cont}). 
For $x \sim x_0$, e.g., for autocorrelation functions, this already yields the diffusive scaling $C_Z(x, x_0; t) \sim t^{-\frac{1}{2}}$.
Alternatively, we can again exploit the relevance of the single-particle problem by noticing that $C_Z(j,j_0;t)$ in Eq.~\eqref{eq:CZZ_formal} becomes
\begin{equation} \label{eq:sing_part_C}
     C_Z(j,j_0;t)=\sum_k e^{- E_k t}\phi_k(j)\phi^*_k(j_0),
\end{equation}
with $\phi_k(j) \equiv (j|E_k)$ the normalized single-particle wavefunctions, which correspond to plane waves $\frac{1}{\sqrt{L}} e^{ikj}$ for PBC.
Hence, solving  Eq.~\eqref{eq:U1_CZZ} is equivalent (via separation of variables) to solving the single-particle problem $(j|\mathbb{H}_{\mathrm{U}(1)}|E_k) = E_k \phi_k(j)$, which is equivalent to
\begin{align}
    -&\Delta_x^2 \phi_k(j)=E_k \phi_k(j),
\end{align}
for all $j$, namely the spectrum of a free particle hopping on a ring.
Indeed, the spectrum of $\mathbb{H}_{\textrm{U}(1)}$ in Eq.~\eqref{eq:H_U1} on a system of size $L$ with PBC reads $E_k = 16\left[1 - \cos(k) \right]$, $k=\frac{2\pi n}{L}, n=0,1,\dots,L-1$.
Hence, when $k \ll \pi$, i.e., at long wavelengths, the dispersion becomes $E_k \propto k^2$, consistent with the fact that the second finite difference becomes a second derivative in the continuum $-\Delta_x^2\sim k^2\sim 1/L^2$. 
For OBC, one can deduce the relevant boundary conditions in the continuum at long wavelengths on a (semi-)finite system phenomenologically by noticing that $\frac{d}{dt}\int \, dx\ C_Z(x,x_0;t)=0$ at all times, namely the total U$(1)$ charge is conserved.
Using Eq.~\eqref{eq:U1_cont}, one finds the boundary condition $\left.\partial_xC_Z(x,x_0;t)\right|_{\textrm{boundary}}=0$.
Hence, we recover the diffusion equation with \textit{reflective} boundary conditions, which corresponds to vanishing current at the boundaries, leading to the solution~\cite{Balakrishnan2021}
\begin{equation} \label{eq:reflect_moi}
\begin{aligned}
C_Z(x,x_0;t)&=G(x-x_0,t) + G(x+x_0,t) \\
&= \frac{e^{-\frac{x^2+x_0^2}{4Dt}}}{\sqrt{\pi D t}} \cosh\left(\frac{x x_0}{2Dt}\right),
\end{aligned}
\end{equation}
with $G(x-x_0,t)=\frac{e^{-\frac{(x-x_0)^2}{4Dt}}}{\sqrt{4\pi D t}}$ the solution of the diffusion equation on an infinite system.\footnote{\label{ft:image_method} This is a specific application of the \textit{method of images}, which allows the solution of a differential equation on a semi-infinite system with certain type of boundary conditions in terms of the solution of a different initial condition on an infinite system. 
In Eq.~(\ref{eq:reflect_moi}), the solution of the diffusion equation on a semi-infinite line with an initial source at $x_0$ and reflective boundary conditions is given in terms of the solution of the diffusion equation on an infinite line with two initial sources at $x_0$ and $-x_0$.}

One can also deduce these boundary conditions in a more microscopic approach, by considering the relevant single-particle problem 
\begin{equation}
    E_k \phi_k(j) = -\Delta_x^2 \phi_k(j) \quad \mathrm{for~} j=2,\dots, L-1,
\end{equation}
i.e., in the bulk, and
\begin{align} \label{eq:bound_sp}
   &E_k \phi_k(1) = \phi_k(1) -\phi_k(2) \equiv -\Delta_x \phi_k(1) ,\\
   &E_k \phi_k(L) = \phi_k(L) -\phi_k(L-1) \equiv \Delta_x \phi_k(L-1)
\end{align}
at the boundary sites.
The eigenstates are then given by (see e.g., Ref.~\cite{moudgalya2024symmetries}) $\phi_k(j)=\sqrt{\frac{2}{L}}\cos(k(j-\frac{1}{2}))$ with $k=\frac{\pi n}{L}$,~ $1 \leq n \leq L-1$.
Hence, as anticipated by Eq.~\eqref{eq:bound_sp}, i.e., since  $E_k \propto k^2$, the derivative of $\phi_k(j)$ close to the (left) boundaries for sufficiently small $k$ anomalously scales as $\phi^\prime_k(j) \sim k^2$, which is parametrically smaller than the expected $\phi^\prime_k(j) \sim k$ found for a generic wavefunction with characteristic wave-vector $k$.
This can be understood as a lattice indication of the emergent boundary condition $\partial_x \phi_k(x)|_{x=0,L} = 0$ in the continuum limit at long wavelengths, hence imposing $\left.\partial_x{C_Z(x,x_0;t)}\right|_{x=0,L}=0$.

\subsection{Modifications due to a symmetry-breaking impurity}
\label{sec:U1_break}
We now extend the previous discussion to the case with a local symmetry-breaking impurity near site $j_s$, which, as we will see, can be interpreted as a \textit{sink} for the conserved quantity.
Specifically, we include a term of the form $J_{\text{imp}}^{(t)} V_{j_s}$ in the $U(1)$ conserving Hamiltonian as in Eq.~(\ref{eq:Brownian_circuit}) that forms the Brownian circuit, where $V_{j_s}$ is a term on site $j_s$ that breaks the $U(1)$ symmetry, and has a strength characterized by $\overline{(J_{\text{imp}}^{(t)})^2} \sim g$.
We then illustrate the effect of the impurity on the decay of two-point correlation functions, and uncover the relevant time scales and robustness of symmetry-imposed behavior when the symmetry is broken at a single point by the impurity.
In particular, starting from the bond algebra $ \mathcal{A}_{\mathrm{U}(1)} $ with commutant $\mathcal{C}_{\mathrm{U}(1)}$, we investigate the effects of including $V_{j_s}$ in $\mathcal{A}_{\mathrm{U}(1)}$.
This naively leads to a reduction of the commutant for any finite $g$, which corresponds to a reduction of the ground state degeneracy of the corresponding super-Hamiltonian, indicating that the two-point functions detect the breaking of the $U(1)$ symmetry as $t \rightarrow \infty$.
Nevertheless, as we discuss below, there are interesting dynamical features at large but finite $t$ that stem from the structure of low-energy excitations of the super-Hamiltonian. 
For concreteness, we add a single spin-flip $X_{j_s}$ at site $j_s$ of the chain, such that the bond algebra is now given by
\begin{equation}
    \mathcal{A}_{\mathrm{U}(1)|\mathrm{imp}} = \lgen \{X_j X_{j+1} + Y_{j}Y_{j+1}\}, \{Z_j\},  X_{j_s}\rgen.
\end{equation}
This leads to an additional local perturbation  
\begin{equation}
    X_{j_s;t}^2 + X_{j_s;b}^2 - 2X_{j_s;t}X_{j_s;b} = 
    2(1 - X_{j_s;t}X_{j_s;b}),
\end{equation}
to the super-Hamiltonian Eq.~\eqref{eq:Pcomp_U1}.
This perturbation preserves the composite-spin sector [see Eq.~(\ref{eq:compositespins})]; hence the full super-Hamiltonian within this sector reads\footnote{Note that the orthogonal complement to the composite spin sector retains at least the same gap as before adding the impurity, and we expect low-energy states of the super-Hamiltonian to be within the composite spin sector.}
\begin{equation}\label{eq:Pomp_U1_imp}
\begin{aligned}
    \hat{\mathcal{P}}_{\mathrm{U}(1)|\mathrm{imp}} &= \hat{\mathcal{P}}_{\mathrm{U}(1)|\mathrm{comp}} + 2g(1-\widetilde{X}_{j_s}).
\end{aligned}
\end{equation}
The spin-flip impurity $X_{j_s}$ completely breaks the original physical U$(1)$ symmetry, hence the corresponding term in the super-Hamiltonian lifts the $(L+1)$-fold ground state degeneracy of $\hat{\mathcal{P}}_{\text{U(1)}|\text{comp}}$ (which corresponds to the commutant algebra $\mathcal{C}_{U(1)}$), and only the identity operator $|\mathbb{1})$ remains as a ground state, since that this is the only operator in the commutant of $\mathcal{A}_{U(1)|\textrm{imp}}$.
Similar to the case without an impurity, the time-evolution of the spin-spin correlation is governed by an appropriate (hydro-mode) single-particle spectrum, with the Hamiltonian $\hat{\mathcal{P}}_{\mathrm{U}(1)|\mathrm{imp}}$ taking the form in the single-particle space
\begin{equation}\label{eq:U1_Heff_imp}
\begin{aligned}
    \mathbb{H}_{\mathrm{U}(1)|\text{imp}} &= \mathbb{H}_{\mathrm{U}(1)} + 4g|j_s)(j_s|,
\end{aligned}
\end{equation}
with $\mathbb{H}_{\mathrm{U}(1)}$ as given in Eq.~\eqref{eq:H_U1}.
Note that the impurity acts as an additional on-site potential for the hydro-mode particle.
On a system of size $L$, it is easy to show that this has a gap of at most $\mathcal{O}(L^{-2})$, e.g., by constructing trial wave-functions for the excited states.~\footnote{\label{ft:U1imp_variational_orbital}Specifically, we can construct a trial standing-wave-like state whose amplitude vanishes at $j_s$.
In a finite system of length $L$ the corresponding smallest wavevector $k$ is $O(1/L)$ and the expectation value of the ``kinetic energy'' $\mathbb{H}_{U(1)}$ in the corresponding state is $O(k^2) \sim O(1/L^2)$.

An explicit example for an open chain $[1,\dots,L]$ and an impurity at $j_s=1$ is $\phi_k^{\text{trial}}(j) \propto \sin[k(j-1)]$ with $k(L-1/2) = \frac{\pi}{2} + \pi n$, $n=0,1,\dots$, with $\epsilon^\text{trial} = 2 - 2 \cos(k)$.
Note that this trial energy is independent of $g$, but the mode has very different local amplitudes near the impurity compared to the case without the impurity.
This structure of the low-energy trial states corresponds directly to the emergent absorbing boundary conditions in the corresponding hydrodynamic diffusion equation discussed below.}
While one could analyze the precise nature of the excited states, it is convenient to analyze the dynamics directly from the time-evolution equation in the continuum limit.
Following a similar derivation as in the previous subsection, we then find that $C_Z(j,j_0;t)$ satisfies the following evolution equation on the lattice
\begin{equation}\label{eq:diff_sink}
\begin{aligned}
    \frac{d}{dt} C_Z(j,j_0;t) &= D^{\textrm{latt.}}\Delta_x^2 C_Z(j,j_0;t) - g C_Z(j,j_0;t) \delta_{j,j_s},
\end{aligned}
\end{equation}
where $D^{\textrm{latt.}} = 8$.
This is simply an unbiased random walk (lattice diffusion) in the presence of a sink (finite-rate absorber) at site $j_s$.
We now use Eq.~\eqref{eq:diff_sink} to make predictions about the effect of the symmetry-breaking impurity $X_{j_s}$ on the hydrodynamic tails that appear as a consequence of the original U$(1)$ conserved quantity. To simplify our analysis and aiming to predict the long-time behavior of the system, we take the continuum limit of Eq.~\eqref{eq:diff_sink} by taking $j$ to a continuous variable $x$ such that $j_s\to x_s, j_0\to x_0$, $D^{\textrm{latt.}}\Delta^2_ x\to D\partial_x^2$, and $g\delta_{j,j_s} \to \mathrm{g} \delta(x - x_s)$, where we distinguish the lattice diffusion constant and impurity strength from their continuum version $D$ and $\mathrm{g}$ respectively.
This leads to the following diffusion with an absorbing sink equation 
\begin{equation}\label{eq:diff_sink_cont}
\begin{aligned}
    \partial_t C_Z(x,x_0;t) &= D\partial_x^2 C_Z(x,x_0;t) - \mathrm{g} C_Z(x,x_0;t) \delta(x-x_s),
\end{aligned}
\end{equation}
where we have absorbed various factors, including the lattice spacing, in $D$ and $\mathrm{g}$.
In the following subsections, we analyze this continuum equation in various settings.
\subsection{Finite-strength impurity in an infinite system}\label{subsec:finiteimpinfinitesys}
Consider Eq.~\eqref{eq:diff_sink_cont} for an infinite system.
Mathematically, this equation means solving the free diffusion equation for $x>x_s$ and $x<x_s$ joining them with conditions $\left. C_Z(x, x_0; t)\right|_{x=x_s^+} = \left. C_Z(x, x_0; t)\right|_{x=x_s^-}$  and $D (\left.\partial_x C_Z(x, x_0; t)\right|_{x=x_s^+} - \left.\partial_x C_Z(x, x_0; t)\right|_{x=x_s^-}) = \left. \mathrm{g} C_Z(x, x_0; t)\right|_{x=x_s}$.
Similar to the case without the impurity, it is appropriate to choose the initial condition as $C_Z(x,x_0;t=0) = \delta(x-x_0)$.
With this, one can find (see e.g., Ref.~\cite{CHHABRA2020124573}) a closed-form solution given by
\begin{equation} \label{eq:Pxt_exact}
\begin{aligned}
C_Z(x,x_0;t)&=\frac{1}{\sqrt{4\pi Dt}}e^{-\frac{(x-x_0)^2}{4Dt}}\\&-\frac{1}{4\ell_{\mathrm{g}}}e^{-\frac{(|x-x_s|+|x_s-x_0|)^2}{4Dt}}e^{z^2}\text{Erfc}(z),
\end{aligned}
\end{equation} 
with $z \defn \frac{\sqrt{Dt}}{2\ell_{\mathrm{g}}}+\frac{|x-x_s|+|x_s-x_0|}{\sqrt{4Dt}}$ and $\text{Erfc}(z) \defn 1-\frac{2}{\sqrt{\pi}}\int_0^z \mathrm{d}u\ e^{-u^2}$. 
Notice that the solution depends on the length scale $\ell_{\mathrm{g}} \defn D/\mathrm{g}$ set by the diffusion constant $D$ and the impurity strength $\mathrm{g}$.
This length scale diverges when no impurity is present ($\mathrm{g}\to 0$), while it vanishes in the limit of a very strong impurity ($\mathrm{g}\to\infty$).
In the former case, Eq.~\eqref{eq:diff_sink} simply becomes the usual diffusion equation, while in the latter, it corresponds to the diffusion equation with an absorbing boundary condition at the sink, i.e., $\left.C_Z(x,x_0;t)\right|_{x=x_s}=0$. 
In the following, we fix $x_s=0$ and assume $x_0 > 0$, and characterize the distinct spatiotemporal regimes of dynamics in the presence of the symmetry-breaking impurity with a finite strength $\mathrm{g}$.

Before we proceed, we note that Eq.~(\ref{eq:Pxt_exact}) simplifies for $z \gg 1$---satisfied when either $\sqrt{Dt} \gg \ell_{\mathrm{g}}$ or $|x|+|x_0| \gg \sqrt{Dt}$, or when $|x_0| \gg \ell_\mathrm{g}$ for all times---using $e^{z^2} \text{Erfc}(z) \approx \frac{1}{\sqrt{\pi} z}$:
\begin{equation}
\begin{aligned}
C_Z(x, x_0; t) &\overset{z \gg 1}{\approx} \frac{e^{-\frac{(x^2 + x_0^2)}{4Dt}}}{\sqrt{\pi Dt}} \sinh(\frac{x x_0}{2Dt}) \Theta(x) \\
&+ \frac{e^{-\frac{(|x| + |x_0|)^2}{4Dt}}}{\sqrt{4\pi}(Dt)^{3/2}} \frac{(|x| + |x_0|)\ell_{\mathrm{g}}}{1 + \frac{(|x| + |x_0|)\ell_{\mathrm{g}}}{Dt}} ~.
\label{eq:zgg1}
\end{aligned}
\end{equation}
Here $\Theta(x)$ is the Heaviside step function, and the first term is present only for $x > 0$ (i.e., on the same side as $x_0 > 0$).
Using these expressions, we analyze three regimes of interest: early times, intermediate times, and late times. 
Note that there are two possible intermediate regimes depending on the precise value of the impurity strength, i.e., whether $|x|, |x_0| \gg \ell_{\mathrm{g}}$ or $|x|, |x_0| \ll \ell_{\mathrm{g}}$. 
In the following, we discuss these regimes separately. 
\subsubsection{Early times: $\sqrt{Dt} \ll |x_0|$}
This is an ``early-time" regime that occurs for any impurity strength, and corresponds to the case when $z \gg 1$.
From Eq.~(\ref{eq:zgg1}) it follows that the $C_Z(x, x_0; t)$ is significant only when $|x - x_0| \lesssim \mathcal{O}(\sqrt{Dt})$.
We obtain the usual solution of the diffusion equation form from Eq.~(\ref{eq:zgg1}).
When $|x - x_0| \ll \sqrt{Dt}$(e.g., when we are interested in the autocorrelation function), this is just the conventional diffusion power law decay $\approx \frac{1}{\sqrt{4\pi D t}}$.
The physics in this regime is that a significant part of the correlation function $C_Z(x, x_0; t)$ has not had time to feel the effect of the impurity yet.

\subsubsection{Intermediate times 1: $|x|, |x_0| \ll \sqrt{Dt} \ll \ell_{\mathrm{g}}$}
One possible intermediate regime can occur if $|x| + |x_0| \ll \ell_{\mathrm{g}}$.
Note that this regime necessarily requires $|x|,|x_0| \ll \ell_{\mathrm{g}}$, i.e., the probing locations of the autocorrelation function are inside the effective impurity action region of size $\ell_{\mathrm{g}}$ around the impurity.
The times are short such that the impurity effects have not fully developed yet, but they are also sufficiently long such that a particle beginning at $x_0$ has been able to diffuse to the impurity.
This corresponds to $z \ll 1$, and Eq.~(\ref{eq:Pxt_exact}) can be evaluated in this regime using $e^{z^2}\text{Erfc}(z)\approx 1- \frac{2}{\sqrt{\pi}}z$. One then finds that
\begin{equation}
    C_Z(x,x_0;t)\overset{z \ll 1}{\approx} \frac{1}{\sqrt{4\pi Dt}}e^{-\frac{(x-x_0)^2}{4Dt}} -\frac{1}{4\ell_{\mathrm{g}}}.
\label{eq:zll1}
\end{equation}
This leads to essentially regular diffusive behavior, where the origin of the small correction can be understood as follows:
A particle inserted at $x_0$ has had enough time to explore the impurity location at the origin ($\sqrt{Dt} \gg |x_0|$), but not much of the total probability has been lost yet until this time because of the relatively small absorption rate $g$. In this case we can treat the probability loss perturbatively in $\mathrm{g}$.
The rate of decrease in probability at time $t'$ is $\mathrm{g} C_Z(0, x_0; t')$, which for $t' < t$ and $\sqrt{Dt'} \gg |x_0|$ is roughly $\sim \mathrm{g}/\sqrt{Dt'}$.
Integrating this up to time $t$, the lost total probability is roughly $\mathrm{g}\sqrt{t/D}$.
When distributed over length of order $\sqrt{Dt}$, this corresponds to lost probability density of $\sim \mathrm{g}/D = 1/\ell_\mathrm{g}$.
\subsubsection{Intermediate times 2: $|x|, |x_0|, \sqrt{Dt} \gg \ell_{\mathrm{g}}$}
This is another possible ``intermediate-time" regime that corresponds to $z \gg 1$ and hence Eq.~\eqref{eq:zgg1} applies. 
It is convenient to start with the extreme case $\mathrm{g} \to \infty$ ($\ell_{\mathrm{g}} \to 0$) where for arbitrary $\sqrt{Dt}, x, x_0$, Eq.~(\ref{eq:zgg1}) reduces to just the first line.
In that limit, $C_Z(x, x_0; t)$ vanishes for $x < 0$, while for $x > 0$ it becomes the solution of the diffusion equation with absorbing boundary condition $C_Z(0, x_0; t) = 0$. 
Alternatively, this can be directly solved using the method of images (see Footnote~\ref{ft:image_method}) applied to the solution of Eq.~\eqref{eq:U1_cont} on $\mathbb{R}$ to obtain the result on $\mathbb{R}^+$.\footnote{\label{ft:absorbing_image}Unlike the reflective boundary condition case in Eq.~\eqref{eq:reflect_moi} and Footnote~\ref{ft:image_method}, we can impose the absorbing boundary condition $C_Z(0, x_0; t) = 0$ by using the difference of the appropriately-sourced infinite-size solutions $C_Z(x, x_0; t) = G(x-x_0,t) - G(x+x_0,t)$.}
Physically, such an infinitely strong impurity cuts the system into two disconnected parts. 
We can then also analyze behaviors for a finite impurity strength $\mathrm{g}$, i.e., $\ell_{\mathrm{g}} \neq 0$ in this regime.
In this case $|\sinh(\frac{x x_0}{2Dt})| > |\frac{x x_0}{2Dt}| \gg \frac{(|x| + |x_0|) \ell_{\mathrm{g}}}{Dt + (|x| + |x_0|) \ell_{\mathrm{g}}}$, and the first line is parametrically larger than the second line.
In this regime we also analyze the cases $x > 0$ and $x < 0$ separately.
For $x > 0$, the solution is dominated by the first line in Eq.~(\ref{eq:zgg1}), which has no $\mathrm{g}$ dependence and is of the $\mathrm{g}_{\text{eff}} = \infty$ fixed point form, i.e, diffusion in the $x > 0$ region with the absorbing boundary condition at $x = 0$.
This is natural since $\mathrm{g}$ is a relevant perturbation flowing to large values under renormalization group (see Sec.~\ref{subsec:RG} below) and for the considered regime the impurity effects have had time to fully develop and are probed at distances much larger than the effective impurity length scale $\ell_{\mathrm{g}}$.
Turning to the case when $x < 0$, i.e., on the other side of the impurity, only the second line in Eq.~\eqref{eq:zgg1} is present.
It manifestly depends on the bare impurity strength, and, as already discussed, in this regime 
it is parametrically smaller than for $x$ on the same side of the impurity. 
Note that even though for $x > 0$ we have the $\mathrm{g}_{\text{eff}} = \infty$ fixed point form, the full solution shows non-zero probability for $x < 0$ because for any finite $\mathrm{g}$ there is some leakage of the probability to the left of the impurity, and how a given deposited probability subsequently evolves is similar on the two sides of the impurity.
\subsubsection{Late-times: $\sqrt{Dt} \gg \ell_{\mathrm{g}}, |x|, |x_0|$}
Finally, we have the late-time regime, where $\sqrt{Dt}$ is the largest length scale in the problem. 
Here we again have $z \gg 1$, and we can use Eq.~(\ref{eq:zgg1}) to obtain
\begin{align} \label{eq:C_t32}
C_Z(x, x_0; t) &\approx \frac{x x_0}{\sqrt{4\pi} (Dt)^{3/2}} \Theta(x) + \frac{(|x| + |x_0|)\ell_{\mathrm{g}}}{\sqrt{4\pi}(Dt)^{3/2}}.
\end{align}
This exhibits the power-law decay of $t^{-3/2}$ for both $x > 0$ and $x < 0$ for any finite impurity strength $\mathrm{g}$.
This is distinct from the diffusive power-law in the early-time regime, and yet another sign that the impurity is relevant at late-times in the renormalization group sense.
The correlation function on the $x < 0$ side manifestly depends on the impurity strength and vanishes in the limit $\mathrm{g} \to \infty$ (complete disconnection of the two sides).
Furthermore, the parametric dependence on $|x|$ and $|x_0|$ is qualitatively different on the two sides, which we can view as an indirect manifestation of the two sides effectively becoming disconnected from each other at long times and long distances.

Finally, we note that on a finite system of size $L$, the finite gap $(\sim L^{-2})$ of the low-lying excitations leading to this hydrodynamic tail will be noticed on a time $\sim L^2$, beyond which the correlation functions $C_Z(x,x;t)$ will decay to zero as $\frac{x^2}{L^{3}}e^{-c t/L^2}$, where $x$ is the distance to the impurity and $c$ is some constant (e.g., $c = D\pi^2/4$ for the lowest-energy orbital on an open chain with impurity at the left edge described in footnote~\ref{ft:U1imp_variational_orbital}).

The previous detailed analysis regarding the effect of a symmetry-breaking impurity was possible due to the exact expression for $C_Z(x,x_0;t)$ displayed in Eq.~\eqref{eq:Pxt_exact}.
This requires adding up the contributions coming from the overlaps of the charge operator $Z_j$ with the approximate conserved quantities in Eq.~\eqref{eq:CZZ_formal}.
However, even when a close-form expression can not be easily found for $C_Z(x,x_0;t)$, one has access to the approximate conserved quantities $\phi_k(x)$ appearing as low-lying excitations of the (continuum) perturbed Hamiltonian $-D\partial_x^2 +\mathrm{g}\delta(x)$.
These correspond to the following eigenmodes
\begin{equation}
\begin{aligned}
& \phi_k^{(a)}(x) = \frac{1}{\sqrt{\pi}} \sin(kx) ~, \\
& \phi_k^{(s)}(x) = 
\frac{1}{\sqrt{\pi}} \frac{ \sin(k|x|) + 2\ell_\mathrm{g} k \cos(kx)}{\sqrt{1 + (2\ell_\mathrm{g}k)^2}}
\end{aligned}
\end{equation}
with energy $\propto k^2$ defined on the domain $x \in (-\infty, \infty)$ and with $k > 0$.
On the one hand, for $\mathrm{g} = 0$ one recovers the standard plane wave solutions as expected in the absence of an impurity.
However, for $\mathrm{g} = \infty$, $\phi_k^{(s)} \approx \frac{1}{\sqrt{\pi}} \sin(k|x|)$, which can be combined with $\phi_k^{(a)}(x)$ to produce eigenmodes which are only non-zero for either $x > 0$ or for $x < 0$.
This is consistent with the effective ``disconnection" of the chain discussed in the previous paragraph. 
In fact, the same conclusion holds for a finite impurity strength $\mathrm{g}$ and at sufficiently low energies ($k \to 0$).
This conclusion is an alternative derivation of the fact that at sufficiently low energies, the approximate conserved quantities correspond to eigenmodes of the free-particle Hamiltonian $-\partial_x^2$ with pinned boundary condition $\left. \phi_k(x)\right|_{x=0} = 0$ as a result of the relevance of the symmetry-breaking impurity.

\subsection{Finite-strength impurity at a boundary}\label{subsec:finiteimpboundary}
\begin{figure}[t!]
    \centering
    \includegraphics[width=1\linewidth]{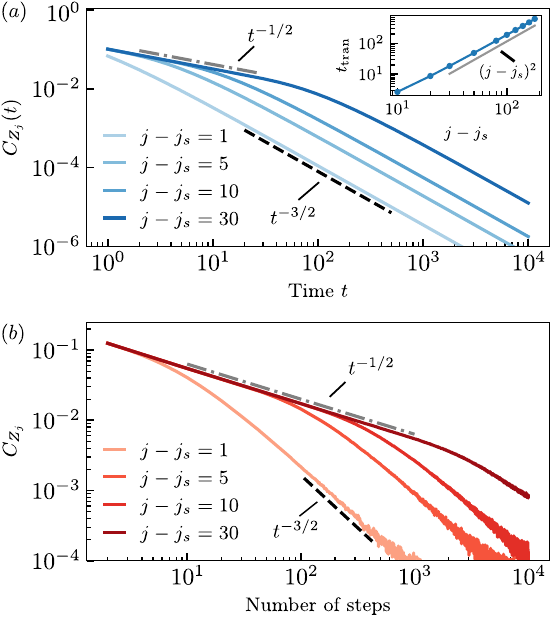}
    \caption{\textbf{Correlation functions $C_{Z_j}$ of U($1$)-symmetric systems with charge-breaking impurity at the boundary $j_s=1$.} 
    (a) Correlation functions with single particle Hamiltonian Eq.~\eqref{eq:U1_Heff_imp} for $L=10000$ and $g=1$. For $j-j_s \geq \mathcal{O}(\sqrt{Dt})$, the correlation functions first decays as $t^{-1/2}$, and then transits to $t^{-3/2}$ when the impurity takes effect. The transition time $t_{\text{tran}}$ for correlation function $C_{Z_j}$ scales as $t_{\text{tran}} \sim (j-j_s)^2$, which is shown in the inset.
    The grey (black) dashed lines indicate the $t^{-1/2} (t^{-3/2})$ scalings.
    (b) Correlation functions obtained from stochastic cellular automata for $L=100$. Each curve is averaged over $10^8$ random realizations. 
    The correlation functions show similar behaviors as given by the single-particle Hamiltonian, which decay as $t^{-1/2}$ and then transit to $t^{-3/2}$. This verifies the analytic calculations.
    }
    \label{fig:U1_imp}
\end{figure}

The fact that correlations $C_Z(x, x_0; t)$ decay in time as $\sim t^{-3/2}$ regardless of the impurity strength $g$ suggests that the impurity is relevant (in the renormalization group sense) at long times and wavelengths, hence the impurity strength flows to infinity $\mathrm{g}_{\textrm{eff}}=\infty$.
Manifestation of this becomes even neater when placing the impurity at one of the boundaries.
Let us consider a semi-infinite system with $x > 0$ and place a U$(1)$-breaking impurity on the left boundary, i.e., $x_s = 0$.
In this setting Eq.~\eqref{eq:diff_sink_cont} means solving the free diffusion equation for $x > 0$ with boundary conditions $\left. D \partial_x C_Z(x, x_0; t)\right|_{x=0^+} = \left. \mathrm{g} C_Z(x, x_0; t)\right|_{x=0^+}$, and admits the exact solution
\begin{equation} \label{eq:Pxt_semi}
\begin{aligned}
C_Z(x,x_0;t)&=\frac{1}{\sqrt{\pi Dt}}e^{-\frac{(x^2+x_0^2)}{4Dt}}\cosh(\frac{xx_0}{2Dt})\\&-\frac{1}{\ell_{\mathrm{g}}}e^{-\frac{(|x|+|x_0|)^2}{4Dt}}e^{w^2}\text{Erfc}(w),
\end{aligned}
\end{equation} 
with  $w \defn \frac{\sqrt{Dt}}{\ell_{\mathrm{g}}}+\frac{|x|+|x_0|}{\sqrt{4Dt}}$.
The first term corresponds to the solution of the one-dimensional diffusion equation with a reflective boundary: Notice that in the absence of an impurity, i.e., for $\mathrm{g}=0$, a reflective boundary condition $\left. D \partial_x C_Z(x, x_0; t)\right|_{x=0} =0$ is imposed, and this contribution solves the standard diffusion equation with a reflective boundary. The second term is similar to that in Eq.~\eqref{eq:Pxt_exact} though with $w$ different from $z$, such that $\left. D \partial_x C_Z(x, x_0; t)\right|_{x=0^+} = \left. \mathrm{g} C_Z(x, x_0; t)\right|_{x=0^+}$ is satisfied.
The behavior at early and intermediate times can be shown to be similar to the case with a bulk impurity.
In the formal case $\mathrm{g}\to \infty$ ($\ell_{\mathrm{g}} \to 0)$, one recovers the absorbing boundary solution, i.e., with $\left. C_Z(x,x_0;t)\right|_{x=0}=0$,
\begin{equation} \label{eq:Pxt_semi_1}
\begin{aligned}
C_Z(x,x_0;t)&=\frac{1}{\sqrt{\pi Dt}}e^{-\frac{(x^2+x_0^2)}{4Dt}}\sinh(\frac{xx_0}{2Dt}),
\end{aligned}
\end{equation}
where we use that $w\overset{\ell_{\mathrm{g}}\to 0}{\longrightarrow}\infty$ and hence $e^{w^2}\text{Erfc}(w)\approx \frac{1}{\sqrt{\pi}w}$. Note that the same expression holds also for finite $\mathrm{g}$ as long as $x, x_0 \sim \sqrt{Dt} \gg \ell_{\mathrm{g}}$.
This is consistent with the system flowing to the new fixed point governed by $\mathrm{g}_{\textrm{eff}} = \infty$ with the absorbing boundary conditions $C_Z(x,x_0;t)|_{x=0} = 0$ at long times and wavelengths.
This also implies that charge correlations decay as $t^{-3/2}$ at late-times (i.e., $\sqrt{Dt} \gg x, x_0$) regardless of the strength of the impurity, which is exactly the same as what we found for the case with the impurity in the bulk.
Similar to the discussion of the previous section, the corresponding approximate conserved quantities correspond to eigenmodes $\phi_k(x) = 
\frac{1}{\sqrt{\pi}} \frac{ \sin(kx) + \ell_\mathrm{g} k \cos(kx)}{\sqrt{1 + (\ell_\mathrm{g}k)^2}}$, $x > 0$ and $k > 0$, which at long wavelengths converge to eigenmodes
of the free-particle Hamiltonian $-D\partial_x^2$ with boundary conditions $\left. \phi_k(x)\right|_{x=0}=0$.
(One can compare these modes also to the variational orbitals for the lattice problem described in footnote~\ref{ft:U1imp_variational_orbital}.)
Figure~\ref{fig:U1_imp} shows numerical results for autocorrelation functions $C_{Z_j} \equiv C_Z(j=j_0,j_0;t)$ obtained directly from the hydro-mode single-particle lattice Hamiltonian in Eq.~\eqref{eq:U1_Heff_imp} using Eq.~\eqref{eq:sing_part_C} with $g=1$ in panel (a), and using cellular automaton dynamics in panel (b).
The simulations are performed with open boundaries, and the impurity is located on the first site $j_s=1$.
Both figures show that as long as 
$\sqrt{Dt} < \mathcal{O}(j - j_s)$, the effect of the impurity remains unnoticed by the correlation functions, and they decay diffusively.
The effect of the impurity kicks in when 
$\sqrt{Dt} \sim \mathcal{O}(j - j_s)$, and at later times, the autocorrelations decay in time as $\sim t^{-3/2}$.
Indeed, we numerically find that the transition time $t_{\mathrm{tran}}$ between the two different scalings scales as $t_{\mathrm{tran}} \sim (j-j_s)^2$ with the distance to the impurity.
Finally, at the latest times and for a finite system size on a lattice, correlations decay exponentially in time, due to the finite-size gap $\mathcal{O}(L^{-2})$ of the super-Hamiltonian.
Using the analogy between the correlation function and the total charge (see Footnote~\ref{ft:charge_correlation}), we can quantify the breaking of the total U$(1)$ charge due to the impurity as
\begin{equation}
\begin{aligned}
   & \int_0^\infty dx\,  C_Z(x,x_0;t) \\ &= \frac{1}{2}\left[\textrm{Erfc}\left(\frac{-x_0}{2\sqrt{Dt}}\right)-\textrm{Erfc}\left(\frac{x_0}{2\sqrt{Dt}}\right)\right] \approx \frac{x_0}{\sqrt{\pi D t}}
\label{eq:chargebreaking}
\end{aligned}
\end{equation}
for $x_0 \ll \sqrt{Dt}$, where we use the expression for $C_Z(x,x_0;t)$ in Eq.~\eqref{eq:Pxt_semi_1}.
We can also loosely interpret this as the average remaining charge for a charge inserted at $x_0$ at time $0$, with charge lost due to the symmetry-breaking impurity at the boundary at $x=0$.
\subsection{Renormalization group arguments and generalization to higher dimensions}\label{subsec:RG}

The problem of diffusion with the absorbing sink at a point is mathematically equivalent to a free particle (e.g., described by a Schr\"odinger equation with kinetic energy $-\partial_x^2$) subject to a localized potential perturbation.
Simple scaling dimension analysis for such a free-particle problem shows that the potential is a relevant perturbation in one spatial dimension.
Indeed, assuming that the diffusion equation in the absence of an impurity is a fixed point under scale transformations in both space $x\to x/e^{\delta l}$ and time (with dynamical exponent $z=2$, i.e., $t \to t/e^{2\delta l}$), one finds that $\mathrm{g}$ grows as $\frac{d \mathrm{g}}{d \delta l}= \left.(2-d)\right|_{d=1}\mathrm{g}$, which relates to the fact that the $\delta$-potential in Eq.~(\ref{eq:diff_sink_cont}) has dimensions of inverse volume.
This is the basis of our statements on the relevance of $\mathrm{g}$ at long-wavelengths/late-times in the renormalization group (RG) sense in the preceding subsections.
While at this level the RG only says that $\mathrm{g}$ flows away from zero, the presented exact solutions in one spatial dimension ($d=1$) show that the system ultimately flows to infinite $\mathrm{g}$, and that the corresponding fixed point can be essentially stated as corresponding to new boundary conditions.
Similar scaling dimension analysis in higher dimensions suggests that the potential term is marginal in $d = 2$ and irrelevant in $d = 3$.\footnote{This is intimately related to the fact that while in $3$ (and higher)-spatial dimensions, a random walk has a finite probability not to return to the origin, i.e., of not being absorbed, it will do surely and almost surely in $1$ and $2$ spatial dimensions respectively.
We can also exploit once again the equivalent single-particle description where the impurity maps to an on-site potential barrier.
While the relevant physics for a potential well is to understand whether a bound eigenstate exists, which is the case for $d=1,2$, we associate the relevance or irrelevance of the barrier to the transmission probability of going through it at vanishing energies.}
Hence, in $d=3$ we expect that the autocorrelations decay in time with the same power law as in the absence of the impurity.
In $d = 2$, we expect that the perturbation is marginally irrelevant:
The scattering $S$-wave phase-shift goes to zero at low energy and the zero-energy wavefunction has the same asymptotic form at large distances as in the absence of the impurity, see, e.g., Ref.~\cite{Khuri_2009:lowenergy}.
We leave detailed explorations in higher dimensions to future work.
\section{Locally breaking dipole conservation}\label{sec:dipole}
We now move on to the case of a dipole $P=\sum_j j S_j^z$ (and hence also charge $Q=\sum_j S_j^z$) conserving systems in the presence of a symmetry-breaking impurity.
Systems with such conservation laws are expected to show subdiffusion~\cite{Morningstar_2020, Gromov_2020, guardadosanchez2020subdiffusion, Zhang_2020_subdiffusion, 2020_Pablo_automato, moudgalya2021spectral}, and we review the derivation of this fact in the superoperator framework.
In the previous section we found that symmetry-breaking impurities (in one dimension) are relevant, and at long distances (larger than the length scale $\ell_\mathrm{g}$ given by the impurity strength) and at long times, the dynamics are controlled by the renormalization group fixed point which pins specific boundary conditions.
Hence, although in that case we could obtain the full non-perturbative expression for the correlation function for any finite impurity strength $g$, the leading behavior (up to subleading corrections) turned out to be independent of $g$. 
Motivated by these findings, in the following section we follow a similar approach, and solve the long time behavior of correlation functions as given by the emergent boundary conditions imposed by a symmetry-breaking impurity.

\subsection{Subdiffusion in the Superoperator Formalism}
\label{sub:dipole_cons}
Similar to the charge-conserving case, we will analyze the lattice super-Hamiltonian first, and then discuss taking the continuum limit, which will be more convenient for analyzing the case with impurities.
\subsubsection{Super-Hamiltonian on a Lattice}
We consider an analytically treatable microscopic problem of a dipole-conserving spin-$1/2$ chain, starting from the bond algebra
\begin{equation} \label{eq:A_dip}
\begin{aligned}
   \mathcal{A}_{\textrm{dip}}=\lgen \{S_j^z\}, \{h_W \defn W_{ijkl} + \textrm{h.c.}\} \} \rgen,
\end{aligned}
\end{equation}
with $W_{ijkl} \defn S_i^- S_j^+ S_k^+ S_l^-$, such that the dipole moment is conserved when $i+l=j+k$.
Similarly to the charge conserving case in the previous section, $\hat{\mathcal{P}}_{Z_j}|O)=0$ enforces all ground states to lie in the composite spin subspace [see Eq.~(\ref{eq:compositespins})].
Hence, as shown in App.~\ref{app:superH_gen}, the super-Hamiltonian takes the form
\begin{equation} \label{eq:P_dip}
    \hat{\mathcal{P}}_{\textrm{dip}}= \sum \left(\ket{\psi^{\textrm{1-flip}}_{ijkl}}+\ket{\psi^{\textrm{3-flips}}_{ijkl}} \right)\left(\bra{\psi^{\textrm{1-flip}}_{ijkl}}+\bra{\psi^{\textrm{3-flips}}_{ijkl}} \right),
\end{equation}
with
$\ket{\psi^{\textrm{1-flip}}_{ijkl}} = \ket{\widetilde{\rightarrow} \widetilde{\rightarrow} \widetilde{\rightarrow} \widetilde{\leftarrow}} 
- \ket{\widetilde{\rightarrow} \widetilde{\rightarrow}  \widetilde{\leftarrow} \widetilde{\rightarrow}}
- \ket{\widetilde{\rightarrow} \widetilde{\leftarrow}    \widetilde{\rightarrow} \widetilde{\rightarrow}}
+ \ket{\widetilde{\leftarrow} \widetilde{\rightarrow} \widetilde{\rightarrow} \widetilde{\rightarrow}}$
and $\ket{\psi^{\textrm{3-flips}}_{ijkl}} = \ket{\widetilde{\leftarrow} \widetilde{\leftarrow} \widetilde{\leftarrow} \widetilde{\rightarrow}}
- \ket{\widetilde{\leftarrow} \widetilde{\leftarrow}  \widetilde{\rightarrow} \widetilde{\leftarrow}}
- \ket{\widetilde{\leftarrow} \widetilde{\rightarrow}    \widetilde{\leftarrow} \widetilde{\leftarrow}}
+ \ket{\widetilde{\rightarrow} \widetilde{\leftarrow} \widetilde{\leftarrow} \widetilde{\leftarrow}}$,
where the spins are defined in Eq.~\eqref{eq:X_spins}.
In general, as long as $i+l=j+k$ in $W_{ijkl}$, both the charge $Q$ and the dipole $P$ are conserved quantities, and hence $|Q)=\sum_j |j)$ and $|P)=\sum_j j|j)$ are ground states of every local term.

The most important difference compared to the super-Hamiltonian in Eq.~\eqref{eq:Pcomp_U1} in the charge-conserving case is that the single-particle subspace of $\widetilde{\leftarrow}$ spins in a background of $\widetilde{\rightarrow}$ spins is not invariant (because of the presence of the ``3-flips'' terms).
This implies that the evolution of the correlation function $C_Z(j,j_0;t)$ is not simply given by a closed-form equation for $C_Z(j,j_0;t)$ alone, but rather involves correlation functions of multi-$Z$ operators.
Indeed, let us denote by $P^{\textrm{1-flip}}$ the projector onto the single spin-flip subspace (spanned by the $\{|j)\}$ basis). 
While we cannot solve for exact low-energy excitations of $\hat{\mathcal{P}}_{\text{dip}}$, we can solve its projection to the single spin-flip subspace
\begin{equation} \label{eq:H_ijkl}
\begin{aligned}
& \mathbb{H}_{\textrm{dip}} \equiv P^{\textrm{1-flip}} \hat{\mathcal{P}}_{\textrm{dip}} P^{\textrm{1-flip}} = \\
& = \sum \left(|l)-|k)-|j) + |i) \right) \left((l|-(k|-(j| + (i| \right) ~.
\end{aligned}
\end{equation}
Each eigenstate $\phi_\alpha$ of $\mathbb{H}_{\textrm{dip}}$ can be used as a trial state of $\hat{\mathcal{P}}_{\textrm{dip}}$, and its energy $\epsilon_\alpha$ under $\mathbb{H}_{\textrm{dip}}$ is an upper bound of the true excitation energy under $\hat{\mathcal{P}}_{\textrm{dip}}$.
Hence, motivated by the success of the single-particle picture at predicting the long-time dynamics of U$(1)$-conserving systems as well as its modification under a symmetry-breaking impurity, it is then reasonable to approximate the autocorrelation function by\footnote{
This is also consistent with approximations performed in earlier works in order to treat the charge-conserving and dipole-conserving problems on similar footing~\cite{2020_Pablo_automato, moudgalya2021spectral, ogunnaike2023unifying}.}
\begin{equation}
C^{\textrm{approx.}}_Z(j,j_0;t) \approx \sum_\alpha e^{- \epsilon_\alpha t} \phi_\alpha(j) \phi_\alpha^*(j_0)~,
\end{equation}
namely, it approximately satisfies the evolution equation
\begin{equation} \label{eq:evol_complete}
    \frac{d}{dt}C^{\textrm{approx.}}_Z(j,j_0;t)=-\sum_{j_1} (j| \mathbb{H}_{\textrm{dip}}|j_1) C^{\textrm{approx.}}_Z(j_1,j_0;t).
\end{equation}
In the following we denote $C^{\textrm{approx.}}_Z$ simply by $C_Z$.
Note that in the case of PBC, there is exactly one state in this sector per momentum, $\alpha = k$, which provides a variational upper bound for an exact eigenstate of $\hat{\mathcal{P}}_{\textrm{dip}}$ with that same momentum, and we expect such variational states to have $\mathcal{O}(1)$ overlap with true eigenstates of $\hat{\mathcal{P}}_{\textrm{dip}}$.
Hence the above approximation is particularly reasonable.
While we do not have momentum quantum numbers for OBC, we expect the approximations to be reasonable in such cases as well for large systems.
We will proceed with our analysis by choosing a specific dipole conserving model, where we take $W_{ijkl}$ in Eq.~\eqref{eq:A_dip} to be geometrically $4$- and $5$-local dipole-conserving terms given by $W^4_{i,i+1,i+2,i+3}$ and $W^5_{i,i+1,i+3,i+4}$ respectively.
The inclusion of the $W^5$ term makes the resulting model weakly fragmented~\cite{2020_sala_ergodicity-breaking}, and also eliminates undesirable symmetries such as the non-local ones due to fragmentation~\cite{2020_SLIOMs, moudgalya_hilbert_2022} and sublattice charge conservation~\cite{Moudgalya_2021_thermalization} that appear in $W^4$.
The correlation $C_Z(j,j_0;t)$ then decays at long times to a system size dependent value that essentially only depends on charge and dipole symmetries of the system~\footnote{Some contributions would still be expected from the non-local symmetries that accompany fragmentation~\cite{moudgalya_hilbert_2022}, but they would be small since the fragmentation is weak.}
The resulting super-Hamiltonian within the single-particle subspace can be straightforwardly derived as a sum of Eq.~(\ref{eq:H_ijkl}) for both the $W^4$ and $W^5$ terms, and reads
\begin{widetext}
\begin{equation} \label{eq:H_dipole_s}
\begin{aligned} 
    \mathbb{H}^{\text{bulk}}_{\text{dip}}& = J_4 \sum_{j}[|j-3) -2|j-2) - |j-1) +4|j)-|j+1)-2|j+2)+|j+3)](j| \\ 
    &+ J_5 \sum_{j}[|j-4) -2|j-3) +|j-2) - 2|j-1) +4|j)-2|j+1)+|j+2)-2|j+3)+|j+4)](j|
\end{aligned}
\end{equation}
\end{widetext}
with the former (latter) term corresponding to the $4$ ($5$)-local contributions.
For OBC, this expression is valid for $j \geq 4$ near the left edge for the $J_4$ term and for $j \geq 5$ for the $J_5$ term.
The remaining boundary contributions can be found in App.~\ref{app:dip_bound}.
\subsubsection{Continuum Limits}
\begin{figure*}
    \centering
    \includegraphics[width=1\linewidth]{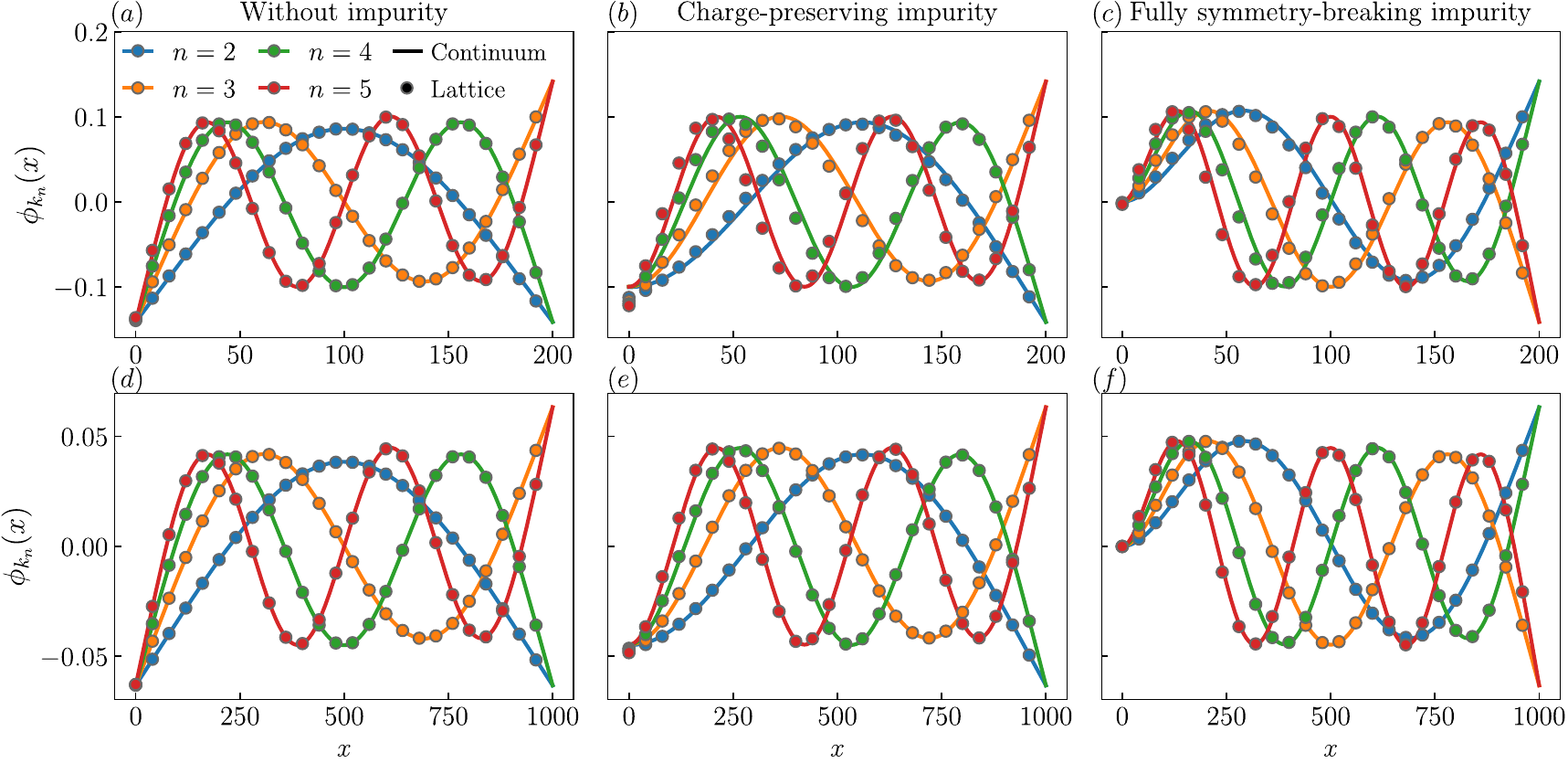}
    \caption{\textbf{Eigenstates $\phi_{k_n}(x)$ of dipole-conserving systems with or without impurities.}
    Comparison of solutions of the continuum equations with boundary conditions (solid lines) and eigenstates of the single-particle Hamiltonian for the hydro-mode in the super-Hamiltonian approach (circles) for dipole-conserving systems with system size $L=200$ (upper panel) and $L=1000$ (lower panel), for the cases (a) and (d) without impurity, (b) and (e) with charge-preserving impurity on the left boundary, and (c) and (f) with fully symmetry-breaking impurity.
    The impurity strength of the single-particle Hamiltonian is $g=1$. For the case of eigenstates on the lattice, we only plot the value of $\phi_{k_n}(x)$ for $25$ equispaced data points.
    The solutions of the continuum equation with boundary conditions in Eq.~\eqref{eq:Dipole_boundary_no_imp} for the case without impurity, Eq.~\eqref{eq:Dipole_boundary_breakP} for charge-preserving impurity, and Eq.~\eqref{eq:Dipole_boundary_breakPQ} for fully symmetry-breaking impurity are compatible with eigenstates of the single-particle Hamiltonian, validating the choice of the boundary conditions for the continuum equations. }
    \label{fig:Dipole_eigenmodes_no_imp}
\end{figure*}

Similar to the charge conservation case, an analytical approach is simpler in the continuum limit. 
To do so, we can rephrase Eq.~(\ref{eq:evol_complete}) with the Hamiltonian of Eq.~(\ref{eq:H_ijkl}) for the choice of $W^4$ and $W^5$ terms as
\begin{equation} \label{eq:evol_sub_dis}
    \frac{d}{dt}C_Z(j,j_0;t)=-D_{\text{dip},\text{latt.}}\Delta_x^4 C_Z(j,j_0;t),
\end{equation}
where in this case $\Delta_x^4$ denotes the linear combination of centered fourth finite-difference operators acting on the first coordinate $j$, and $D_{\text{dip},\text{latt.}}$ is a constant that depends on precise details of the couplings.
In the continuum, this becomes the subdiffusive equation
\begin{equation} \label{eq:evol_sub_cont}
    \partial_tC_Z(x,x_0;t)=-D\partial_x^4 C_Z(x,x_0;t).
\end{equation}
This is similar to the diffusion equation of Eq.~(\ref{eq:U1_cont}), and we wish to solve it with an initial condition $C_Z(x,x_0;t) = \delta(x - x_0)$.
Similar to Eq.~(\ref{eq:CZ_diff_inf}), the solution for an infinite-system can be written as
\begin{equation} \label{eq:G_cont}
    C_Z(x,x_0;t) = \frac{1}{2\pi} \int_{-\infty}^\infty dk\, e^{-Dk^4t} \cos(k(x-x_0)),
\end{equation}
This recovers the well-known subdiffusive behavior of dipole-conserving system, since for $x \sim x_0$, we obtain the scaling $C_Z(x, x_0;t) \sim t^{-\frac{1}{4}}$.
For a finite system, we first need to verify that the continuum equation is compatible with the single-particle Hamiltonian $\mathbb{H}$ that describes the corresponding hydrodynamic mode on a discrete lattice, we can study the spectral representation in Eq.~\eqref{eq:sing_part_C} and compare the continuum eigenmodes $\phi_k(x)$ with eigenfunctions of $\mathbb{H}_{\textrm{dip}}$.
First, the continuum equation on a finite length $L$, and impose the boundary conditions, 
\begin{equation} \label{eq:Q_P_cons_BC}
    \left.\partial^2_x C_Z(x,x_0;t)=  \partial^3_x C_Z(x,x_0;t)\right|_{x=0,L}=0,
\end{equation}
ensuring the conservation of the charge and the dipole moment.
These conditions follow phenomenologically from using that $\partial_t \int dx\,C_Z(x,x_0;t)=0 $ together with $\partial_t \int dx\,x\,C_Z(x,x_0;t)=0$, which follow from charge and dipole conservation. 
Then the eigenvalue problem on a finite segment $[0,L]$ is
\begin{equation}\label{eq:Dipole_diffusive_eq}
    \partial_x^4 \phi_k(x) = E_k \phi_k(x),
\end{equation}
with boundary conditions
\begin{equation}\label{eq:Dipole_boundary_no_imp}
    \left.\partial^2_x \phi_k(x)=  \partial^3_x \phi_k(x)\right|_{x=0,L}=0.
\end{equation}
General solutions are given by 
\begin{equation}
\begin{aligned} \label{eq:exact_phi}
\phi_k(x) = \mathcal{N}_k [&\cos(kx) + \cosh(kx) \\
&+ \gamma_k \left(\sin(kx) +\sinh(kx)\right)],
\end{aligned}
\end{equation}
with energy $E_k \propto k^4$,
where $\mathcal{N}_k$ is a normalization constant, $\gamma_k=(\cos(kL)-\cosh(kL))/(\sinh(kL) - \sin(kL))$, and possible values of $k$ are quantized by the condition $\cos(kL)\cosh(kL)=1$.
Note that the eigenfunctions also involve hyperbolic functions due to the finite size $L$; indeed they would not be normalizable for an infinite system without boundaries.\footnote{In addition, there are two zero-energy eigenfunctions $\phi_{0,1}(x)=a+bx$ that also satisfy the boundary conditions of Eq.~(\ref{eq:Dipole_boundary_no_imp}) -- these correspond to the charge and dipole conservation (in our treatment, the corresponding states are part of the ground state manifold, i.e., the exact symmetries, while we are focusing on the approximate symmetries and their contributions to the long-time properties).}
In Fig.~\ref{fig:Dipole_eigenmodes_no_imp}, we show that the solutions of Eq.~\eqref{eq:Dipole_diffusive_eq} with boundary conditions Eq.~\eqref{eq:Dipole_boundary_no_imp} correspond very accurately to the eigenfunctions of the single particle Hamiltonian in Eq.~\eqref{eq:H_dipole_s}.
This confirms the identification of the appropriate boundary conditions in the continuum limit.

With this setup, we can use a few ``empirical'' observations to obtain an analytical solution for $C_Z(x,x_0;t)$.
While finding the first few modes requires a numerical solution (e.g., $k_1 \approx 1.50562 \pi/L$, $k_2 \approx 2.49975 \pi/L$), we notice that Eq.~\eqref{eq:exact_phi} highly simplifies for sufficiently large $k_n L$. 
There we have essentially analytical forms for $k_n \approx (n+1/2)\pi/L$ [with accuracy $\mathcal{O}(e^{-k_n L})$] and $\phi_{k_n}(x) \approx [\cos(k_nx) - \sin(k_nx) + e^{-k_nx} - (-1)^n e^{-k_n L + k_n x}]/\sqrt{L}$. 
For a semi-infinite system $[0, \infty)$ (or when $x \ll L$), we can drop the last term and arrive at the eigenfunctions
\begin{equation}
\phi_k^{[0,\infty)}(x) = \sqrt{\frac{1}{\pi}} [\cos(k x) - \sin(k x) + e^{-k x}], 
\end{equation}
defined for continuous $k \in (0,\infty)$ and satisfying the normalization (up to an overall constant)
\begin{equation} \label{eq:orthogo}
\int_0^\infty dx\; \phi_k(x) \phi_{k'}(x) = \delta(k-k'),
\end{equation}
The spectral decomposition \eqref{eq:sing_part_C} for the semi-infinite system becomes
\begin{align}
C_Z(x,x_0;t) =  \int_0^\infty dk e^{-Dk^4 t} \phi_k(x) \phi_k^*(x_0) ~.
\end{align}
For $x,x_0 \gg (Dt)^{1/4}$ we recover the infinite-system result Eq.~\eqref{eq:G_cont}.
Indeed, in this regime we can neglect $e^{-kx}$ and $e^{-kx_0}$ terms as exponentially small for important $k \sim (Dt)^{-1/4}$, and for $|x - x_0| \sim \mathcal{O}((Dt)^{1/4})$, we can also drop $\sin(k(x+x_0))$ due to its rapid oscillations and retain only $\cos(k(x-x_0))$.
On the other hand, for $x, x_0 \ll (Dt)^{1/4}$ (such that $kx \approx 0, kx_0 \approx 0$ for the relevant contributing modes with $k \sim (Dt)^{-1/4}$), we obtain
\begin{equation} \label{eq:auto_consQP}
C_Z(x,x_0;t) \approx  \int_0^\infty dk\, e^{-Dk^4 t} \, \frac{4}{\pi} = \frac{\Gamma(\frac{1}{4})}{\pi (Dt)^{1/4}} ~.
\end{equation}
This shows the same power-law decay of autocorrelations as in the bulk, but with an amplitude which is $4$ times that in the bulk (as given by Eq.~\eqref{eq:G_cont}), where the enhanced amplitude arises because of the reflections from the boundary~\footnote{Note that for the regular diffusion (i.e., U($1$) case) the enhancement from a symmetric boundary is $2$ (manifest from the method of images solution).}.
The intuition for the larger enhancement factor in the presence of both the charge and dipole conservation is that to preserve the dipole moment more charge needs to be held back from diffusing away from the boundary, which is encoded in the detailed distribution $C_Z(x,x_0;t)$.
In the following subsections, we move on to the analysis of the effect of impurities on the two-point correlation functions in a dipole conserving system.
There are two different ways in which the symmetry can be locally broken in this system. 
Either the impurity completely breaks both the charge $Q$ and dipole $P$  conservation, or it breaks the dipole conservation while preserving the charge conservation.
We investigate the latter and former situations and show that they exhibit qualitatively different behaviors.

\subsection{Charge-preserving impurity breaking dipole conservation}
We start by considering a charge-preserving impurity of strength $g$.
Given the intertwinement between the charge and dipole conservation, we need a $2$-site impurity to preserve the former while breaking the latter.
For example, we can accomplish this by considering the microscopic impurity $S_{1}^+S_{2}^-+\textrm{h.c.}$ acting on sites $1$ and $2$ at the left boundary.
The resulting super-Hamiltonian takes the form
\begin{equation}
    \label{eq:P_dip_imp_S}
    \hat{\mathcal{P}}_{\textrm{dip}|\textrm{imp}} = \hat{\mathcal{P}}_{\textrm{dip}} + g\left(S_{1,t}^+ S_{2,t}^- - S_{1,b}^+ S_{2,b}^- +\textrm{h.c.} \right)^2
\end{equation}
with $\hat{\mathcal{P}}_{\textrm{dip}}$ given in Eq.~\eqref {eq:P_dip}.
The impurity-generated term is identical to terms in the U$(1)$-symmetric case in Eq.~\eqref{eq:Pcomp_U1}; in particular it acts within the  composite spin subspace and further in its single-particle subspace.
Within the single-particle subspace projection approximation for $\hat{\mathcal{P}}_{\textrm{dip}}$, the super-Hamiltonian then takes the form
\begin{equation}\label{eq:H_breakP}
    \mathbb{H}_g=\mathbb{H}_{\textrm{dip}} + g(|1)-|2))((1|-(2|).
\end{equation}
\subsubsection{Boundary conditions in the continuum limit}
Given our conclusion in Sec.~\ref{sec:U1_break} regarding the effect of boundaries and impurities for U$(1)$ conserving dynamics, we now directly analyze how the boundary conditions are modified in the continuum theory for the dipole case in the presence of such a boundary impurity.
As discussed in Sec.~\ref{subsubsec:U1continuum}, a generic charge-preserving impurity placed at the left boundary requires that $\left.\partial_x C_Z(x,x_0;t)\right|_{x=0}=0$, namely that charge is reflected at the boundary due to charge conservation.
On the other hand, the fact that the bulk is still dipole-preserving ---hence the local charge density $\rho$ is subdiffusively transported via the $\partial_x^4 \rho $---, together with a charge-preserving boundary implies that $\left.\partial_x^3 C_Z(x,x_0;t)\right|_{x=0}=0$. 
Therefore, we propose the following long-time and long-wavelength continuum description
\begin{equation}\label{eq:Dipole_diff_breakP}
\begin{aligned}
&~\partial_t C_Z(x,x_0;t) = - D\partial_x^4 C_Z(x,x_0;t),\\
&\left.\partial_x C_Z(x,x_0;t)\right|_{x=0}=\left.\partial^3_x C_Z(x,x_0;t)\right|_{x=0}=0.
\end{aligned}
\end{equation}
As before, this directly implies that the desired eigenfunctions $\phi_k(x)$ of the differential operator $\partial_x^4$ need to satisfy the following boundary conditions at the left boundary
\begin{equation}\label{eq:Dipole_boundary_breakP}
\left.\partial_x \phi_k(x)\right|_{x=0}=\left.\partial^3_x \phi_k(x)\right|_{x=0}=0,
\end{equation}
together with the symmetry-preserving boundary conditions at the right boundary: $\left.\partial^2_x \phi_k(x)\right|_{x=L}=\left.\partial^3_x \phi_k(x)\right|_{x=L}=0$.
A comparison between the exact expressions of the corresponding eigenmodes and their agreement with eigenfunctions of the single particle Hamiltonian Eq.~\eqref{eq:H_breakP} can be found in Fig.~\ref{fig:Dipole_eigenmodes_no_imp}(b).
\subsubsection{Dynamics of a semi-infinite system}
Here we consider approximate expressions in the limit of large $L$ and for sufficiently large $kL$.
In this regime one finds the approximate expression $\phi_{k_n}(x) \approx \sqrt{\frac{2}{L}}\cos(k_nx)$
with $k_n\approx \pi (n-\frac{1}{4})/L$, together with a zero-energy solution corresponding to charge conservation. 
For a semi-infinite system these eigenfunctions simply become
\begin{equation}
\phi_k^{[0,\infty)}(x) = \sqrt{\frac{2}{\pi}} \cos(kx) ~,
\end{equation}
which gives the correlation function
\begin{equation}\label{eq:G_cont_breakP}
\begin{aligned}
    C_Z(x,x_0;t)&= \frac{2}{\pi} \int_0^\infty dk \, e^{-Dk^4 t} \cos(kx) \cos(kx_0) \\
    &\approx \frac{\Gamma(\frac{1}{4})}{2\pi (Dt)^{1/4}} \quad \text{~for~} (Dt)^{1/4} \gg x,x_0 ~.
\end{aligned}
\end{equation}
The last equation in particular also applies for the autocorrelation function ($x=x_0$) at very long times ($t \gg x_0^4/D$), and $\Gamma(z)$ is the Gamma function.
One can alternatively obtain the exact expression in Eq.~\eqref{eq:G_cont_breakP} using the method of images by imposing $\left.\partial_x C_Z(x,x_0;t)\right|_{x=0}=0$ (see Footnote~\ref{ft:image_method}).
Hence, a charge-conserving but dipole-breaking impurity does not modify the power-law exponent of the charge autocorrelation functions.
However, the amplitude at late times is half of that of the case with boundary that preserves both the charge and dipole, Eq.~(\ref{eq:auto_consQP}); and double that of the bulk far away from the boundary, which is similar to the case of the regular diffusion.
We also note that for a finite system of size $L$, this hydrodynamic tail would stop at a time scale $\mathcal{O}(L^4)$, beyond which the correlation will saturate to the Mazur bound ($\sim 1/L$) given by the total charge conservation.
As mentioned in Footnote~\ref{ft:charge_correlation}, the correlation function $C_Z(j,j_0,t)$ is related to the charge density.
Hence the we have that $\sum_{j}C_Z(j,j_0,t)=\langle  Q(t)S_{j_0}^z \rangle$ and $\sum_{j}jC_Z(j,j_0,t)=\langle  P(t)S_{j_0}^z \rangle$ are conserved in a system with charge and dipole moment conservation. 
Hence, similar to Eq.~(\ref{eq:chargebreaking}), we can quantify the breaking of the dipole moment conservation by calculating
\begin{equation} 
\begin{aligned}
&\frac{d}{dt}\int_0^\infty dx\, x C_Z(x,x_0;t) \\ =& -D \left. \partial^2_x C_Z(x,x_0;t)\right|_{x=0} \\
 =& D \frac{2}{\pi} \int_0^\infty dk \, e^{-Dk^4 t} k^2 \cos(kx_0) \\
\approx& D \frac{\Gamma(\frac{3}{4})}{2\pi (Dt)^{3/4}}, \quad  \text{~for~} (Dt)^{1/4} \gg x_0 ~,
\label{eq:dipole_break}
\end{aligned}
\end{equation}
which vanishes for dipole-conserving boundary conditions.
Hence, we find that the system center of mass grows as
\begin{equation}
    \int_0^\infty dx\, x C_Z(x,x_0;t) \sim (Dt)^{1/4}
\end{equation}
The physical intuition is that the conserved charge now spreads under dipole-conserving dynamics to distances $\sim (Dt)^{1/4}$ with no special structures in the charge distribution to preserve the dipole moment. This is in contrast to the spatial modulations with alternating positive and negative charges appearing for a fully-symmetric boundary that ensures the dipole moment conservation.

\subsubsection{Numerical verifications using cellular automata}
\begin{figure}
    \centering
    \includegraphics[width=1\linewidth]{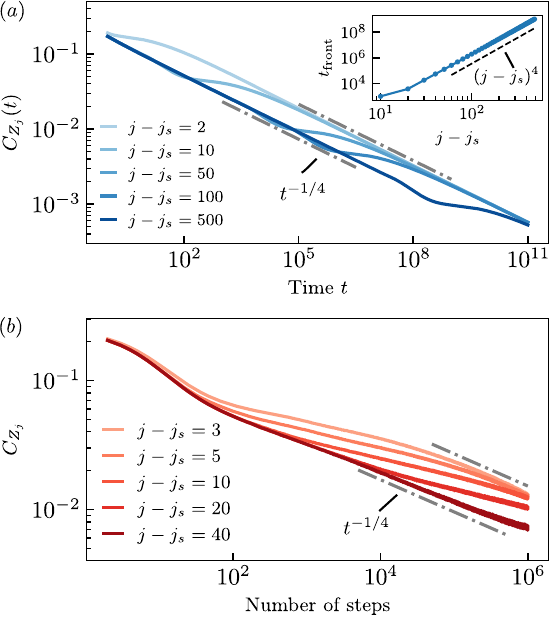}
    \caption{\textbf{Bulk autocorrelation functions of the dipole-conserving system with a local dipole-breaking but charge-preserving impurity at the left boundary $j_s = 1$.}  
    (a) Correlation functions obtained from the single particle Hamiltonian Eq.~\eqref{eq:H_breakP} with dipole-breaking but charge-conserving impurity, for system size $L=10000$ and $g=1$ 
    The correlation functions then decay as $t^{-1/4}$ (grey dashed lines) at long times before saturation (due to charge conservation).
    In addition, a front appears at intermediate times, as the particles reflect back from the boundary due to particle conservation.
    The inset shows that the front appears in $C_{Z_j}$ at time $t_{\text{front}}$ that scales as $t_{\mathrm{front}} \sim (j-j_s)^4$, which is compatible with the reflective boundary prediction. 
    (b) Correlation functions from stochastic dynamics with spin-$1/2$ and $4$- and $5$-local dipole-conserving terms, with system size $L=100$. 
    The correlation functions show a front at intermediate times and $t^{-1/4}$ scaling at long times, similar to the results from the single particle Hamiltonian.}
    \label{fig:Dipole_Ct_breakP}
\end{figure}

We numerically calculate the autocorrelation functions $C_{Z_{j}}(t)\equiv C_Z(j,j;t)$ from the hydro-mode single particle Hamiltonian in the super-Hamiltonian approach and with stochastic dynamics for the impurity at the edge ($j_s = 1$). 
Results for classical cellular automaton dynamics implementing local moves as generated by $W^4_{i,i+1,i+2,i+3}$ and $W^5_{i,i+1,i+3,i+4}$, are shown in Fig.~\ref{fig:Dipole_Ct_breakP} panels (a) and (b) respectively.
Here, the impurity is given by a two-site local gate exchanging the local configurations $\{\uparrow,\downarrow\} \leftrightarrow \{\downarrow,\uparrow\}$, while leaving $\{\uparrow,\uparrow\}$ and $\{\downarrow,\downarrow\}$ unchanged.
The numerical results show that there are two regimes where the autocorrelation $C_{Z_{j}}(t)$ decays as $\sim t^{-1/4}$, which is compatible with the previous discussion of dynamics in a semi-infinite system. 
The earlier time decay is due to regular subdiffusion before the impurity is ``detected", whereas at later times there is a
quantitative amplitude crossover to a different regime which indicates the increase of charge density.
This crossover is as predicted, due to the diffusion and reflection of local charges at the charge-preserving boundary.
The time of this crossover scales with the distance of the particle to the boundary impurity as $t_{\mathrm{front}} \sim (j-j_s)^4$ [see inset in Fig.~\ref{fig:Dipole_Ct_breakP}(a)], which is compatible with the timescale of impurity detection due to the subdiffusive behavior in a dipole conserving system.
Also, at long times beyond $\mathcal{O}(L^4)$, the correlation functions saturate to a finite value ($\sim 1/L$) due to the charge conservation and finite system size.
Figure~\ref{fig:Dipole_Ct_breakP}(b) shows the stochastic CA dynamics with $4$-local and $5$-local dipole-conserving terms on spin-$1/2$, which shows the same scaling $t^{-1/4}$ and the front at intermediate time, validating the previous analysis.
Finally, despite the fact that we use a specific microscopic setting, namely a two-dimensional local Hilbert space where the dynamics are generated by $W^4$ and $W^5$ local terms, we observe the same qualitative behavior (in fact the same scaling with system size) when increasing the local Hilbert space dimension or using different local charge- and dipole-preserving dynamics.
We include the ensuing stochastic dynamics for a combination of a $3$-local and $4$-local dipole-conserving models on spin-$1$ in App.~\ref{app:dip_bound}.
This shows the universality of our analysis for charge and dipole-conserving models with weak fragmentation.

\subsection{Impurity breaking charge and dipole symmetry}\label{subsec:dipole_breakPQ}
\begin{figure}
    \centering
    \includegraphics[width=1\linewidth]{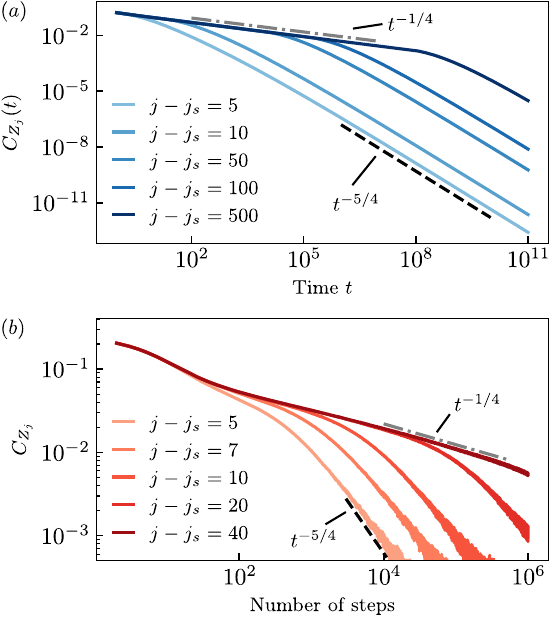}
    \caption{\textbf{Bulk autocorrelation functions of the dipole-conserving system with fully symmetry-breaking impurities at the left boundary on three consecutive sites $j_s = 1$, $j_s+1$, $j_s+2$.} 
    (a) Correlation functions obtained from hydro-mode single particle Hamiltonian for the spin-$1/2$ system described by Eq.~\eqref{eq:H_breakPQ}. The system size is $L=10000$ with $g=1$. 
    The correlation functions decay as $t^{-1/4}$ (grey dashed line) at early times until the boundary impurity takes effect, then decay as $t^{-5/4}$ (black dashed line) at long times.
    (b) Correlation functions from stochastic dynamics with $4$-local and $5$-local dipole-conserving gates and fully symmetry-breaking local impurities (three-site spin flips at the boundary), for system size $L=100$ and spin-$1/2$. The stochastic dynamics show similar behaviors as the single particle Hamiltonian. Each curve is averaged over more than $10^8$ random samples.
    }
    \label{fig:Dipole_Ct_breakPQ}
\end{figure}

We now consider the case of an impurity breaking both the charge and dipole conservation at the left boundary.
We numerically find that neither a single- nor a two-site impurity is sufficient to break all conservation laws due to the weak Hilbert space fragmentation for the bond algebra generated by spin-$1/2$ $4$- and $5$-local terms.
Hence, for numerical studies we instead consider three impurity terms $X_1$, $X_2$ and $X_3$ acting on three consecutive sites, which restores full ergodicity. This leads to the super-Hamiltonian
\begin{equation}\label{eq:P_dip_imp_X}
\begin{aligned}
\hat{\mathcal{P}}_{\textrm{dip}|\textrm{imp}} &= \hat{\mathcal{P}}_{\textrm{dip}} + g\left[(X_{1,t} - X_{1,b})^2  \right.\\
&\left.+ (X_{2,t} - X_{2,b})^2 + (X_{3,t} - X_{3,b})^2\right]
\end{aligned}
\end{equation}
with $\hat{\mathcal{P}}_{\textrm{dip}}$ given in Eq.~\eqref {eq:P_dip}. 
Within the single-particle subspace, the super-Hamiltonian then becomes
\begin{equation}\label{eq:H_breakPQ}
    \mathbb{H}_g=\mathbb{H}_{\textrm{dip}} + 4 g (|1)(1|+|2)(2| +|3)(3|),
\end{equation}
similar to Eq.~\eqref{eq:U1_Heff_imp} with a single-site impurity.
If one naively follows the arguments similar to the charge-conserving case [e.g., see around Eq.~\eqref{eq:Pxt_semi}], one would expect that in the continuum limit and at long times/wavelengths an emergent absorbing boundary condition would be $C_Z(x=0,x_0;t)=0$.
After ignoring the right boundary conditions, e.g., for a semi-infinite system and applying the method of images to the infinite-size solution $C_Z^{\infty}(x,x_0;t)$ in Eq.~\eqref{eq:G_cont} as $C_Z^{\infty}(x,x_0;t)-C_Z^{\infty}(x,-x_0;t)$ (see Footnotes~\ref{ft:image_method} and \ref{ft:absorbing_image}), we obtain
\begin{equation}
\begin{aligned}
    C_Z^{\text{naive}}(x,x_0;t)&= 
   \frac{1}{\pi}
    \int_{-\infty}^\infty dk \, e^{-Dk^4 t} \sin(kx)\sin(kx_0).
\label{eq:Cnaive}
\end{aligned}
\end{equation}
This, however, leads to the wrong conclusion that autocorrelations close to the boundary decay as $C_Z(x,x_0;t) \sim (Dt)^{-3/4}$ at long times.
Indeed, Fig.~\ref{fig:Dipole_Ct_breakPQ} clearly shows that  $C_Z(x,x_0;t)$ decays with a power-law exponent larger than $3/4$.

However, since the subdiffusion equation is a fourth-order differential equation, it is clear that imposing $\left.C_Z(x,x_0;t)\right|_{x=0} = 0$ alone (together with the boundary conditions at the other boundary) is not sufficient to obtain a unique solution given an initial condition.
Lacking a fully consistent phenomenological description to obtain the missing emergent boundary condition at long times/wavelengths, we numerically analyze the dependence of the single-particle eigenstates $\phi_{k_n}(j)$ of $\mathbb{H}_g$ in Eq.~\eqref{eq:H_breakPQ} at low energies (see App.~\ref{subapp:QP_break}). We find that indeed, for any value of $g$, $\phi_{k_n}(j)$ becomes zero at the boundary with increasing system size. Moreover, by numerically evaluating the first order finite difference $\phi_{k_n}(j-1)-\phi_{k_n}(j)$ between consecutive sites, we find that the boundary condition $\left.\partial_x \phi_k(x)\right|_{x=0}=0$ emerges at long wave lengths. 
Hence, we propose the following emergent (appropriate fixed-point) boundary conditions
\begin{equation} \label{eq:BC_QPbreak}
\left.C_Z(x,x_0;t)\right|_{x=0} = 0, \quad
\left.\partial_x C_Z(x,x_0;t)\right|_{x=0} = 0.
\end{equation}
Alternatively, we can phenomenologically argue that Eq.~\eqref{eq:Dipole_boundary_breakPQ} provides the relevant boundary conditions as follows:
In the absence of an impurity the charge $J_c \sim \partial_x^3 \rho$ and dipole $J_d \sim \partial_x^2 \rho$ currents need to vanish at the boundary.
When either of charge or dipole are not conserved, these currents will be sourced by the more relevant (linear) operators $\rho$ and $\partial_x \rho$ at the corresponding boundary, i.e., $J_c, J_d \sim \left. \mathrm{g}^{\prime}_{c,d} \rho + \mathrm{g}^{\prime \prime}_{c,d} \partial_x \rho\right|_{x=0}$.\footnote{Note that the microscopic lattice hydro-mode super-Hamiltonian, where the effects of the boundary and impurities are encoded in the terms operating near the boundary as in Eq.~\eqref{eq:H_45_bound} and Eq.~\eqref{eq:H_breakP}, is a linear problem.
Hence it is natural to expect that in the continuum description, the corresponding physics near the boundary can be encoded via linear relations among $\rho, \partial_x \rho, \partial^2_x \rho, \partial^3_x \rho$ evaluated at the boundary, with $\mathrm{g}^{ \prime}_{c,d}$ and $\mathrm{g}^{\prime \prime}_{c,d}$ some non-universal coefficients.} 
However, at long wavelengths these equations imply that both $\rho$ and $\partial_x \rho$ vanish at the boundary.

We can numerically verify this choice of the boundary conditions by comparing the corresponding eigenmodes for the continuum equation and the hydro-mode single particle Hamiltonian on the lattice.
Similar to the previous discussion, we consider the continuum equation on finite size $L$ with left boundary conditions, 
\begin{equation}\label{eq:Dipole_boundary_breakPQ}
    \phi_k(x)|_{x=0} = \partial_x \phi_k(x)|_{x=0} = 0,
\end{equation}
together with the charge- and dipole- conserving right boundary conditions.
We exactly solve for the continuum problem eigenmodes $\phi_k(x)$ with energy $E_k \propto k^4$, which are shown to agree with the lattice eigenfunctions of the single particle Hamiltonian in Fig.~\ref{fig:Dipole_eigenmodes_no_imp}(c).
The eigenmodes are given by
\begin{equation}
\begin{aligned}
\phi_k(x) = \mathcal{N}_k \big[&\cos(kx) -\cosh(kx) \\ 
&+ \gamma_k(\sin(kx) - \sinh(kx)) \big]
\end{aligned}
\end{equation}
with $\gamma_k=(\cos(kL)-\cosh(kL))/(\sinh(kL) - \sin(kL))$, and $k$ quantized by the condition $\cos(kL)\cosh(kL)=-1$.
For sufficiently large $k_nL$, these can be approximated by $\phi_{k_n}(x) \approx \sqrt{\frac{1}{L}} (\cos(k_nx)-\sin(k_nx)-e^{-k_nx}+(-1)^ne^{-k_n(L-x)})$, with $k_n\approx \frac{\pi}{L}(n+\frac{1}{2})$. 
For a semi-infinite system these become $\phi_{k}(x) \approx \sqrt{\frac{1}{L}} (\cos(kx)-\sin(kx)-e^{-k_nx})$.

We conclude that appropriate eigenfunctions for a semi-infinite system $[0, \infty)$ are
\begin{equation}
\phi_k^{[0,\infty)}(x) = \sqrt{\frac{1}{\pi}} [\cos(k x) - \sin(k x) - e^{-k x}],  \quad k \in (0,\infty),
\end{equation}
which satisfy the normalization conditions of the form Eq.~\eqref{eq:orthogo}.
Plugging this into the spectral decomposition then gives the closed-form expression for the correlation function $C_z(x,x_0;t)$.
Utilizing these expressions,\footnote{In this case, we cannot apply the method of images to the infinite-system solution to impose both boundary conditions simultaneously.} we then find the late-time dynamics of the correlation function $C_Z(x,x_0;t)$ for $(Dt)^{1/4} \gg x, x_0$,
\begin{equation}
C_Z(x,x_0;t) \approx \frac{\Gamma(\frac{5}{4}) x^2 x_0^2}{4\pi(Dt)^{5/4}},
\end{equation}
namely with a larger power law exponent than we naively obtained in Eq.~(\ref{eq:Cnaive}). 
Figure~\ref{fig:Dipole_Ct_breakPQ}(a) shows that this prediction is indeed consistent with the numerical results obtained from the single-particle simulations even for $\mathrm{g}=1$.
Since this symmetry-breaking impurity is relevant at long wavelengths and long times, we conjecture that this prediction does in fact hold for all $g$, and that the power-law exponent becomes $5/4$ at sufficiently long times and large system sizes for all $g$.

In Fig.~\ref{fig:Dipole_Ct_breakPQ}(b), we also show numerical results for the stochastic dynamics for a spin-$1/2$ dipole-conserving model with $4$-local and $5$-local dipole-conserving models. We see similar behaviors as in the super-Hamiltonian approach and as predicted by the analytical treatment, thus providing an unbiased support for these predictions.
Similar to the discussion in previous sections, we note that the finiteness of the system will be noticed at a time that scales as the inverse of the gap $(\sim L^{-4})$. For times beyond $L^4$, we expect that correlation functions $C_Z(x,x;t)$ decay to zero as $\frac{x^4}{L^{5}} e^{-ct/L^4}$, which follows from examining the lowest-energy orbital $\phi_{k_n}(x)$ in this case.

Finally, following a similar discussion as around Eq.~\eqref{eq:dipole_break}, and using the full expression for $C_Z(x,x_0;t)$, we can directly calculate the decay of the total charge and center of mass of the system in this case.
At long times $(Dt)^{1/4} \gg x_0$, we obtain
\begin{align}
& \int_0^\infty dx \, C_Z(x,x_0;t)\sim \frac{x_0^2}{(Dt)^{1/2}}, \\
& \int_0^\infty dx \, x C_Z(x,x_0;t) \sim \frac{x_0^2}{(Dt)^{1/4}} ~.
\end{align}
The first contribution can be interpreted as the remaining charge after been initially inserted at $x_0$ at time $0$.
The scaling of the second line then follows from the fact that the surviving charge is spread over a length scale of order $\mathcal{O}((Dt)^{1/4})$.
\section{Locally breaking strong fragmentation}\label{sec:fragmentation}
We now turn to systems with strong Hilbert space fragmentation~\cite{2020_sala_ergodicity-breaking, 2020_khemani_local}, which are known to possess exponentially many conserved quantities~\cite{moudgalya_hilbert_2022}.
Strong Hilbert space fragmentation can lead to anomalously large Mazur bounds for two-point correlation functions of local operators (e.g., $\mathcal{O}(L^{-1/2})$ in the bulk of the $t-J_z$ model discussed below, compared to $\mathcal{O}(L^{-1})$ for ergodic systems with $U(1)$ symmetry)~\cite{2020_SLIOMs,moudgalya_hilbert_2022, 2024_hart2023exact_PF}, or even non-vanishing correlation functions in the long time limit~\cite{2020_sala_ergodicity-breaking, 2022_Lehmann_Pablo_Markov}
Moreover, it has been observed that strongly fragmented systems with a local impurity can exhibit an exponentially long thermalization time, which was explained using graph theoretic methods~\cite{Balasubramanian2024glassyword, han2024exponentially, wang2025exponentiallyslowthermalization1d}to study the Hilbert space connectivity. 
In this section, we investigate the long-time dynamics of strongly fragmented systems in similar settings using superoperator formalism.
In particular, we consider two examples. 
First, the $t-J_z$ model exhibits strong fragmentation due to simple pattern conservation, whose conserved quantities are well studied~\cite{2020_SLIOMs, moudgalya_hilbert_2022}. 
Second, we consider the spin-$1$ $3$-local dipole conserving model with strong fragmentation~\cite{2020_khemani_local, 2020_sala_ergodicity-breaking}, and compare with $3$- and $4$-local dipole-conserving model, where longer-range interactions lead to weak fragmentation, similar to the dipole-conserving models discussed in Sec.~\ref{sec:dipole}.
We find that strong fragmentation can indeed lead to exponentially slow thermalization, resulting from certain approximate conserved quantities that are only weakly perturbed by the local impurities.
Moreover, the correlation functions can exhibit prethermal plateaus which last for an exponentially long time that scales with the distance to the impurity.
\subsection{$t-J_z$ model}  \label{sec:tJz}
The $t-J_z$ model is a chain with local Hilbert space spanned by $\ket{\uparrow}$, $\ket{\downarrow}$, and $\ket{0}$ on each site~\cite{zhang1997tJz, Batista2000tJz}.
We can think of states $\ket{\uparrow}$ and $\ket{\downarrow}$ as spin states of a spin-1/2 particle on a given site (without double occupancy) and will often refer to these as simply \textit{particle spins}, and call state $\ket{0}$ as an empty site. 
The bond algebra can be specified by the following local generators
\begin{equation}
\begin{aligned}
    \mathcal{A}_{t-J_z} &= \lgen \{T_{j,j+1}^\sigma \}, \{S_j^z\} \rgen, \,\,\, \textrm{with}\\
    T_{j,j+1}^\sigma &= (|\sigma\ 0 \rangle \langle 0\ \sigma|)_{j,j+1} + \mathrm{h.c.},\,\,\, \sigma \in \{\uparrow, \downarrow\}, \\
    \textrm{and~} S_j^z &= (\ket{\uparrow}\bra{\uparrow}-\ket{\downarrow}\bra{\downarrow})_j.
\end{aligned}
\end{equation}
Under the $t-J_z$ dynamics, the particle spins $\uparrow$ and $\downarrow$ can hop to a neighboring site if it is empty, but they are not allowed to cross each other.
For example, a state can evolve as
\begin{equation}
    \ket{\uparrow\ \downarrow\ 0\ 0} \leftrightarrow \ket{\uparrow\ 0\ \downarrow\ 0} \leftrightarrow \ldots \leftrightarrow \ket{0\ 0\ \uparrow\ \downarrow},
\end{equation}
with the particle spin pattern $(\uparrow, \downarrow)$ remaining invariant.
Therefore, the particle spin patterns, $(\sigma_1, \sigma_2, \ldots, \sigma_l)$, with $l$ the number of particles, are conserved under the dynamics.
For a system size $L$, the Hilbert space separates(fragments) into $\sum_{l=0}^L 2^l = 2^{L+1}-1$ Krylov subspaces, each labeled by a spin pattern. 
The commutant algebra $\mathcal{C}_{t-J_z}$ is spanned by the projectors onto these Krylov subspaces~\cite{moudgalya_hilbert_2022}.
In the following, we will first show that the $t-J_z$ model thermalizes exponentially slow due to certain approximately conserved quantities under perturbation. Then, we will show that there can be prethermal plateaus of spin-spin correlation functions both at the boundary and in the bulk, with its decay timescales connected to the approximately conserved quantities.
\subsubsection{Super-Hamiltonian and its perturbations}
For the super-Hamiltonian corresponding to the $t-J_z$ model, similar to the U$(1)$ case, the $S_j^z$ generator gives rise to the term $(S_{j;t}^z - S_{j;b}^z)^2$, which enforces the ground states to have $\ket{m}_{j;t} = \ket{m}_{j;b}, m \in \{\uparrow, \downarrow, 0 \}$ at each location $j$.
Therefore, the ground states and low-lying excited states lie in the composite spin subspace with local Hilbert space spanned by $|\widetilde{m}\rangle \defn |m,m\rangle_{t,b}$ with $\widetilde{m} \in \{\widetilde{\uparrow}, \widetilde{\downarrow}, \widetilde{0} \}$, similar to Eq.~(\ref{eq:compositespins}).
(Here and below, we are mildly abusing the language referring to all three states $\widetilde{\uparrow}, \widetilde{\downarrow}, \widetilde{0}$ as composite spin states; this should not cause any confusion but sometimes to be more specific we will refer to the first two states as composite particle spins.)
The super-Hamiltonian restricted to the composite spin subspace is given by~\cite{moudgalya2024symmetries}
\begin{equation}
    \hat{\mathcal{P}}_{t-J_z|\mathrm{comp}} = 2\sum_{j} \sum_{\widetilde{\sigma} \in \{\widetilde{\uparrow},\widetilde{\downarrow}\}} (|\widetilde{\sigma}\widetilde{0}\rangle - |\widetilde{0}\widetilde{\sigma}\rangle) (\langle \widetilde{\sigma}\widetilde{0}| - \langle\widetilde{0}\widetilde{\sigma}|)_{j,j+1}.
\label{eq:tJzcomp}
\end{equation}
This $\hat{\mathcal{P}}_{t-J_z|\mathrm{comp}}$ has exponentially many degenerate ground states due to the fragmented character of the $t-J_z$ model.
These are spanned by the projectors on the Krylov subspaces, expressed in the double space language as
\begin{equation}\label{eq:tJz_GS}
    |G^{\sigma_1\sigma_2 \ldots \sigma_l}) = \sum_{j_1 < j_2 < \ldots < j_l} |\widetilde{\sigma}_1(j_1) \widetilde{\sigma}_2(j_2) \ldots \widetilde{\sigma}_l (j_l)\rangle,
\end{equation}
where $\widetilde{\sigma}_k(j_k)$ indicates that the $k$-th composite particle spin $\widetilde{\sigma}_k$ is at site $j_k$, and the other sites $j \notin \{j_1, j_2, \ldots, j_l\}$ are in composite empty state $\widetilde{0}$.
The above ground state is an equal weight superposition of product states with the same composite particle spin pattern, $(\widetilde{\sigma}_1, \widetilde{\sigma}_2, \ldots, \widetilde{\sigma}_l)$.
Now we add an impurity to break the fragmentation structure.
We take a local ``state-flip'' term $F_{j_s}$ acting on site $j_s$, given by
\begin{equation}\label{eq:spin_flip_F}
    F_j = \frac{1}{2}(\ket{\downarrow}\bra{0} + \ket{0}\bra{\uparrow} + \ket{\uparrow}\bra{\downarrow})_j + \mathrm{h.c.}
\end{equation}
Notice that this is a different operator from the local particle spin operator $S^x_j$, as it allows all possible transitions between local basis states. 
The perturbed super-Hamiltonian projected into the composite spin subspace (which we expect is sufficient to understand the structure of the low-energy excitations) is given by
\begin{equation}\label{eq:Pimp_tJz}
    \hat{\mathcal{P}}_{t-J_z|\mathrm{imp}} \equiv \hat{\mathcal{P}}_{t-J_z|\mathrm{comp}} + \hat{\mathcal{V}}_{j_s},\;\;\hat{\mathcal{V}}_{j_s} \equiv g (1-\widetilde{F}_{j_s}),
\end{equation}
with $g$ as the coupling strength of the impurity~\footnote{We could have also chosen 
to add separate terms $\{ (\ket{\downarrow}\bra{0} + \mathrm{h.c.})_{j_s},
(\ket{0}\bra{\uparrow} + \mathrm{h.c.})_{j_s},
(\ket{\uparrow}\bra{\downarrow} + \mathrm{h.c.})_{j_s}$ to the bond algebra, see Appendix~\ref{app:superH_gen}.
This is a somewhat different impurity model, but it should capture the same qualitative physics; the corresponding contribution to the super-Hamiltonian would leave the composite spin subspace exactly invariant, hence avoiding any approximation at this step.}.
As noted in our discussion in Sec.~\ref{sec:review}, this strength directly relates to the variance of the Hamiltonian coupling $J_{\text{imp}}^{(t)}$ via  $\overline{(J_{\text{imp}}^{(t)})^2} \sim g$, where $\overline{\cdots}$ is the average over different realizations of the dynamics.
For the unperturbed $t-J_z$ system, the Krylov subspaces with different spin patterns are dynamically disconnected, which is reflected in its ground state structure.
The state-flip impurity changes the particle spin numbers and the particle spin patterns and hence connects all the subspaces, restoring full ergodicity. 
This leads to a unique ground state $|\mathbb{1})/\sqrt{3^L}$ of the perturbed super-Hamiltonian corresponding to the identity, signifying that all the symmetries of the $t-J_z$ model are broken under the addition of this impurity.
\subsubsection{Exponentially slow thermalization}
We will now show that the thermalization time, i.e., the timescale where the system relaxes to full thermalization, is exponentially long in system size.
Similar results were recently obtained using graph-theoretic methods in Refs.~\cite{Balasubramanian2024glassyword, han2024exponentially, wang2025exponentiallyslowthermalization1d}.
In contrast, here we show that this large time-scale arises solely due to the fact that certain conserved quantities of the $t-J_z$ model remain approximately conserved under the impurity perturbation.
Indeed, this large timescale can be shown to persist even \textit{without} fragmentation, as long as the relevant approximate conserved quantities are preserved.

To employ the conventions of previous works on the strong Hilbert space fragmentation of this model~\cite{2020_SLIOMs, moudgalya_hilbert_2022} and without loss of generality, we consider a the spin-flip impurity on the right boundary of the chain, i.e., $j_s=L$. 
Intuitively, at short times, the boundary impurity only changes the rightmost particle spins, while the particle spin patterns in the bulk remain unchanged.
The leftmost particle spin remains unchanged for a long time until all the other physical spins are scrambled; the only way for this to happen is if all the particles move to the right of the chain, are removed, and inserted back into the chain again.
To understand the effect of the local impurity, we consider operators that measure the $k$-th particle spin \cite{2020_SLIOMs}
\begin{equation}
    q_k = \sum_j P_j^{k} S_j^z,
\end{equation}
where $P_j^k$ is a non-local projector onto all the states where the $k$-th particle spin is located at site $j$.
These are the \textit{statistically localized integrals of motions} (SLIOMs) for the unperturbed $t-J_z$ model, which are important conserved quantities of this model~\cite{2020_SLIOMs}.
They can generate all conserved quantities~\cite{2020_SLIOMs, moudgalya_hilbert_2022}, and they admit a \textit{statistical} notion of spatial localization that was elucidated in Ref.~\cite{2020_SLIOMs}. 
In particular, the leftmost SLIOM 
\begin{equation}
    q_{\textrm{left}} \equiv q_1 = \sum_{j=1}^L \left(\bigotimes_{k=1}^{j-1} |0\rangle\langle0|_k \right) S_j^z 
\end{equation}
is exponentially localized near the left boundary, and this is manifest in the expression in the double space language:
\begin{equation}
|q_{\text{left}}) = \sum_{j=1}^L \left[\bigotimes_{k=1}^{j-1} |\widetilde{0})_k \right] \otimes [|\widetilde{\uparrow}) - |\widetilde{\downarrow})]_j \sqrt{3}^{L-j} \left[\bigotimes_{k=j+1}^L |\widetilde{e})_k \right] ~, 
\end{equation}
where $\sqrt{3}|\widetilde{e}) \equiv |\widetilde{\uparrow}) + |\widetilde{0}) + |\widetilde{\downarrow})$ is the double space representation of a local identity operator with normalization $(\widetilde{e}|\widetilde{e}) = 1$.

Consider the perturbed super-Hamiltonian $\hat{\mathcal{P}}_{t-J_z|\mathrm{imp}}$ in Eq.~\eqref{eq:Pimp_tJz}.
The thermalization time $t_{\mathrm{th}}$ is determined by the gap $\Delta$ of $\hat{\mathcal{P}}_{t-J_z|\mathrm{imp}}$, with $t_{\mathrm{th}} \sim 1/\Delta$. 
Without loss of generality, consider the state-flip impurity located in the right half of the chain, $j_s \geq L/2$.
From the argument above, the state $|q_{\text{left}})$ is a good trial state for the first excited state, as it measures the leftmost particle spin that remains unchanged for a long time.  
The energy gap $\Delta$ is upper-bounded by the energy of the trial state $|q_\text{left})$,
\begin{equation}\label{eq:left_sliom_energy_gap}
    \Delta \leq \frac{(q_{\text{left}}|\hat{\mathcal{V}}_{j_s}|q_{\text{left}})}{(q_{\text{left}}|q_{\text{left}})} 
    = g\frac{4\times 3^{L-j_s} - 1}{3^L-1} \sim \mathcal{O}( g \times 3^{-j_s}).
\end{equation}
We used $\hat{\mathcal{P}}_{t-J_z|\text{comp}}|q_{\text{left}}) = 0$, as $q_{\text{left}}$ is a conserved quantity of the unperturbed $t-J_z$ model.
Analogous arguments can be applied to an impurity located in the left half of the chain $j_s \leq L/2$, which gives  $\Delta \sim \mathcal{O}(g \times 3^{-(L-j_s)})$.
Therefore, the thermalization time is exponentially slow with respect to system size $L$, as $t_{\mathrm{th}} \sim \Delta^{-1} \geq \mathcal{O}(3^{L/2}/g)$, regardless of the location of the local impurity.
This result applies to other fragmented systems with similar mechanisms, e.g., with statistically localized integrals of motions or pattern conservation~\cite{dhar1993conservation, barma1993jamming, menon1995irreducible, menon1997conservation, 2020_SLIOMs, moudgalya_hilbert_2022}, which includes the $3$-local dipole-conserving model and the Pair-Flip model~\cite{barma1993jamming, 2018_PFmodel}.
The exponentially slow thermalization time under local impurities for the Pair-Flip model, together with other models with strong Hilbert space fragmentation, can also be explained by the graph theory, where the constrained dynamics under perturbation can be mapped to a graph with a strong bottleneck or a large diameter~\cite{han2024exponentially, wang2025exponentiallyslowthermalization1d} 
In fact, using the first-order perturbation theory, the effective super-Hamiltonian of the $t-J_z$ model with an impurity at the boundary is exactly a normalized Laplacian~\cite{chung1997spectral, aldous-fill-2014}, which is associated with a graph with strong bottleneck and thus exponentially slow thermalization.
We demonstrate this connection in App.~\ref{app:rel_graph}.
It is interesting to note that adding arbitrary perturbation terms that commute with $q_{\text{left}}$ does not affect the exponentially long thermalization, even when the fragmentation is fully violated in the bulk.
This is because the energy gap remains upper bounded by the trial state $|q_{\text{left}})$ given by Eq.~\eqref{eq:left_sliom_energy_gap}. 
An example of such a perturbation is 
$V = \sum_j (S_j^z)^2 F_{j+1}$, where $F_j$ is defined in Eq.~(\ref{eq:spin_flip_F}) preserves the leftmost physical spin while being capable of flipping all the other physical spins~\cite{2020_SLIOMs}. 
In the rest of this subsection,  we perform a detailed study of the late-time dynamics of spin-spin correlation functions at different locations across the system, by analyzing the low-energy spectrum of $\hat{\mathcal{P}}_{t-J_z|\mathrm{imp}}$ and their connection to SLIOMs across the system.

\subsubsection{Decay timescales for the bulk SLIOMs}
We have shown that the boundary SLIOM $|q_{\mathrm{left}})$ is connected to the exponentially slow thermalization timescales. 
Here, we further highlight some properties of the bulk SLIOMs, which exhibit similar behaviors as $|q_{\mathrm{left}})$. 
In the latter discussions, we will show that SLIOMs play a crucial role in understanding both boundary and bulk correlation functions.

Similarly to the leftmost SLIOM, a general SLIOM $q_k$ admits the following expression in the double space language
\begin{equation} \label{eq:qk_super}
\begin{aligned}
|q_k) = &\sum_{j_1<\dots <j_k}^L \left[\bigotimes_{n\notin \{j_i\}_{i=1}^k, n<j_k} |\widetilde{0})_n \right] \bigotimes_{n=1}^{k-1}[|\widetilde{\uparrow}) + |\widetilde{\downarrow})]_{j_n}\\
&\otimes[|\widetilde{\uparrow}) - |\widetilde{\downarrow})]_{j_k} \sqrt{3}^{L-j_k} \left[\bigotimes_{n=j_k+1}^L |\widetilde{e})_n \right] ~.
\end{aligned}
\end{equation}
It is easy to calculate the trial energy of $|q_k)$ in the super-Hamiltonian $\hat{\mathcal{P}}_{t-J_z|\mathrm{imp}}$ (with impurity at the right boundary $j_s=L$)
\begin{equation} \label{eq:exact_Delta_k}
\Delta^{\text{var}}_k = \frac{(q_k|\hat{\mathcal{V}}_L|q_k)}{(q_k|q_k)}=\frac{3g}{2}\frac{k}{L} \frac{2^k \binom{L}{k}}{\sum_{l=k}^L 2^l \binom{L}{l}},
\end{equation}
since $(q_k|\hat{\mathcal{P}}_{t-J_z|\mathrm{imp}}|q_k) = (q_k|\hat{\mathcal{V}}_L|q_k)$.
Note that it is not clear a priori what this variational energy implies for general $k$, e.g., it is already not useful as an upper bound on the ground state energy compared to $\Delta^{\text{var}}_1$ given by Eq.~\eqref{eq:left_sliom_energy_gap}, and there can be many states with energies between the ground state and $\Delta^{\text{var}}_k$.
However, the very construction of the super-Hamiltonian has important physics in it that makes $\Delta^{\text{var}}_k$ useful as follows.
We can consider the time autocorrelation function of a SLIOM $q_k$ with itself as defined in Eq.~\eqref{eq:Inn_prod}.
Adopting a different normalization convenient for discussing Mazur bounds and using Eq.~\eqref{eq:CBAt}, we have
\begin{align} \label{eq:Cq_t}
C_{q_k}(t) \equiv \frac{(q_k| e^{-\hat{\mathcal{P}}_{t-J_z|\mathrm{imp}}t} |q_k)}{(q_k|q_k)},
\end{align}
which starts at $C_{q_k}(t=0) = 1$ and whose time derivative is given by
\begin{equation}
    \frac{d}{dt} C_{q_k}(t) = - \frac{(q_k| e^{-\hat{\mathcal{P}}_{t-J_z|\mathrm{imp}}t}\hat{\mathcal{P}}_{t-J_z|\mathrm{imp}} |q_k)}{(q_k|q_k)}.
\end{equation}
Because of the positive-semidefiniteness of $\hat{\mathcal{P}}_{t-J_z|\mathrm{imp}}$,
at any time $t>0$ the right-hand side is negative and its absolute value is upper-bounded by its value at $t=0$:
\begin{align} \nonumber
\sum_\mu \frac{|(q_k|E_\mu)|^2}{(q_k|q_k)} e^{-E_\mu t} E_\mu  \leq \sum_\mu \frac{|(q_k|E_\mu)|^2}{(q_k|q_k)} E_\mu,
\end{align}
where we have used the orthonormal basis $\{|E_\mu)\}$ of eigenstates of the super-Hamiltonian, similar to Eq.~\eqref{eq:CBAt}.
 Hence,
\begin{align}\label{eq:Cqk_decay_rate}
\left|\frac{d}{dt} C_{q_k}(t) \right| \leq  \frac{(q_k| \hat{\mathcal{P}}_{t-J_z|\mathrm{imp}} |q_k)}{(q_k|q_k)}=\Delta_k^{\text{var}}.
\end{align}
Hence, the decay rate of the SLIOM (as a quasi-conserved quantity) autocorrelation is upper-bounded by the variational energy $\Delta_k^{\text{var}}$ at all times.

Of interest to us are consequences for autocorrelations of local observables like $Z_j$.
Intuitively, if a given observable receives a sizable contribution to its unperturbed Mazur bound from such $q_k$'s because of the sizable overlaps with them, we expect that the slow relaxation of $q_k$'s also implies slow relaxation of the original Mazur bound.
For example, we could expand $|Z_j)$ in an orthogonal basis obtained by extending the orthogonal basis $\{|q_k)\}$ and write
\begin{equation}
\begin{aligned}
    &C_Z(j,j;t) = \frac{1}{3^L}\left(\sum_{k}\frac{|(q_k|Z_j)|^2}{(q_k|q_k)}C_{q_k}(t)\right. \\+ & \left.\sum_{k\neq k'} \frac{(q_k| e^{- \hat{\mathcal{P}}_{t-J_z|\mathrm{imp}}t} |q_{k'})}{\sqrt{(q_k|q_{k})(q_{k'}|q_{k'})}} \frac{(Z_j|q_k)}{\sqrt{(q_k|q_k)}} \frac{(q_{k'}|Z_j)}{\sqrt{(q_{k'}|q_{k'})}} + \dots \right) 
    \end{aligned}
\end{equation}
where ``$\dots$'' contains also non-SLIOM contributions.
While we cannot prove this expectation rigorously even for the exhibited terms involving only the SLIOMs, it seems plausible that the slow decay exhibited by the first line will be at least part of the overall slow decay of the whole autocorrelation function.
[One may be able to find rigorous arguments when the SLIOMs are exponentially localized operators (i.e., essentially LIOMs), as happens for the boundary SLIOMs in the $t-J_z$ model or bulk blockade SLIOMs in the $H_3$ model discussed in the next section, such that the unperturbed Mazur bound $\frac{1}{3^L} \sum_{k} \frac{|(q_k|Z_j)|^2}{(q_k|q_k)}$ is finite.
In fact, in such cases, the LIOMs themselves could be viewed as essentially local observables, so their proved slow decay also proves slow decay of autocorrelations of local observables.
We leave more rigorous such pursuits for future studies.]

Let us examine $\Delta_k^{\text{var}}$ in Eq.~\eqref{eq:exact_Delta_k} in more detail.
Assuming that $k < \frac{2}{3} L$, i.e., for SLIOMs measuring the charges of the first $\frac{2L}{3}$ particles, and using the approximation $(q_k|q_k) \approx 3^L$ (which is valid with an exponentially small relative error, e.g., in the regime $L \to \infty$ keeping fixed $k/L < 2/3$), we find the approximate form
\begin{equation} \label{eq:Deltak}
\Delta_k^{\text{var}}\approx \frac{3g}{\sqrt{8\pi}}\frac{\sqrt{k}e^{-f(\frac{k}{L})L}}{\sqrt{L(L-k)}},
 \end{equation}
with 
\begin{equation}
f(\alpha) =\ln(3) - \alpha\ln(2) + \alpha \ln(\alpha) + (1-\alpha) \ln(1-\alpha),
\end{equation}
such that $0\leq f(\frac{k}{L}) \leq \ln(3)$.
Hence, \textit{all} SLIOMs $q_k$ with $k<\frac{2}{3}L$, not only the leftmost one, take an exponentially long time to feel the impurity, which as we show in the following leads to an exponentially long time non-trivial Mazur bounds even in the bulk of the system.
\subsubsection{Structure of the super-Hamiltonian in degenerate perturbation theory}
The use of the commutant algebra, in the super-Hamiltonian formulation, allows us to make the previous observation more precise.
In particular, we consider the effect of the impurity on the highly degenerate ground state subspace given by the exponentially many conserved quantities of the unperturbed system using degenerate perturbation theory.
Although such an application of degenerate perturbation theory will be limited,~\footnote{The success depends on the relation between the gap above this degenerate subspace, and the gap within this subspace predicted by the degenerate perturbation theory. 
Perturbation theory is expected to work only if the former is much larger than the latter.
Indeed it fails for the charge and/or dipole symmetries discussed in previous sections.} we will argue that we can accurately capture the late-time physics with such a projection.
Recall that the unperturbed ground states of the super-Hamiltonian $\hat{\mathcal{P}}_{t-J_z|\mathrm{comp}}$ are given by $\{|G^{\sigma_1\sigma_2\dots\sigma_l}) \}$ in Eq.~\eqref{eq:tJz_GS}, which correspond to spin patterns $\{ (\widetilde{\sigma}_1, \widetilde{\sigma}_2, \ldots, \widetilde{\sigma}_l)\}$.
Moreover, low-lying excitations on top of these ground states have an energy gap that scales as $L^{-2}$ with system size, resulting all together in tracer diffusion~\cite{moudgalya2024symmetries}.
We can also choose a different basis for the ground state manifold, which is associated with composite spin patterns in the local basis of $\widetilde{\sigma} \in \{\widetilde{\rightarrow}, \widetilde{\leftarrow}\}$ as introduced in Eq.~\eqref{eq:X_spins}.
That is, the same expression Eq.~\eqref{eq:tJz_GS} but with $\ket{\widetilde{\sigma}} \in \{\ket{\widetilde{\rightarrow}}, \ket{\widetilde{\leftarrow}}\}$ gives an (unnormalized) orthogonal basis of the ground state manifold.\footnote{This can be seen directly using the corresponding $|G^{\{\sigma\}})$'s by expressing $\ket{\widetilde{\rightarrow}},\ket{\widetilde{\leftarrow}}$ in terms of $|\widetilde{\uparrow}\rangle, |\widetilde{\downarrow}\rangle$ and vice versa ~\cite{moudgalya_hilbert_2022}, or indirectly by noting that $\hat{\mathcal{P}}_{t-J_z|\mathrm{comp}}$ happens to have a formal SU(2) symmetry for the composite particle spin variables, while its ground state manifold is completely spanned by the $|G^{\{\sigma\}})$-states~\cite{moudgalya2024symmetries}.}
In this basis, the impurity-generated operator $\widetilde{F}_{j_s}$ entering Eq.~\eqref{eq:Pimp_tJz} takes the form 
\begin{equation}
    \widetilde{F}_{j_s} = \frac{1}{2} \left(\sqrt{2}|\widetilde{0}\rangle \langle \widetilde{\rightarrow}| + \sqrt{2}|\widetilde{\rightarrow}\rangle \langle \widetilde{0}| + |\widetilde{\rightarrow}\rangle \langle \widetilde{\rightarrow}| - |\widetilde{\leftarrow}\rangle \langle \widetilde{\leftarrow}| \right)_{j_s} \!\!,
\label{eq:imp_Hamil}
\end{equation}
which physically signifies that the impurity can add or remove a $|\widetilde{\rightarrow}\rangle$ composite spin, leaving the $|\widetilde{\leftarrow}\rangle$ spins unchanged.
Moreover, reminscent to the U(1) case in Sec.~\ref{sec:U1}, the local spin operators $\{|S_j^z)\}$ only have overlaps with $|G^{\{\sigma\}})$ whose composite spin patterns have a single ``flip'' $\widetilde{\leftarrow}$, i.e., $(\widetilde{\rightarrow}_1 \widetilde{\rightarrow}_2 \ldots \widetilde{\rightarrow}_{k-1} \widetilde{\leftarrow}_k \widetilde{\rightarrow}_{k+1} \ldots \widetilde{\rightarrow}_{l})$ with the flip on the $k$-th physical composite spin state as counted from the left.
We denote the (normalized) ground states $|G^{\{\sigma\}})$ with such spin patterns as $|k,l)$, with $1 \leq k \leq l \leq L$ and $(k,l|k',l')=\delta_{k,k'}\delta_{l,l'}$.
Due to the form of the impurity Hamiltonian of Eq.~(\ref{eq:imp_Hamil}), such a term can connect a given $|k,l)$ only to $|k,l \pm 1)$.
Therefore, the effective Hamiltonian that emerges when projecting the action of the impurity within the subspace spanned by $\{|k,l)\}$ block-diagonalizes into  $\mathbb{H}_{\mathrm{eff}} = g(\oplus_{k=1}^L \mathbb{H}_{k}) \oplus \mathbb{H}^{\perp}$, where $\mathbb{H}_k$ acts in the subspace spanned by $\{|k,l), l=k,k+1,\ldots,L \}$, and $\mathbb{H}^{\perp}$ acts in the subspace associated with spin patterns with more than one flip.
We will focus on the former ``single-flip" subspaces for two reasons:
First, they contain the SLIOMs, $|q_k) \in \text{span}\{|k,l), l = k, \dots, L\}$.
Second, they also contain the projections of the spin operators onto the unperturbed commutant, $\Pi_{\mathcal{C}_{t-J_z}} |S_j^z) \in \text{span}\{|k,l),~k=1,\dots,j;~l=k,\dots, L\}$, which are relevant for the discussion of the unperturbed Mazur bounds and how the spin autocorrelations at long times are affected by the impurity.
In the subspaces of interest, we have
\begin{equation}\label{eq:tJz_single_H_eff}
\begin{aligned}
        \mathbb{H}_k &= \mu_k^{+} |k,k)(k,k| + \sum_{l=k+1}^L \mu_l^{-} |k,l)(k,l| \\
        &- \sum_{l=k}^{L-1} t_l \big(|k,l+1)(k, l| + \mathrm{h.c.} \big), \\
        \mu_k^{+} &= 1 + \frac{k}{2L},\, \mu_l^{-} = 1 - \frac{l}{2L},\, t_l = \frac{\sqrt{(l+1)(L-l)}} {\sqrt{2} L}.
\end{aligned}
\end{equation}
Note that each $\mathbb{H}_{k}$ is a tri-diagonal matrix---a kind of inhomogeneous nearest-neighbor ``hopping problem" in the $l$ parameter---in the appropriately organized basis of single-flip patterns, whose eigenstates thus govern the long-time dynamics of autocorrelation functions in our original physical problem.

\begin{figure}[h]
\includegraphics[width=1\linewidth]{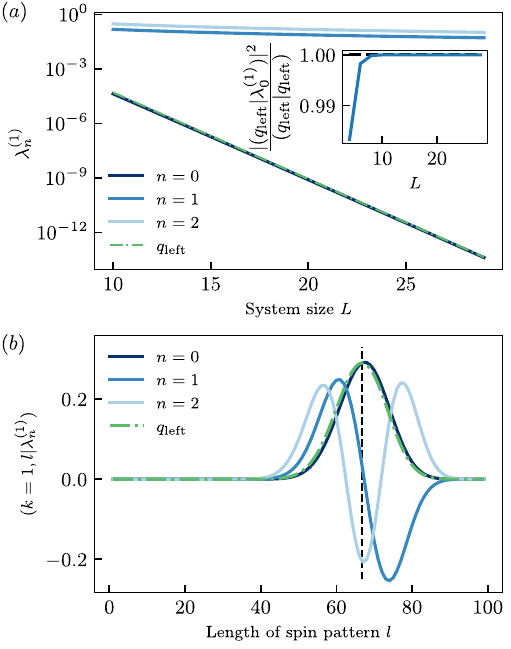}
\caption{\label{fig:tJz_Hk_spectrum} \textbf{Effective Hamiltonian $\mathbb{H}_1$ for the $t-J_z$ model with an impurity on the right boundary.}
(a) The low energy spectrum $\lambda_n^{(1)}$ of $\mathbb{H}_1$ for varying system size $L$.
The lowest eigenvalue $\lambda_0^{(1)}\sim \mathcal{O}(3^{-L})$ is upper bounded by $\Delta_1^{\mathrm{var}}$ given by $|q_{\mathrm{left}})$, with the corresponding eigenstate $|\lambda_0^{(1)})$ becoming essentially $\propto |q_{\text{left}})$ as shown in the inset.
The higher eigenvalues decay as $\mathcal{O}(L^{-1})$.
(b) Profiles of several low-lying eigenstates $(k,l|\lambda_n^{(k)})$ of $\mathbb{H}_k$ for $k=1$ and system size $L=100$.
The eigenstates are localized around $l_{\mathrm{max}} = \frac{2}{3}L$ (dashed black vertical line). 
Moreover, the lowest eigenstate $n=0$ of $\mathbb{H}_1$ can be well approximated by the leftmost SLIOM $|q_{\mathrm{left}})$.}
\end{figure}

\subsubsection{Low-energy spectrum of the super-Hamiltonian}\label{subsubsec:superhamlow}
We now study the low-energy spectrum of $\mathbb{H}_1$, which is relevant for the autocorrelation function ${C_{S_1^z}}(t)$ at the left boundary.
The leftmost SLIOM $|q_{\textrm{left}})$ resides completely in the space spanned by $\{|k=1,l)\}$ and takes the form
\begin{equation} \label{eq:qlef_kl}
    |q_{\textrm{left}}) = \sum_{l=1}^L \sqrt{2^{l} \binom{L}{l}} |k=1, l).
\end{equation}
As shown in Ref.~\cite{2020_SLIOMs}, $q_{\textrm{left}}$ is sufficient to prove a finite Mazur bound for the unperturbed system. 
Hence, we expect $|q_{\textrm{left}})$ to play an important role in the late-time behavior of this correlation function.
Recall that in Eq.~(\ref{eq:left_sliom_energy_gap}) we found that $\Delta_1^{\mathrm{var}}$ 
is exponentially small in system size.
This result is more clear when computed using $\mathbb{H}_1$ and in the $\{|1,l), 1 \leq l \leq L\}$ basis.
On the one hand, using Eq.~(\ref{eq:tJz_single_H_eff}) one finds that the contribution of the impurity $\hat{\mathcal{V}}_L$ only comes from the left `boundary' $l=1$ of the formal hopping problem $\mathbb{H}_1$, i.e., $(q_{\textrm{left}}|\mathbb{H}_1|q_{\textrm{left}})=(q_{\textrm{left}}|1,1)(1,1|\mathbb{H}_1|q_{\textrm{left}})$, although this is not directly obvious from the form of $\mathbb{H}_1$.
On the other hand, from Eq.~\eqref{eq:qlef_kl} and in the limit of large $L$, it is easy to explicitly evaluate that $|q_{\textrm{left}})$ is localized around the basis state $|1,l)$ with $l_{\textrm{max}}=\frac{2}{3}L$, and with most of its weight contained on a window of $l$'s of width $\sqrt{L}$ around $l_{\textrm{max}}$~\footnote{This is the localization in this basis $\{|k, l)\}$ in the super-Hamiltonian, and should not be confused with the real-space localization on the original chain.}.
This explains the exponentially small expectation value of $\mathbb{H}_1$ evaluated on $|q_{\textrm{left}})$. 
Given that $\mathbb{H}_1$ is non-negative, one then expects that the (properly normalized) leftmost SLIOM $|q_{\textrm{left}})$ is a good trial state of the ground state $|\lambda^{(1)}_0)$ of $\mathbb{H}_1$.
In Fig.~\ref{fig:tJz_Hk_spectrum}(a), we show the low energy spectrum of $\mathbb{H}_1$, where the lowest energy scales as $\mathcal{O}(3^{-L})$ as predicted by Eq.~\eqref{eq:left_sliom_energy_gap} for an impurity at $j_s=L$.
Also, the inset shows that $|q_{\mathrm{left}})$ is indeed a good approximation of the lowest eigenstate, with the approximation improving with increasing system size, i.e., $\lim_{L\rightarrow\infty} |(\lambda^{(1)}_0|q_{\textrm{left}})|^2 /(q_{\textrm{left}}|q_{\textrm{left}}) \rightarrow 1$.
Figure~\ref{fig:tJz_Hk_spectrum}(b) shows the amplitude $(1,l|\lambda_0^{(1)})$ of the ground state $|\lambda_0^{(1)})$ (dark green data set corresponding to $n=0$) in the $|1,l)$ basis, in comparison to that of the leftmost SLIOM $|q_{\textrm{left}})$ showing a good quantitative agreement with the same $l$-parameter localization around $l=\frac{2L}{3}$ (gray dashed vertical line). 
In App.~\ref{app:num_check}, we also compare the ground state $|\lambda_0^{(1)})$ of $\mathbb{H}_1$ with the first excited state $|E_1)$ of the super-Hamiltonian $\hat{\mathcal{P}}_{t-J_z|\mathrm{imp}}$ in Eq.~\eqref{eq:Pimp_tJz} in the full composite-spin subspace (i.e., without any perturbative approximations) for small system sizes.
We find that $|\lambda_0^{(1)})$ is indeed a good approximation of $|E_1)$, and the approximation improves with increasing system sizes such that $|(\lambda_0^{(1)}|E_1)|^2 \rightarrow 1$.\footnote{Note that the ground state $|\mathbb{1})$ of $\hat{\mathcal{P}}_{t-J_z|\mathrm{imp}}$
is outside of the single-flip subspace, therefore the ground state $|\lambda_0^{(1)})$ of $\mathbb{H}_1$, which is restricted to the single-flip subspace, approximates the first excited state $|E_1)$.}
We also numerically find that the low-lying excited states of $\mathbb{H}_1$ have an energy gap of $\lambda_1^{(1)}\sim \mathcal{O}(L^{-1})$.
This observation can also be directly understood from the localization properties of the ground state or, similarly, those of the leftmost SLIOM $|q_{\textrm{left}})$.
Since only the basis elements $|1,l)$ with $l$ in a window of size $\sqrt{L}$ around $l=\frac{2L}{3}$ contribute to $|q_{\textrm{left}})$, we can approximate $\mathbb{H}_1$ by a Hamiltonian with uniform hopping $t_l\approx \frac{1}{3}$ and uniform chemical potential $\mu^-_l\approx \frac{4}{3}$ on a chain of length $\sqrt{L}$ (with OBC).
The lowest energy of the local hopping band is $\mu^{-} - 4t$ and vanishes at this location, giving quadratic local dispersion.
Hence, low-lying excitations correspond to modes localized within such a region of size $\sim \sqrt{L}$, have momentum of $\sim 1/\sqrt{L}$, and thus have energy $\lambda_1^{(1)}\propto L^{-1}$. 
However, unlike for the exponentially small splitting of the ground state degeneracy given by $\lambda_0^{(1)}$, the low-lying excitations $\lambda^{(1)}_1 \propto L^{-1}$ of $\mathbb{H}_1$ are parametrically larger than the unperturbed low-lying excitations with energies $\sim L^{-2}$. 
Hence, this limits the applicability of our perturbative approach regarding the time scales of correlation functions. 
An instructive analysis comparing similar perturbative treatment to the non-perturbative approach discussed in Sec.~\ref{sec:U1_break} for a U$(1)$ symmetric system in the presence of a symmetry-breaking impurity, can be found in App.~\ref{app:U1_pert}. 
Nonetheless, as we discuss in the following section, the energies $\lambda_1^{(1)}$ lower bound the timescales at which charge correlations functions are expected to be qualitatively modified.
Moreover, the previous discussion clearly shows that the relevant physics is governed by a region far from the left end in the parameter $l$ in the effective Hamiltonian $\mathbb{H}_1$.
Hence, one can instead consider a closely related ``parent" Hamiltonian with an added ``site'' $l=0$, from which we can directly extract the low-energy properties of $\mathbb{H}_k$ as we discuss in App.~\ref{subsec:parent_H0}.

\begin{figure}[tb]
\includegraphics[width=1\linewidth]{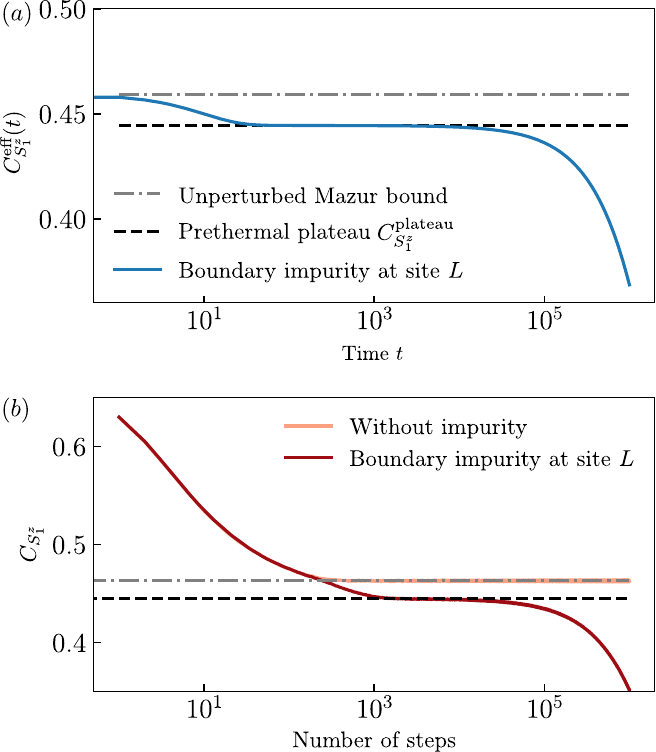}
\caption{\label{fig:tJz_AC_edgecorre} \textbf{Boundary autocorrelation functions of the $t-J_z$ model with a state-flip impurity at the right boundary $L$.} 
(a) Effective correlation functions obtained from the effective Hamiltonian $\mathbb{H}_{\mathrm{eff}}$ given by Eq.~\eqref{eq:tJz_single_H_eff} for $L=15$.
Note that small system sizes are chosen to observe the prethermal plateaus.
The effective correlation function starts from the Mazur bound of the unperturbed $t-J_z$ model (grey dash-doted line),
then evolves to a prethermal plateau (black dashed line) contributed by $|\lambda_0^{(1)}) \approx \mathcal{N} |q_{\mathrm{left}})$ (with $\mathcal{N}$ a normalization factor) at intermediate times, as described by Eq.~\eqref{eq:tJz_Cz1_timescale}.
The correlation function eventually vanishes as $e^{-g \lambda_0^{(1)}t}$ at late times $t \sim (g \lambda_0^{(1)})^{-1} \sim g^{-1} 3^{-L}$.
(b) The correlation functions obtained from stochastic dynamics for $L=12$.
Note that here we also have access to the very initial dynamics governed primarily by the unperturbed $t-J_z$ model, while the effective correlation in panel (a) starts after the dynamics has effectively settled to the unperturbed Mazur bound.
Without impurity, the correlation function saturates to the unperturbed Mazur bound.
With the local impurity, the correlation function shows similar behaviors as in (a).
}
\end{figure}
\subsubsection{Behavior of boundary correlation functions}
Based on this analysis of the spectrum of $\mathbb{H}_1$, we can now understand the late-time dynamics of the boundary spin autocorrelation $C_{S_1^z}(t)$.
As we have argued, this is governed by the exponentially slow mode of $\mathbb{H}_1$, namely $|\lambda_0^{(1)}) \sim 
|q_{\textrm{left}})$, with $\lambda_0^{(1)} \sim  3^{-(L-1)}$ as calculated in Eq.~\eqref{eq:left_sliom_energy_gap}
%
Then at long times, the correlation function is given by
\begin{equation}\label{eq:tJz_Cz1_timescale}
\begin{aligned}
C_{S_1^z}(t)\approx 
\begin{cases} 
 \frac{4}{9(1-3^{-L})} &\text{for~} t_{\mathrm{start}} \ll t \ll (g \lambda^{(1)}_0)^{-1}  \\
 \frac{4e^{-g \lambda^{(1)}_0 t}}{9(1-3^{-L})}  &\text{for~} t \gg (g \lambda^{(1)}_0)^{-1} ~,
\end{cases}
\end{aligned}
\end{equation}
where the first line corresponds to an exponentially long intermediate time window governed by the leftmost SLIOM $|q_{\textrm{left}})$, while other originally conserved quantities contributing to the unperturbed Mazur bound have decayed.
Note that higher modes (contributing to the boundary correlation) have a gap that is polynomial in system size and upper bounded by the gap of the $\mathbb{H}_1$ Hamiltonian.
Hence, the starting time of the prethermal plateau satisfies $t_{\mathrm{start}} > (g \lambda_1^{(1)})^{-1} \sim \mathcal{O}(L/g)$.
However, $t_{\mathrm{start}}$ can scale as $\mathcal{O}(L^p)$ with $p>1$, since
As discussed before, there can be lower-energy modes than $\lambda_1^{(1)}$ derived from perturbation theory.
This indicates that there is a prethermal plateau of the boundary correlation functions. 
We numerically evaluate the \textit{effective} boundary autocorrelation functions obtained from the effective Hamiltonian $\mathbb{H}_1$.
Note that unlike the U$(1)$ case studied in Sec.~\ref{sec:U1_break}, here we can define an effective correlation function obtained from the application of degenerate perturbation theory, i.e., completely from the subspace spanned by the ground states of the unperturbed super-Hamiltonian.
This is different from the true correlation function, whose behavior in time would require studying low-energy excitations for the full super-Hamiltonian, while our treatment focuses only on the original ground state manifold and effects of the impurity there.
Nevertheless, as we discuss below, the former captures the important late-time features of the latter.
Figure~\ref{fig:tJz_AC_edgecorre}(a) shows the dynamics of the effective boundary autocorrelation function $C_{S_1^z}^{\mathrm{eff}}(t)$, which is given by
\begin{equation}  \label{eq:tJz_Ceff_edge_evol}
    C^{\textrm{eff}}_{S_1^z}(t) = \frac{1}{3^L}\sum_n e^{- E_n^{(1)} t}|(S_1^z|\lambda^{(1)}_n)|^2,
\end{equation}
with $|\lambda_n^{(1)})$ the full set of eigenstates of $\mathbb{H}_1$ and energy $E_n^{(1)}$.\footnote{Note that the overall factor $\frac{1}{3^L}$ relates to our choice of the inner product in the double space such that $ (\mathbb{1}|\mathbb{1}) = 3^L$, see Eq.~\eqref{eq:Inn_prod}.}
At $t=0$, $C^{\textrm{eff}}_{S_1^z}(t=0)$ corresponds to the full Mazur bound $2(2L+1)/9L$ of the unperturbed $t-J_z$ model at the left boundary (shown with dash-doted gray line).
After a time scale $t\sim \mathcal{O}(L)$, which for the specific object in Eq.~\eqref{eq:tJz_Ceff_edge_evol} provides an exact rather than a loose lower bound, $C^{\textrm{eff}}_{S_1^z}(t)$ decays to the prethermal plateau $C_{S_1^z}^{\mathrm{plateau}} = \frac{4}{9(1-3^{-L})}$ (black dashed line), which is the contribution by the exponentially slow mode $|q_{\mathrm{left}})$ (see Fig.~\ref{fig:tJz_Hk_spectrum}).
Eventually, at times of $t \sim \mathcal{O}(e^L)$, $C^{\textrm{eff}}_{S_1^z}(t)$ vanishes exponentially due to the finite gap of the $\sim |q_{\text{left}})$ in finite system size.
In Fig.~\ref{fig:tJz_AC_edgecorre}(b), we show the true boundary autocorrelation function obtained from stochastic cellular automaton dynamics. 
Without impurity, the autocorrelation function saturates to the unperturbed Mazur bound. 
On the other hand, in the presence of an impurity at the right boundary, the autocorrelation function $C_{S_1^z}$ shows only a very weak shoulder at the unperturbed Mazur bound and quickly decays to a prethermal plateau $C_{S_1^z}^{\textrm{plateau}}$, and eventually vanishes at very long times.
The stochastic dynamics of the true correlation function at times $t \gtrsim \mathcal{O}(L^{p})$ (more general polynomial that we do not try to determine at this point) is captured by the behavior of $C_{S_1^z}^{\mathrm{eff}}(t)$.

\subsubsection{Behavior of bulk correlation functions}
We can perform a similar analysis for bulk correlations $C_{S_j^z}(t)$, which, in degenerate perturbation theory, are controlled by the spectra of $\mathbb{H}_k$ with $k\leq j$. 
Analogous to our discussion for $\mathbb{H}_1$ in Sec.~\ref{subsubsec:superhamlow}, the SLIOM $|q_k)$, Eq.~\eqref{eq:qk_super}, lies within the subspace spanned by the basis $\{|k,l),~ l=k,k+1,\dots,L\}$, and is given by
\begin{equation}\label{eq:qk_kl}
    |q_k) = \sum_{l=k}^L \sqrt{2^l \binom{L}{l}} |k, l).
\end{equation} 
Following the same reasoning as there, we find that $|q_k)$ for $k<\frac{2L}{3}$, is localized in a region of size $\sqrt{L}$ around the basis state $|k,l)$ with $l=\frac{2}{3}L$, and that it provides a good approximation for the ground state of $\mathbb{H}_k$ by the same line of reasoning.
Its variational energy $\Delta_{k}^{\mathrm{var}}$ is given by Eq.~\eqref{eq:exact_Delta_k}. %
We have numerically confirmed that indeed for $k< \frac{2}{3} L$, $|q_k)$ converges to the ground state $|\lambda_0^{(k)})$ of $\mathbb{H}_k$, i.e., $|(q_k|\lambda_0^{(k)})|^2/(q_k|q_k) \rightarrow 1$ for increasing system size $L$, leading to similar results as those shown in Fig.~\ref{fig:tJz_Hk_spectrum} for $\mathbb{H}_1$.
The reason why $\frac{k}{L}$ is restricted to the domain $[0,\frac{2}{3})$ relates to the following two facts which were shown in Ref.~\cite{2020_SLIOMs}: (i) finding more than $\frac{2}{3}L$ particles in a Haar random state is unlikely, and (ii) that $|q_k)$ is statistically localized around location $\frac{3}{2}k$, which goes beyond the system size for $k \geq \frac{2}{3}L$. 
And the low-lying excited states $|\lambda_0^{(k)})$ are separated by a gap scaling as $L^{-1}$ with system size.

Writing $k=\frac{2}{3}(1-\beta)L$, such that $\beta L$ is the distance from the location of the main weight of $q_k$ on the chain (most probable location of the $k$th charge) to the impurity, with $\lambda_0^{(k)} \sim \Delta_{k}^{\mathrm{var}}$ in Eq.~(\ref{eq:Deltak}), acquires a more insightful functional form, such that
\begin{equation}\label{eq:tJz_E0k_scaling}
\lambda_0^{(k)}\sim e^{-\beta ^2L}
\end{equation}
for small $\beta$. 
Namely, the time scale on which a site at the distance $\beta L$ from the right boundary impurity notices its effect, scales exponentially with the distance.

Therefore, we find that the ground states $|\lambda_0^{(k)})$  of $\mathbb{H}_k$, which are well approximated by the SLIOMs $|q_k)$, have a gap that approximately scales exponentially with system size as $e^{-f(\frac{k}{L})L}$, where $f(\frac{k}{L})\geq 0$ is a monotonously decreasing function of $\frac{k}{L}$.
Combining these results together with our discussion below Eq.~\eqref{eq:Cq_t}, we then predict that a prethermal plateau also survives for bulk correlation functions $C_{S_j^z}(t)$.
In particular, 
we can estimate that
\begin{equation}
    C_{S_j^z}(t)\approx \begin{cases} 
 C_{S_j^z}^{\text{plateau}}\quad &\text{for~} t_{\mathrm{start}} \ll t \ll (g \lambda_0^{(2j/3)})^{-1}  \\
 0 \quad &\text{for~} t \gg  (g \lambda_0^{(2j/3)})^{-1},
\end{cases}
\end{equation}
with $(\lambda_0^{(2j/3)})^{-1} \approx e^{(1-j/L)^2 L}$, $t_{\mathrm{start}} > (g \lambda_1^{(2j/3)})^{-1} \sim \mathcal{O}(L/g)$, although it can have a different scaling $t_{\mathrm{start}} \sim \mathcal{O}(L^p)$, similar to the case in Eq.~\eqref{eq:tJz_Cz1_timescale}, and
\begin{equation}\label{eq:tJz_bulk_plateau}
\begin{aligned}
        C_{S_j^z}^{\text{plateau}} &\approx \frac{1}{3^L}\sum_{k=1}^j \frac{|(S_j^z|q_k)|^2}{(q_k|q_k)}\sim \frac{1}{\sqrt{L}}
\end{aligned}
\end{equation}
for $j$ lying in the bulk as found in Ref.~\cite{2020_SLIOMs}.
Notice that when correlation functions are evaluated at a finite distance from the impurity, this prethermal plateau is drastically shortened or even non-existent.

\begin{figure}[t!]
\includegraphics[width=1\linewidth, scale=1]{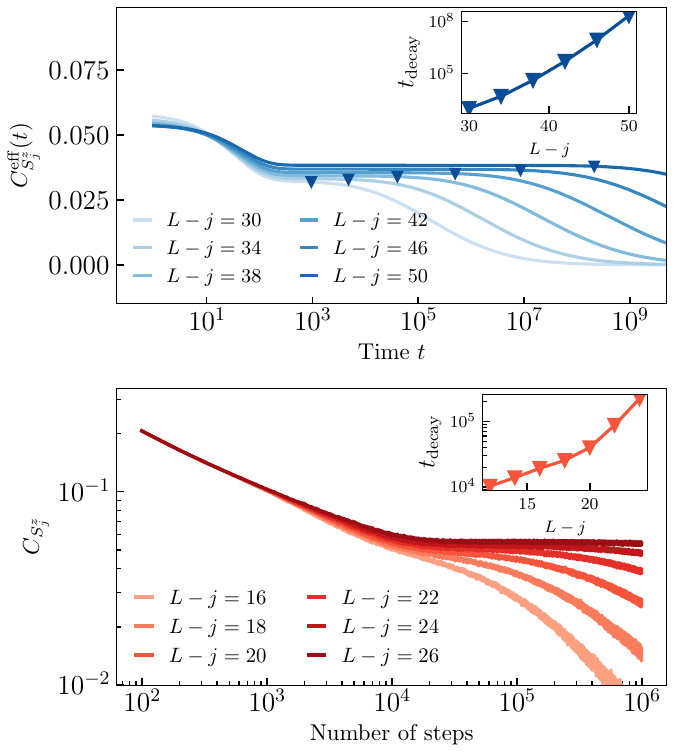}
\caption{\label{fig:tJz_AC_bulkcorre} \textbf{Bulk autocorrelation functions of the $t-J_z$ model with a state-flip impurity at the right boundary $L$.}
(a) Effective bulk correlation functions $C_{S_j^z}^{\mathrm{eff}}(t)$ obtained from the effective Hamiltonian $\mathbb{H}_{\mathrm{eff}}$ in Eq.~\eqref{eq:tJz_single_H_eff}, for system size $L=100$. The prethermal plateau lasts longer with the increase of the distance to the impurity $L-j$. 
We estimate the decay times $t_{\mathrm{decay}}$ (blue triangles) where the correlation functions decay by $0.5$\% of the corresponding plateau values.
The decay time $t_{\mathrm{decay}}$ grows faster than exponentially of $L-j$ as shown in the inset, which is predicted by Eq.~\eqref{eq:tJz_E0k_scaling}.
The correlation functions eventually vanish due to the restored ergodicity for finite $L$.
(b) Bulk autocorrelations of the $t-J_z$ model obtained from stochastic dynamics for system size $L=50$.
The bulk correlations decay to the prethermal plateau, and the timescales of the prethermal plateau increase when increasing the distance to the impurity. 
We also estimate the decaying time $t_{\mathrm{decay}}$ where the correlation functions decay by $5$\% times the plateau values, which scales faster than exponentially with $L-j$ as shown in the inset.
Each curve is averaged over $10^8$ random realizations. 
}
\end{figure}

Similar to our discussion for the boundary correlations, we confirm these scalings by numerically computing the evolution of the effective bulk autocorrelation functions $C^{\textrm{eff}}_{S_j^z}(t)$ as obtained from the effective Hamiltonians $\mathbb{H}_{k\leq j}$ in Eq.~\eqref{eq:tJz_single_H_eff}. 
The effective bulk autocorrelation function is given by 
\begin{equation}  \label{eq:tJz_Ceff_bulk_evol}
    C^{\textrm{eff}}_{S_j^z}(t)= \frac{1}{3^L}\sum_{k=1}^j\sum_n e^{- \lambda_n^{(k)} t}|(S_j^z|\lambda^{(k)}_n)|^2,
\end{equation}
whose dependence on time is shown in Fig.~\ref{fig:tJz_AC_bulkcorre}(a).
The effective bulk correlation functions evolve to the prethermal plateau given by Eq.~\eqref{eq:tJz_bulk_plateau} and then decay exponentially.
The inset shows that the prethermal plateau lasts for a time $t_{\mathrm{decay}}$ that scales exponentially with the distance to the impurity.
We numerically confirm these predictions with cellular automaton simulations in Fig.~\ref{fig:tJz_AC_bulkcorre}(b).
We observe that before the impurity takes effect, there exists an initial regime showcasing subdiffusive behavior of $C_{S_j^z}(t) \sim t^{-1/4}$.
Such initial power law decay is expected from the tracer diffusion picture due to pattern conservation in the unperturbed $t-J_z$ model~\cite{barma1994slow, menon1995irreducible, Feldmeier_2022_tracer, moudgalya2024symmetries}.
Then we see a similar prethermal plateau to that in the effective autocorrelation from the super-Hamiltonian approach that lasts exponentially long with the distance to the impurity.
Finally, we conjecture that various hydrodynamic tails (potentially also on top of observed prethermal plateaus) will emerge, including crossovers among various regimes at times that depend on the distance to the impurity, analogous to effect of impurities breaking conventional conserved quantities.
\subsection{$3$-local dipole-conserving model} \label{sec:H3}
Many aspects of the analysis of the $t-J_z$ model extends to the spin-$1$ dipole-conserving model $H_3$, which exhibits strong fragmentation~\cite{2020_khemani_local, 2020_sala_ergodicity-breaking}.
The bond algebra is given by $3$-local dipole-conserving local terms, 
\begin{equation}
    \mathcal{A}_{H_3} = \lgen \{S_j^+ (S_{j+1}^-)^2 S_{j+2}^+ + \mathrm{h.c.}\}, \{S_j^z\}, \mathbb{1}\rgen, 
\label{eq:AH3}
\end{equation}
Note that the $\{S^\alpha_j\}$ are spin-1 operators, and the local Hilbert space is spanned by $\{0, +, -\}$; we will often refer to states $+$ and $-$ as positive and negative charges (site ``occupied" by a ``physical charge''), while $0$ as having no charge (``empty'' site).
The above three-site operators implement the following transitions:
\begin{align}
&\ket{0+0} ~\leftrightarrow~ \ket{+-+}, \quad
&\ket{0+-} ~\leftrightarrow~ \ket{+-0}, \nonumber \\
&\ket{0-0} ~\leftrightarrow~ \ket{-+-}, \quad
&\ket{0-+} ~\leftrightarrow~ \ket{-+0}.
\end{align}
We focus on a one-dimensional chain with OBC.
The $H_3$ model conserves both the total charge $Q=\sum_j S_j^z$ and dipole moment $P=\sum_j j S_j^z$.
In addition, there are exponentially many conserved quantities due to dynamical constraints~\cite{2020_SLIOMs, moudgalya_hilbert_2022}.
\subsubsection{Structure of the invariant subspaces and SLIOMs}
Reference~\cite{2020_SLIOMs} identified the resulting Krylov subspace structure by introducing defect degrees of freedom as follows.
For each product state configuration, if a given physical charge has the same sign as the physical charge to its left, this charge is marked as a defect with the corresponding charge.
The number $N_d$ and the charge pattern of such defects are conserved.
Also, the leftmost and rightmost physical charges are conserved (one needs to specify additionally whether the rightmost defect and the rightmost charge refer to the same physical charge or not, since both cases can happen).
Moreover, $N_d$ defects separate the chain into $N_d+1$ regions, and the dipole moment for each region is conserved, specifically: $P_0$ associated with the region from the left end to the first defect (exclusive); $P_k$ associated with the region between the $k$th defect (inclusive) and $(k+1)$-th defect (exclusive), $1 \leq k < N_d$; and  $P_{N_d}$ associated with the region between the $N_d$-th defect (inclusive) and the right end of the chain.
For an alternative visualization of the Krylov subspaces and the commutant algebra perspective, see Ref.~\cite{moudgalya_hilbert_2022}.
The Krylov subspace labels in terms of the above conserved patterns of defects (including also the leftmost and rightmost conserved charges), as well as the conserved dipole moments of the intervening regions, can be stated in terms of a set of SLIOMs, similar to the $t-J_z$ model.
For example, the operator that measures the leftmost physical charge is given by
\begin{equation}\label{eq:H3_ql}
\begin{aligned}
\hat{\sigma}_{\text{left}} &= \hat{\Pi}_{\text{left},+} - \hat{\Pi}_{\text{left},-}, \\
\hat{\Pi}_{\text{left},c} &= \sum_{j=1}^L \left(\bigotimes_{k=1}^{j-1} \ketbra{0}_k \right) \ketbra{c}_j, \quad c \in \{+,-\},
\end{aligned}
\end{equation}
where for later convenience we have also introduced the projector $\hat{\Pi}_{\text{left},c}$ onto the subspace spanned by configurations with the  leftmost physical charge being $c$.
Similar to the $t-J_z$ model, $\hat{\sigma}_{\text{left}}$ is exponentially localized near the left boundary and is already sufficient to show non-zero Mazur bound for the autocorrelations of the boundary spin $S_1^z$.
One can similarly write expressions for the $k$th defect charge and the $P_k$ dipole moment using appropriate projectors, but they are not as simple or useful as in the $t-J_z$ model.
In addition, we also do not have a description of the full commutant algebra for $\mathcal{A}_{H_3}$ in terms of a basis with simple expressions via spin operators.
We do have a description in terms of projectors to the Krylov subspaces, but it is not easy to work with either~\cite{moudgalya_hilbert_2022}.
Nevertheless, we can understand important fragmentation physics here that is distinct from the $t-J_z$ model by more direct means.
For the purposes of our analysis, a dramatic aspect of the fragmentation in the $H_3$ model is the so-called blockade phenomenon~\cite{2020_khemani_local,2020_sala_ergodicity-breaking}:
There are dynamically preserved subspaces such that states in a given blockade region are completely frozen, while the regions to the right and to the left are effectively disconnected from each other.
As a simple example, consider configurations of the form $|\cdots + 0 \cdots 0\ \fbox{+ +}\ 0 \cdots 0 + \cdots \rangle$, where $\fbox{+ +}$ refers to two fixed consecutive sites $j_0, j_0+1$, and the closest non-zero charges to the right and to the left are also $+$.
It is easy to see that the $H_3$ dynamics preserves the form of such configurations; in particular, the charges at $j_0, j_0+1$ remain completely frozen, while the allowed configurations on the right of the blockade are independent of those on the left.
Hence, the projector to the space of such configurations is a conserved quantity of the $H_3$ terms.
We can write it as
\begin{equation}
\label{eq:B++SLIOM}
\hat{B}_{j_0,j_0+1}^{++} = \hat{\Pi}_{\text{right},+}^{[1, j_0-1]} \otimes \ketbra{++}_{j_0,j_0+1} \otimes \hat{\Pi}_{\text{left},+}^{[j_0+2, L]} ~,
\end{equation}
where $\hat{\Pi}_{\text{left},+}^{[j_0+2, L]}$ is analogous to $\hat{\Pi}_{\text{left},+}$ in Eq.~(\ref{eq:H3_ql}) except that it is defined on the segment $[j_0+2, L]$ instead of the whole chain, and similarly for the projector $\hat{\Pi}_{\text{right},+}^{[1, j_0-1]}$ onto the subspace spanned by configurations with the rightmost charge on the segment $[1, j_0-1]$ being $+$. 
For any $j_0$, including sites in the bulk, this SLIOM is exponentially localized near $j_0$, just like the rightmost and leftmost charges are localized near the corresponding boundaries.
In particular, this gives a non-zero contribution to the Mazur bound in the spin autocorrelations at $j_0$ that persists also in the thermodynamic limit.
Note that an accurate accounting of the full Mazur bound from all blockades, including larger blockaded regions over $j_0$ and nearby blockades, is complicated---see Ref.~\cite{moudgalya_hilbert_2022} for a more detailed discussion on this. 
Nevertheless, using the above simplest blockade SLIOM will already be sufficient for illustrative discussion of the effects of fragmentation-breaking impurities below.

\subsubsection{Impurity effects and timescales}
\begin{figure}[t!]
    \centering
    \includegraphics[width=1\linewidth]{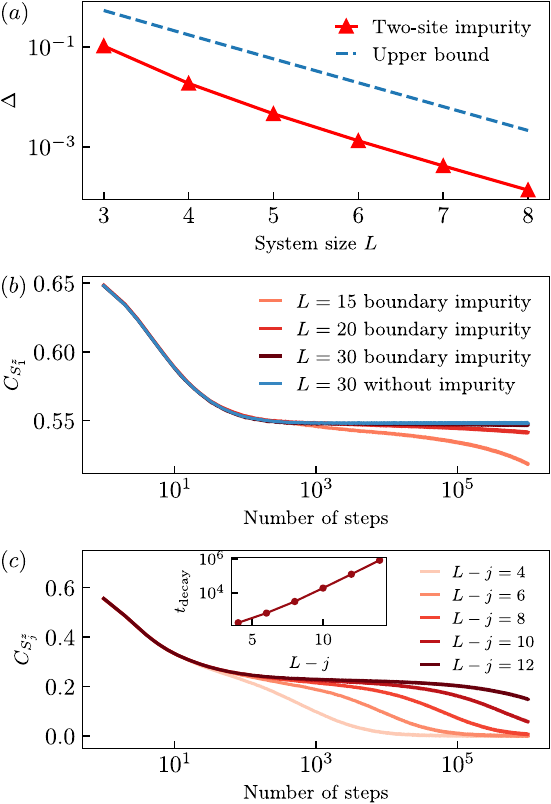}
    \caption{
    \textbf{Energy gap of super-Hamiltonian and autocorrelation functions of the strongly fragmented spin-$1$ $3$-local dipole-conserving model with a two-site state-flip impurity at the boundary.}
    (a) Energy gap of the super-Hamiltonian Eq.~\eqref{eq:H3imp_superH} with the two-site state-flip impurity at the right boundary, i.e., at sites $L-1$ and $L$, with impurity strength $g=1$. 
    The energy gap is upper bounded by the energy of the trial state $|\sigma_{\mathrm{left}})$, which scales as $\mathcal{O}(3^{-L})$.
    (b) Correlation function $C_{S_1^z}$ at the left boundary with or without impurity from stochastic dynamics.
    The prethermal plateau lasts longer with the increase of the system size.
    (c) Bulk correlation functions $C_{S_j^{z}}$ with different sites $j$ from stochastic dynamics for $L=50$. The prethermal plateau lasts longer with the increase of distance to the impurity.
    The decay time $t_{\mathrm{decay}}$ is estimated by the correlation functions decay to $C_{S_j^z}^{\mathrm{decay}} = 0.22$ and is shown in the inset; we see that $t_{\mathrm{decay}}$ scales exponentially with the distance to the impurity $L-j$.
    Each curve of correlations from stochastic dynamics is averaged over $10^8$ random realizations.
    }
    \label{fig:H3_deg_gap_corre}
\end{figure}

For quantitative studies of the impurity effect time scales, we now turn to the super-Hamiltonian formalism. 
Due to the inclusion of $S^z_j$ in the bond algebra of Eq.~(\ref{eq:AH3}), the super-Hamiltonian for the $H_3$ model can 
be restricted to the space of composite spins on a two-legged ladder, which are defined similarly to Eq.~\eqref{eq:compositespins} in the U($1$) conserving case.
The unperturbed super-Hamiltonian in the composite-spin subspace is thus given by 
\begin{equation}
\begin{aligned}
    \hat{\mathcal{P}}_{H_3|\mathrm{comp}} &= 2 \sum_j \Big[\left(|\widetilde{0}\widetilde{+}\widetilde{0}\rangle - |\widetilde{+}\widetilde{-}\widetilde{+}\rangle \right)\left(\langle \widetilde{0}\widetilde{+}\widetilde{0}| - \langle\widetilde{+}\widetilde{-}\widetilde{+}| \right) \\
    &+ \left(|\widetilde{0}\widetilde{+}\widetilde{-}\rangle - |\widetilde{+}\widetilde{-}\widetilde{0}\rangle \right)\left(\langle\widetilde{0}\widetilde{+}\widetilde{-}| - \langle\widetilde{+}\widetilde{-}\widetilde{0}| \right) \\
    &+ \left(|\widetilde{-}\widetilde{+}\widetilde{-}\rangle - |\widetilde{0}\widetilde{-}\widetilde{0}\rangle \right)\left(\langle\widetilde{-}\widetilde{+}\widetilde{-}| - \langle\widetilde{0}\widetilde{-}\widetilde{0}| \right) \\
    &+\left(|\widetilde{-}\widetilde{+}\widetilde{0}\rangle - |\widetilde{0}\widetilde{-}\widetilde{+}\rangle \right)\left(\langle\widetilde{-}\widetilde{+}\widetilde{0}| - \langle\widetilde{0}\widetilde{-}\widetilde{+}| \right)\Big]_{[j,j+2]}.
\end{aligned}
\end{equation}
By construction, all conserved quantities of $H_3$ discussed above are zero-energy ground states of this super-Hamiltonian.
For concreteness, we then consider a local spin-flip impurity 
\begin{equation}
    G_j = \frac{1}{2}\big(\ket{+}\bra{0} + \ket{0}\bra{-} + \ket{-}\bra{+} + \mathrm{h.c.} \big)_j
\end{equation}
added to the bond algebra terms generating the Brownian circuits.
The perturbed super-Hamiltonian restricted to the composite spin subspace is given by
\begin{equation}\label{eq:H3imp_superH}
    \hat{\mathcal{P}}_{H_3|\mathrm{imp}} = \hat{\mathcal{P}}_{H_3|\mathrm{comp}} + \hat{\mathcal{V}}_G,
\end{equation}
where $\hat{\mathcal{V}}_G$ already denotes restriction to the composite spin subspace and takes the form $\hat{\mathcal{V}}_{G_j} = g(1 - \widetilde{G}_j)$ with 
\begin{equation}
  \widetilde{G}_j= \frac{1}{2}\left( |\widetilde{+}\rangle \langle\widetilde{0}| + |\widetilde{0}\rangle \langle\widetilde{-}| + |\widetilde{-}\rangle \langle\widetilde{+}| + \mathrm{h.c.} \right)_j.
\end{equation}
Note that the range-$3$ dipole-conserving model exhibits more ``robust" fragmentation than the $t-J_z$ model due to the conservation of the dipole moment in between the defects in addition to the pattern of defects.
Thus, we find that adding an impurity on only {\it one} site leaves the system fragmented.
Indeed, for an impurity on the right boundary, we can show that only the rightmost defect can be modified for $N_d\geq2$, as we detail in App.~\ref{app:frag_1site}.
The full breaking of fragmentation can be ensured by adding two impurities at sites $L-1$ and $L$, which can now work as a dipole source/sink, and similarly for charge, with 
\begin{equation}
\label{eq:V2imp_4_H3}
\hat{\mathcal{V}}^{\text{2imp.}} = \hat{\mathcal{V}}_{G_{L-1}} + \hat{\mathcal{V}}_{G_L},
\end{equation}
where we find that this leads to a unique ground state $\sim |\mathbb{1})$ of $\hat{\mathcal{P}}_{H_3|\mathrm{imp}}$.
We now discuss our numerical study of the super-Hamiltonian Eq.~(\ref{eq:H3imp_superH}) with the two-site impurity Eq.~(\ref{eq:V2imp_4_H3}).
We set $g=1$ and use exact diagonalization (ED) for system sizes up to $L = 8$.
Figure~\ref{fig:H3_deg_gap_corre}(a) shows the lowest excitation gap $\Delta$ as a function of $L$, where we see that it decreases exponentially with $L$.
Similar to the $t-J_z$ model, the state $|\sigma_{\text{left}})$ corresponding to the leftmost physical charge SLIOM can be a good trial state, which gives an upper bound on the energy gap,
\begin{equation}\label{eq:H3imp_superH_gap}
    \Delta \leq \frac{(\sigma_{\text{left}}| \hat{\mathcal{V}}^{\text{2imp.}}|\sigma_{\text{left}})}{(\sigma_{\text{left}}|\sigma_{\text{left}})} = \frac{14 g}{3^L-1} = \mathcal{O}(g \times 3^{-L}),
\end{equation}
This is also shown in Fig.~\ref{fig:H3_deg_gap_corre}(a).
While we see that the variational bound is significantly larger than the gap $\Delta$, both show the same exponential decay rate with $L$.
Therefore, this simple consideration of the simplest SLIOM already shows that the $H_3$ model also exhibits exponentially slow thermalization under boundary impurities.

The bulk autocorrelations are mainly controlled by the SLIOMs located in the bulk of the chain, similar to the $t-J_z$ model.
As mentioned earlier, in particular the physics of blockades, these SLIOMs are more spatially localized than in the $t-J_z$ model, which can lead to non-vanishing Mazur bounds for bulk correlations in the unperturbed scenario~\cite{2020_sala_ergodicity-breaking,2020_SLIOMs, moudgalya_hilbert_2022}. 
Therefore, we also intuitively expect that these SLIOMs are less sensitive to the impurity far away, contributing to the prethermal plateau with exponentially long timescales. 
Moreover, they can even produce a non-zero prethermal plateau even in the thermodynamic limit, as also observed numerically with classical stochastic dynamics.
Unlike the $t-J_z$ model, we do not have analytical structures in the unperturbed commutant that would allow large system studies of the impurity effect, and the available sizes in our direct ED study are too small to study bulk physics.
Nevertheless, we can variationally employ blockade SLIOMs to develop some idea about such physics.
Specifically, we can calculate the variational energy of the simplest blockade SLIOM in Eq.~(\ref{eq:B++SLIOM}):
\begin{equation}
\frac{(\hat{B}_{j_0,j_0+1}^{++}| \hat{\mathcal{V}}^{\text{2imp}}|\hat{B}_{j_0,j_0+1}^{++})}{(\hat{B}_{j_0,j_0+1}^{++}|\hat{B}_{j_0,j_0+1}^{++})} = \frac{8g}{3^{L-j_0-1} - 1}.
\end{equation}
Here we have assumed that $j_0$ is far enough from the right boundary, and because of the factorization of $|\hat{B}_{j_0,j_0}^{++})$, the calculation is completely analogous to that with $|\sigma_{\text{left}})$ except for the restriction to the region $[j_0+2, L]$ and some matrix element details.\footnote{Strictly speaking, $|\hat{B}_{j_0,j_0+1}^{++})$ is not orthogonal to the ground state $\sim |\mathbb{1})$, but this is easy to fix by considering analogous $|\hat{B}_{j_0,j_0+1}^{--})$ obtained by flipping the signs of all charges and then taking the combination $|\hat{B}_{j_0,j_0+1}^{++}) - |\hat{B}_{j_0,j_0+1}^{--})$, which gives the same variational energy.}
As mentioned in the previous section [see around Eq.~\eqref{eq:Cqk_decay_rate}], such a super-Hamiltonian variational energy of the SLIOM is already useful since it upper bounds the decay rate of SLIOM's autocorrelation functions, and hence contributes to the spin correlation functions.
Thus we see that near $j_0$ we have a time scale that depends on the distance to the boundary impurity as $\sim \frac{1}{g} 3^{L-j_0}$.
As discussed earlier, this simplest blockade SLIOM already gives non-zero contribution to the Mazur bound even in the thermodynamic limit, which means that the prethermal plateau will also have non-zero value even in the thermodynamic limit. 
Note that it is expected~\cite{2020_SLIOMs, moudgalya_hilbert_2022} (though not proven) that the unperturbed Mazur bound near $j_0$ is completely determined by SLIOMs capturing blockades that occur close to $j_0$ (either larger blockades covering $j_0$ or blockades few sites away from $j_0$).
Hence, since the super-Hamiltonian variational energies for all of them are $\sim g 3^{-(L-j_0)}$, differing only in numerical prefactors, we expect that the time scale for the prethermal plateau duration in the autocorrelations at $j_0$ will be fairly sharply given by $\sim \frac{1}{g} 3^{L-j_0}$.
This suggests fixed exponential growth of the time scale with $L-j_0$, i.e., $\sim e^{(L-j_0)/\xi}$ with fixed $\xi = 1/\ln(3)$, which can be viewed as some characteristic length scale for describing spreading of the breaking of the fragmentation structure.
This is unlike the $t-J_z$ case, where the dependence of the time scale on $L-j_0$ was more complicated, the difference coming from different ``tail properties'' for SLIOMs concentrated at different locations along the chain.

\subsubsection{Numerical verifications cellular automaton dynamics}
We also numerically study the time evolution of true correlation functions using stochastic dynamics described in Sec.~\ref{subsec:CA} and show the results in Fig.~\ref{fig:H3_deg_gap_corre} in panels (b) and (c).
We use two-site state-flip impurity at the rightmost two sites of the chain to fully break all the symmetries.
For the $S_1^z$ boundary autocorrelation in panel (b), the correlation decays to a prethermal plateau, which lasts longer as we increase the system size.
For $L=30$, the prethermal plateau value in the presence of the right boundary impurity is close to the unperturbed Mazur bound given by the case without the impurity.
It appears that, in contrast to the $t-J_z$ model, where only the leftmost SLIOM $|q_{\mathrm{left}})$ survives for exponentially long time and leads to an $\mathcal{O}(1/L)$ distinction between the unperturbed Mazur bound and the prethermal plateau, in the $3$-local dipole-conserving model all or almost all of the originally conserved quantities that contributed to the unperturbed Mazur bound for $S_1^z$ persist for exponentially long times, such that the prethermal plateau almost overlaps with the unperturbed Mazur bound~\footnote{It could also be that the non-SLIOM contributions are parametrically smaller in the $H_3$ model than in the $t-J_z$ model and are not as detectable.}.
The bulk correlations also decay to a prethermal plateau, which lasts longer with the increase of the distance to the impurity $L-j$, as shown in Fig.~\ref{fig:H3_deg_gap_corre}(c).
We estimate the time of decay $t_{\mathrm{decay}}$ by calculating the time when correlation functions reach to $C_{S_j^z}^{\mathrm{decay}} = 0.22$ (the bulk Mazur bound in the unperturbed model is $\sim 0.24$~\cite{2020_sala_ergodicity-breaking}), which scales exponentially with $L-j$ as shown in the inset.
Therefore, for the $3$-local dipole-conserving model with strong fragmentation, the bulk correlation functions can also exhibit prethermal plateaus under fully-symmetry breaking impurities, similar to the $t-J_z$ model but with finite plateau values that are independent of $L$.
\subsection{Comparison of strong and weak fragmentation}
\begin{figure}[t!]
    \centering
    \includegraphics[width=1\linewidth]{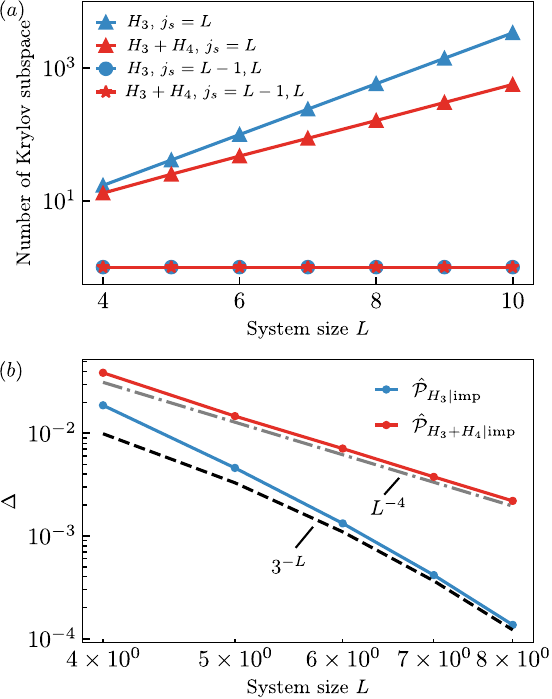}
    \caption{\textbf{Comparision of spin-$1$ dipole-conserving models with strong or weak fragmentation with local state-flip impurities.}
    (a) Spin-$1$ dipole-conserving models with an impurity at the boundary $j_s = L$ or impurities at two sites $j_s = L-1,L$.
    An impurity at one site does not break the Hilbert space fragmentation fully in either the spin-$1$ $H_3$ or $H_3 + H_4$ systems, as evidenced by the number of Krylov subspaces scaling exponentially with system size.
    On the other hand, a two-site impurity completely restores ergodicity.
    (b) For $\hat{P}_{H_3|\mathrm{imp}}$ with strong fragmentation, the energy gap is upper bounded by $\mathcal{O}(3^{-L})$ as given by Eq.~(\ref{eq:H3imp_superH_gap}).
    On the other hand, for $\hat{P}_{H_3+H_4|\mathrm{imp}}$ with weak fragmentation, the energy gap scales polynomially as $\mathcal{O}(L^{-4})$. 
    }
    \label{fig:H3_H34_gap}
\end{figure}

The combination of charge and dipole conservation can lead to Hilbert space fragmentation of two types, where strong fragmentation can lead to non-thermal behaviors, such as an anomalously large or non-vanishing Mazur bound, while for weak fragmentation, the system thermalizes despite exponentially large disconnected sectors~\cite{2020_sala_ergodicity-breaking, 2020_khemani_local,Morningstar_2020,2023_Feng_HSF_boson_weak_strong_transition}. 
Under local impurities, they also exhibit distinct behaviors, as we discussed with examples of weak fragmentation in Sec.~\ref{sec:dipole} (focusing on the anomalous diffusion in dipole-conserving systems, for which the models we studied actually exhibit weak fragmentation), as well as strong fragmentation in the preceding parts of Sec.~\ref{sec:H3}.
Here, we compare the $3$-local spin-$1$ dipole-conserving model (referred to simply as $H_3$) with strong fragmentation as discussed in the previous subsection, to the $3$- and $4$-local dipole-conserving model ($H_3 + H_4$) with weak fragmentation~\cite{2020_sala_ergodicity-breaking}, where the $4$-local terms are given by $S_i^+ S_{i+1}^- S_{i+2}^- S_{i+3}^+ + \mathrm{h.c.}$.
Both models exhibit a certain level of robustness against local impurities.
For example, both systems remain fragmented under a one-site state-flip impurity, which we demonstrate in Fig.~\ref{fig:H3_H34_gap}. 
(This is also the case for the spin-$1/2$ dipole-conserving model with $4$- and $5$-local terms in Sec.~\ref{sec:dipole}.)
The number of Krylov subspaces for both systems under one-site state-flip impurity at boundary $j_s=L$ scales exponentially with the system size, indicating that at least a part of the fragmentation survives.
In App.~\ref{app:frag_1site}, we provide some analytical understanding of the persistence of fragmentation in this case and show that it persists also when the one-site impurity is in the bulk.

On the other hand, we find that two-site state-flip impurities at the right boundary restore full ergodicity for both models, which is shown in Fig.~\ref{fig:H3_H34_gap}(a).
Focusing on such impurities from now on, the $3$-local dipole-conserving model with strong fragmentation is still much more robust due to the SLIOMs as discussed earlier in the text.
First, the difference in the effect of such ergodicity-restoring impurities manifests in the gap of the respective super-Hamiltonians.
We perform ED calculations of the spectra of $\hat{\mathcal{P}}_{H_3|\mathrm{imp}}$ given by Eq.~\eqref{eq:H3imp_superH} and appropriately constructed $\hat{\mathcal{P}}_{H_3+H_4|\mathrm{imp}}$, for system sizes up to $L=8$.
Figure~\ref{fig:H3_H34_gap} shows the results for the lowest energy gap $\Delta$ for the impurity strength $g=1$.
We see that the energy gap of $\hat{\mathcal{P}}_{H_3|\mathrm{imp}}$ scales exponentially as $\mathcal{O}(3^{-L})$, while the energy gap of $\hat{\mathcal{P}}_{H_3+H_4|\mathrm{imp}}$ appears to scale polynomially as $\mathcal{O}(L^{-4})$.
Moreover, for weak fragmentation, where the effect of fragmentation is expected to be negligible in the thermodynamic limit (when evaluating infinite temperature expectation values)~\cite{2020_sala_ergodicity-breaking}, correlation functions decay polynomially with time as dominated by the charge and dipole conservation. 
When performing stochastic dynamics simulations of the $H_3 + H_4$ spin-$1$ system in the presence of impurities on two consecutive sites at the boundary, we find that correlation functions show similar behavior as the $4$- and $5$-local spin-$1/2$ dipole-conserving model with spin-$1/2$ discussed in Sec.~\ref{sec:dipole}, which we include in App.~\ref{app:H34_CA_Cz}.
This is consistent with previous works using stochastic cellular automaton dynamics to study the hydrodynamics of dipole-conserving systems with various on-site local Hilbert space dimensions, all showing qualitatively similar (i.e., universal) long-time behavior.
Overall, this suggests that our results in Sec.~\ref{sec:dipole}, in the presence of an impurity, extend to dipole-conserving models with higher spins and weak fragmentation.
In contrast, for the strongly fragmented $H_3$ spin-1 model under two-site impurity discussed in this section, the correlation functions exhibit a prethermal plateau that decays exponentially slowly with system size.

\subsection{Impurities at both boundaries}\label{subsec:tJz_twoimp}
Finally, we consider the effect of adding impurities at both boundaries of the system, which, in fragmented systems, can give rise to qualitatively new physics compared to the case of a single impurity.
For the sake of completeness, we also briefly present the case of two impurities for a U$(1)$ symmetric system.
In this case, in the absence of additional symmetries, the analysis follows directly from our discussion in Sec.~\ref{sec:U1}.
Impurities effectively impose absorbing boundary conditions at long wavelengths at both boundaries, and the profiles of eigenfunctions are modified, but the energy dispersion $\sim k^2$ of low-energy orbitals remains unchanged. 
Therefore, the characteristic time for the ergodicity restoration in the whole system still scales as $\sim L^2$.
Furthermore, in the study of autocorrelations at a distance $d$ from one edge such that $L - d \gg d$, we still expect a crossover from a $t^{-1/2}$ scaling to a $t^{-3/2}$ scaling at time $\sim d^2$, and the latter scaling remaining until time $\sim L^2$.
Similar arguments apply to dipole conserving systems discussed in Sec.~\ref{sec:dipole} in the presence of impurities at both boundaries. 
We verify analytically using the continuum equations and numerically using classical cellular automata, for both the U($1$) charge-conserving systems and the dipole-conserving systems.
The physics of the strongly-fragmented dipole-conserving system discussed in Sec.~\ref{sec:H3} too remains qualitatively similar under the addition of two impurities on the boundaries. 
Recall that the discussed SLIOMs in that case are all spatially localized due to the presence of blockades~\cite{2020_SLIOMs} (they are essentially LIOMs).
Therefore, they are only weakly perturbed by the impurities at both boundaries, and still lead to exponentially slow thermalization timescales and prethermal plateaus for bulk correlation functions away from both boundaries.

However, as pointed out by a recent work~\cite{wang2025exponentiallyslowthermalization1d}, the $t-J_z$ model thermalizes after a time polynomial in system size (instead of exponential) when impurities are placed at both boundaries.  
Intuitively, for the $t-J_z$ model, the exponentially long thermalization timescale with one impurity originates from the physical spin pattern conservation. 
Assuming the impurity at the right boundary,
the physical spins near the left edge remain untouched for a long time, until all spins to their right are changed by the impurity action. 
However, the case with symmetry-breaking impurities placed at both boundaries can be viewed as a $t-J_z$ model with PBC and two impurities sitting at two consecutive sites (see Fig.~\ref{fig:tJz_two_imp} in App.~\ref{app:imp_both_boundaries} for a quantitative comparison).
For PBC, no SLIOMs are spatially localized and only the relative spin pattern is relevant~\cite{2020_SLIOMs}.
This translates into the Mazur bound decaying as $L^{-1}$ with system size instead of $L^{-1/2}$ in the OBC case~\cite{2020_SLIOMs, moudgalya_hilbert_2022}. 
The conserved physical spin patterns in the bulk can move as a whole, and hence the impurities have access to spoil the whole pattern, leading to much faster thermalization. 
Therefore, from this equivalent formulation of the problem, we do not expect the system to take exponentially long (in system size) to thermalize in the presence of impurities at both boundaries.
From the perspective of, say, the left SLIOMs used when a single impurity was at the right edge, unlike the strongly-fragmented dipole-conserving model, the bulk left SLIOMs $q_k$ of the $t-J_z$ model with OBC although localized in a statistical sense, contain Pauli strings that extend to the left boundary.
Hence, in the presence of an additional impurity at the left boundary, the $|q_k)$ acquires a finite $\mathcal{O}(1)$ gap, turning them into a bad ansatz for approximate conserved quantities. 
Hence, a more detailed analysis is required.
We numerically study the energy gap of the perturbed super-Hamiltonian of the $t-J_z$ model with OBC and impurities at both boundaries, where we can still work within the composite subspace. 
As shown in App.~\ref{app:imp_both_boundaries} we find that this gap scales as $\sim \mathcal{O}(L^{-2.6(1)})$, indicating that the system thermalizes polynomially fast. 
This energy gap can be upper bounded by the first-order perturbation theory in the commutant basis similar to the discussion in Sec.~\ref{sec:tJz}, which gives the energy gap scaling approximately as $\mathcal{O}(L^{-2})$ as detailed in Appendix~\ref{app:imp_both_boundaries}.
This $\mathcal{O}(L^{-2})$ scaling can be understood from a hopping model of the SLIOMs with OBC, which approximately becomes a single-particle hopping problem similar to the U$(1)$ case with a boundary impurity, and hence connects directly to the analysis of U$(1)$-conserving systems in the presence of a symmetry-breaking impurity.  
This gives an ansatz of the lowest excited state $|\psi_{\mathrm{ansatz},\kappa}) \sim \sum_k \sin(\kappa k) |q_k)$, $\kappa \sim 1/L$, such that it minimizes the weight on the leftmost SLIOM, and leads to an energy scaling as $\sim \kappa^2 \sim \mathcal{O}(L^{-2})$.

Finally, in Appendix~\ref{app:imp_both_boundaries} we also numerically study the stochastic dynamics of the correlation functions in the presence of impurities at both boundaries.
As we mentioned earlier, we find that the numerical results for an OBC chain and a PBC chain with impurities at sites $j_s=1,L$ are very similar. 
Moreover, we observe that correlation functions first decay polynomially in time and then exponentially, while the prethermal plateau is absent.

We note that while we have so far focused on the evolution of correlation functions, the evolution of the total magnetization $M_Z=\sum_j S_j^z$ numerically evaluated in Ref.~\cite{wang2025exponentiallyslowthermalization1d} from the initial state $\ket{\uparrow}^{\otimes L}$, maps to the evolution of $(\tilde{\uparrow}\cdots \tilde{\uparrow}|e^{-\hat{\mathcal{P}}t}|M_Z)$ in the super-Hamiltonian language.
As we show in  App.~\ref{sec:magnetization}, the time scale $t^*$ before an exponential decay of magnetization kicks in is lower bounded as $t^* > \mathcal{O}(L^2/g)$.
While this is a loose bound, it indicates the presence of polynomial (rather than exponential) in $L$ time scale for the relaxational behavior of magnetization.
We note that the numerical analysis in Ref.~\cite{wang2025exponentiallyslowthermalization1d} found that $t^* \sim \mathcal{O}(L^{3.5})$, which we cannot derive yet and which remains an open question.

\section{Conclusion and Discussion}\label{sec:conclude}
In this work, we introduced a systematic method for studying the long-time relaxation of correlation functions under local symmetry-breaking impurities in noisy Brownian circuits.
Using the superoperator formalism, where the symmetries appear as ground states of a super-Hamiltonian, the dynamics with a symmetry-breaking impurity can be treated using a perturbed super-Hamiltonian.
The low-energy spectra of these super-Hamiltonians correspond to approximate symmetries that govern the long-time dynamics of two-point correlation functions.
Our method reveals that even though the original symmetries are broken by the impurity, there are remnants of them visible in the slow relaxation behavior of the correlation functions. 
We investigated many examples including locally breaking a conventional U($1$) charge conservation, dipole conservation, and strong and weak Hilbert space fragmentation.
Utilizing the superoperator formalism, we characterize the dynamics of two-point correlation functions via exact or variational mappings to single-particle problems.  
When locally breaking a global symmetry such as U($1$) or dipole conservation, autocorrelation functions initially exhibit the standard power-law decay corresponding to diffusion or subdiffusion respectively.
However, at late times, their behaviors cross over to exhibiting faster power-law decays or different amplitudes than those for the unperturbed diffusive or subdiffusive behavior.
We understand this apparent temporal transition as driven by the impurity that is relevant in the renormalization group sense in one spatial dimension.
In the U$(1)$ case, this relevance and its outcome can be explicitly shown by connecting the problem in the continuum limit and at long times and wavelengths to standard diffusion but with an absorbing boundary condition at the impurity.
Specifically, we found that locally breaking a U($1$) charge leads to a transition from standard diffusion, with autocorrelations initially decaying as $C_Z(t) \sim t^{-1/2}$ until the impurity is ``detected" by the diffusing operator front, to a decay $C_Z(t) \sim t^{-3/2}$ at sufficiently long times, until the finiteness of the system is detected at times $\mathcal{O}(L^2)$. 
Moreover, when placed in the bulk, this relevant impurity modifies the spatial dependence of charge correlations across the impurity, which can be interpreted as an effective disconnection between the two sides.
These findings suggest that a renormalization group approach, as in the seminal Kane-Fisher impurity problem~\cite{Kane_Fisher_92}, might be useful to characterize the effect of symmetry-breaking impurities in the long-time dynamics.
Although our discussion started from a specific choice of the bond algebra generators, our conclusions should apply to general systems with impurities locally breaking a U($1$) symmetry, independent of specific microscopic models, as different choices map onto the same continuum models that we ultimately study.
For dipole-conserving models (with weak fragmentation), we considered two scenarios: (i) charge-conserving but dipole-breaking impurities, and (ii) both charge- and dipole-breaking impurities.
In the first case, we found that the subdiffusive behavior $C_Z(t) \sim t^{-1/4}$ is unaffected by the impurity, and is retained until a time that scales as $t \sim \mathcal{O}(L^4)$ with system size, although there is a crossover in the overall amplitude which happens at times $t \sim \mathcal{O}(d^4)$, where $d$ is the distance to the impurity.
This prediction analytically follows from the fact that a charge-conserving impurity acts as a reflective boundary, and the operator front that reaches the boundary becomes reflected and spreads out in a dipole-conserving medium.
On the other hand, charge- and dipole-breaking impurities lead to a modification of the power law decay of autocorrelation functions which show new $C_Z(t)\sim t^{-5/4}$ scaling beyond timescales at which the impurity is detected. 
We provide an analytical derivation of this power-law exponent, by exactly diagonalizing a single-particle problem with an anomalous $\partial_x^4$ kinetic term, in the presence of charge- and dipole-absorbing boundary conditions.
In this context, it will be interesting to understand if, in general, breaking the conservation of higher multipole moments of a U$(1)$ charge while preserving some of the lower ones, maintains the subdiffusive behaviors appearing in the absence of impurities.
On the other hand, when such impurities are fully symmetry-breaking in an $m$-pole (where $m = 1$ and $m = 2$ correspond to charge and dipole respectively) conserving system~\cite{2020_Pablo_automato}, we predict that the emergent boundary conditions driven by this relevant impurity correspond to the vanishing of all spatial derivatives of order lower than $m$.
In this scenario, we then predict that the algebraic decay of autocorrelations $\sim t^{-\frac{1}{2m}}$, turns into $\sim d^{2m} t^{-(1+\frac{1}{2m})}$ at a time $\sim \mathcal{O}(d^{2m})$, where $d$ is the distance to the impurity.
We expect this decay lasts till a time that scales as $\sim \mathcal{O}(L^{2m})$ with system size $L$, where the finiteness of the system is detected, leading to an exponential decay $\frac{d^{2m}}{L^{2m+1}} e^{-t/L^{2m}}$ at later times.
This behavior is displayed in Fig.~\ref{fig:schematic}(c) (where $z=2m$).
While we have carefully studied the $m = 1$ and $m = 2$ cases, it would be useful to verify the above scaling predictions for $m > 2$ with comparable thoroughness.

Finally, we considered two different systems exhibiting strong Hilbert space fragmentation, the $t-J_z$ model and spin-1 dipole-conserving systems.  
We analytically showed that even when explicitly broken by an impurity, the effect of fragmentation persists for exponentially long times in the size of the system, leading to a long-lived anomalous behavior of certain correlation functions. 
Using the fact that these systems exhibit SLIOMs~\cite{2020_SLIOMs}, we analytically showed the hierarchy of time scales that emerge in the behavior of both boundary and bulk autocorrelation functions.
In particular, these correlation functions exhibit anomalous plateaus up to times that scale exponentially with the distance to the impurity, where the plateaus correspond to SLIOMs' contributions to the unperturbed Mazur bounds.
Further, in App.~\ref{app:rel_graph} we showed that graph theoretic considerations previously applied in the context of fragmentation~\cite{han2024exponentially, wang2025exponentiallyslowthermalization1d} can also be obtained by a particular first-order perturbation theory calculation for the super-Hamiltonian.
These time scales are depicted in panel (d) of Fig.~\ref{fig:schematic}.
First, on a time scale $\sim \mathcal{O}(d^{\alpha})$ with $\alpha$ potentially a non-trivial exponent, and $d$ the distance to the impurity, autocorrelation functions decay to a prethermal plateau (after crossing the unperturbed Mazur bound) determined only by the SLIOMs. Here, $d$ can be finite or can be a fraction of the systems size (our results are most controlled in the latter case).
This prethermal saturation value (which may be finite and $L$-independent as in the dipole-conserving models or decrease polynomially with $L$ but still prominent as in the $t-J_z$ model) persists for a time exponentially large in $d$, where the impurity starts to affect the conservation of SLIOMs closer to the impurity.
Finally, at a time exponentially large in the size of the system, none of the SLIOMs are conserved, and correlations decay to zero. 
Even in fragmented systems, we conjecture that the approach to these prethermal plateaus as well as the ensuing decay of correlations past the plateaus, takes place in a hydrodynamic fashion, with autocorrelations governed by the transport of approximate conserved charges.
For example, in the case of strongly-fragmented dipole-conserving systems, one expects autocorrelations to at least subdiffusively saturate to a prethermal plateau as $\sim t^{-1/4}$, and eventually decay to zero showcasing a potentially different scaling in time.
This ``hydrodynamic tail'' (controlled by the symmetry-breaking impurity) remains until a time scale $\sim L^{\beta}$, with $\beta$ depending on the specific hydrodynamic behavior of the corresponding system.
An analogous dependence is expected for the $t-J_z$ model, although now the prethermal value depends on system size. 
We leave a detailed analysis of both of these scenarios, regarding  the temporal decay of correlations and crossovers between different regimes as open questions for future work. 
\subsection*{Open questions}
Our work suggests some additional directions worth exploring.
First, we leave open a systematic renormalization group analysis on the effect of various types of impurities (regarding the conserved quantities they preserve as well as the dimensionality of their support) acting on various charge and higher multipole-moment conserving systems.
For example, our analysis of U(1) breaking impurities predicts that impurities acting on a finite region in more than two spatial dimensions do not modify the scaling of charge correlation functions.
Can we exploit this approach to predict the effect of symmetry-breaking impurities in the long-time dynamics of other quantum many-body systems, e.g., with extended impurities or with different symmetries such as various subsystem symmetries?
Our work has focused on examples of ``classical symmetries," i.e., those which are diagonal in a product state basis, and whose long-time dynamics can be captured by appropriate classical cellular automata.
However, the superoperator formalism is not limited to these~\cite{moudgalya2024symmetries}, and an intriguing direction is to extend our analysis to correlation functions in systems with more general symmetries, e.g., with non-Abelian symmetries and with quantum fragmentation~\cite{2007_Read_Commutant, moudgalya_hilbert_2022,  23_Brighi_particle_conserving_east, 24_Chen_quantum_fragmentation_breakdown, 2021_minimal_HSF_with_local_constraints, 2024_Parameswaran_SY_commutant, li2023hilbert}, or with quantum many-body scars~\cite{serbyn2020review, 2022_Moudgalya_review_HSF_SCAR, papic2021review, chandran2022review, moudgalya2022exhaustive, gotta2023asymptotic, moudgalya2024symmetries, morettini2025unconventional}.

Moreover, while in this work we focused on one-dimensional fragmented models, recent works have analytically~\cite{2022_Lehmann_Pablo_Markov,Quantum_drums, 2022_Neupert_2D_HSF_subsystem_symmetries, melissa_24_2DHSF, 24_Stahl_quantum_loop_frag} and experimentally~\cite{Melissa_24} characterized higher-dimensional fragmented models. 
In particular, Ref.~\cite{2022_Lehmann_Pablo_Markov} analytically showed that for a family of models that can be defined in any dimension, the infinite-temperature correlation functions saturate to a finite value. 
It would then be interesting to understand the effect of impurities on such higher-dimensional fragmented systems
and compare them with renormalization group predictions for more conventional symmetries.
Beyond correlation functions, the superoperator formalism can be generalized to study the propagation of entanglement or higher point correlation functions in quantum many-body systems~\cite{Nahum_17, Roberts2017, Cotler2017, Rakovszky18, Tibor_19,  zhou2019emergent, sunderhauf2019quantum, zhou2020entanglement, bernard2022dynamics, swann2023spacetime, sahu2024phase, vardhan2024entanglementdynamicsuniversallowlying, lastres2024nonuniversalityconservedsuperoperatorsunitary}.
It would be interesting to explore the effect of local impurities on the dynamics of such quantities, which might show interesting behavior even in the absence of symmetries~\cite{fraenkel2024entanglementdualunitaryquantum}.
More generally, it is important to check to what extent these impurity-based behaviors, which are analytically tractable for Brownian circuit dynamics, extend to generic time-independent dynamics of non-integrable systems.
While we expect the qualitative behavior of dynamics to be similar to what we have discussed, powerful techniques such as those based on influence matrices~\cite{sonner2021influence, ye2021constructing, ng_real-time_2023, lerose_influence_2021, thoenniss_nonequilibrium_2023}, which have been useful for understanding various dynamical effects in the presence of impurities, might also be useful for quantitatively analyzing such dynamics. 
In addition, the hydrodynamic effective field theory is another useful tool that applies to the analysis of hydrodynamics behavior from conventional U($1$) symmetries~\cite{crossley_effective_2017, glorioso_effective_2017} to more nontrivial symmetries~\cite{Gromov_2020, huang_hydrodynamic_2022, landry_coset_2020, glorioso_goldstone_2023, jain_dipole_2023}.
It was recently shown that the hydrodynamic modes can be interpreted as Goldstone modes from strong-to-weak spontaneous symmetry breaking~\cite{huang_hydro_SWSSB_2025}.
It would be interesting to explore whether hydrodynamic effective field theory of random circuits can be systematically constructed from the super-Hamiltonian formalism, in particular for systems with local symmetry-breaking impurities.
These could eventually lead to a general framework for hydrodynamics in the presence of impurities, and we defer the exploration of these connections to future works.  
\begin{acknowledgments}
We thank Jason Alicea, Yue Liu, and Sara Murciano for inspiring discussions on defect line actions and their relation to impurity problems on previous collaborations, Daniil Asafov, Shankar Balasubramanian, Alexey Khudorozhkov, Ethan Lake, Ruchira Mishra, Federica Surace, and Sara Vanovac for discussions of Hilbert space fragmentation and hydrodynamic descriptions in various contexts, and Marcos Rigol for discussion of thermalization with perturbation.
P.S.\ acknowledges support from the Caltech Institute for Quantum Information and Matter, an NSF Physics Frontiers Center (NSF Grant No.PHY-1733907), and the Walter Burke Institute for Theoretical Physics at Caltech.
This work was supported by the Deutsche Forschungsgemeinschaft (DFG, German Research Foundation) under Germany’s Excellence Strategy EXC-2111-390814868, TRR 360 (project-id 492547816), FOR 5522 (project-id 499180199), and the Munich Quantum Valley, which is supported by the Bavarian state government with funds from the Hightech Agenda Bayern Plus.
S.M.\ acknowledges support from the Munich Center for Quantum Science and Technology (MCQST). 
O.I.M.\ acknowledges support by the National Science Foundation through Grant No. DMR-2001186.
 
\end{acknowledgments}

\textbf{Data and materials availability.}
Data analysis and simulation codes are available on Zenodo upon reasonable request~\cite{data_set}. 

\providecommand{\noopsort}[1]{}\providecommand{\singleletter}[1]{#1}%
\clearpage
\onecolumngrid
\appendix

\section{Remarks about naive first-order perturbative treatment of impurities}
\label{app:U1_pert}
In the main text, we directly analyzed what happens with the hydrodynamic mode in the presence of a U$(1)$ symmetry-breaking impurity without any approximations.
On the other hand, for fragmented systems, we studied the problem using degenerate perturbation theory on the ground state manifold of the super-Hamiltonian.
Here we intend to compare these two approaches, by analyzing the $U(1)$ problem using perturbation theory as well.
In particular, it is instructive to also ask what happens with the original exact symmetries once we add the impurity perturbation and compare with naive perturbation theory.
The unperturbed $U(1)$ commutant, Eq.~\eqref{eq:comm_alg_U1}, maps to the $(L+1)$-dimensional ferromagnetic manifold of the composite spins\cite{moudgalya2024symmetries}.
The total ``spin'' of the ferromagnet is $\widetilde{J} = L/2$.
The identity operator maps to the polarized state in the $\widetilde{\rightarrow}$ direction.
It is convenient to choose basis in this manifold consisting of eigenstates of $\widetilde{S}^x_{\text{tot}} = \frac{1}{2} \sum_j \widetilde{X}_j$, which we denote as $\{|\widetilde{J} = L/2, \widetilde{M}^x \rangle, \widetilde{M}_x = L/2, L/2-1, \dots, -L/2+1, -L/2 \}$; we will use shorthand $\{|\widetilde{M}^x \rangle \}$ and assume the basis states are normalized.

Let us denote the contribution of the impurity in the super-Hamiltonian Eq.~(\ref{eq:Pomp_U1_imp}) as $\hat{\mathcal{V}}$.
Naive first-order perturbation theory treatment of the ground state manifold simply projects $\hat{\mathcal{V}}$ to this manifold.
It is easy to see that this projection is diagonal in the $\{|\widetilde{M}^x \rangle \}$ basis, with the corresponding matrix elements (which can be also viewed as trial energies)
\begin{equation}
\langle \widetilde{M}_x| \hat{\mathcal{V}} |\widetilde{M}_x \rangle = 2g\left(1 - \frac{2\widetilde{M}_x}{L} \right) ~.
\end{equation}
The identity operator corresponds to $\widetilde{M}_x = L/2$ and remains the exact zero-energy ground state, while the rest of the states appear uniformly split with spacing $4g/L$.
However, we should take this perturbation theory result very critically.

Specifically, in this treatment the state $\widetilde{M}_x = L/2 - 1$ appears split from the ground state by an energy $4g/L$.
On the other hand, this state is actually part of the single-spin-flip manifold treated fully by the single-particle problem in Eq.~(\ref{eq:U1_Heff_imp}).
While $|\widetilde{M}_x = L/2 - 1 \rangle$ is the lowest-energy state in the unperturbed single-spin-flip sector corresponding to $\phi^\text{unperturbed}_{k=0} \propto \text{const}$, its trial energy $\propto 1/L$ is clearly parametrically larger than the new ground state energy $\propto 1/L^2$ in the perturbed problem.
We also see that the new ground state in this problem is qualitatively different from the unperturbed one, particularly near the impurity, cf.~footnote~\ref{ft:U1imp_variational_orbital}.
Thus, the naive perturbation theory focusing on the original ground state manifold \textit{fails qualitatively} in this problem.
Of course this is consistent with the fact that the effect of such an impurity is non-perturbative at low energies/long wavelengths in one dimension, which our full solution essentially incorporates.
Note that even without knowing the actual solution in the presence of the impurity, we could have anticipated an issue with the described naive perturbation theory by noting that the trial energy of $|\widetilde{M}_x = L/2 - 1 \rangle$, being $\propto g/L$, is parametrically larger than the excitation energy $\propto 1/L^2$ in the unperturbed problem. 
While viewed as variational energy it is a valid one predicting relaxation time scale lower-bounded by $O(L/g)$, it is simply a loose prediction compared to what actually happens as solved in the main text.
One can then legitimately ask about the status of the perturbative/variational treatment used in the Hilbert space fragmentation examples in Sec.~\ref{sec:fragmentation}.
Here we do not have fully controlled solutions to compare with, but certainly as variational bounds they are all valid statements.
We should be very critical about the $O(1/L)$ excited states found by analyzing the degenerate perturbation theory in the impurity-perturbed $t-J_z$ model [analysis starting with Eq.~\eqref{eq:tJz_single_H_eff}], as we also mention in the main text.
On the other hand, the variational results associated with the SLIOMs in both the $t-J_z$ model and the 3-local spin-1 dipole conserving model are likely to be more robust.
Thus, we conjecture that they do not require qualitative modifications at least for the SLIOMs that are a finite fraction of $L$ away from the impurity.
The reason is that the corresponding trial energies, being exponentially small in the distance from the impurity, are parametrically much smaller than any of the low-energy excitations in the unperturbed super-Hamiltonian, which scale polynomially with $L$.
For the leftmost SLIOMs that are furthest away from the impurity we have strong numerical evidence in support of this conjecture (see discussion in the main text and App.~\ref{app:num_check}), while it would be useful to explore this in general, which we leave for future studies.

\section{Common structure in algebra generators and the corresponding super-Hamiltonian terms} \label{app:superH_gen}
In this appendix, we discuss some common structures that appear in the bond algebra generators we use in the main text, and the corresponding super-Hamiltonians.
We often have some ``move'' term (operator) $W$ that has desired symmetries but is not hermitian, and we include $h_W \defn W + W^\dagger$ in the bond algebra generators.
The double space expression for the corresponding contribution to the super-Hamiltonian (omitting the overall coupling constant) reads:
\begin{equation}
\begin{aligned}
\mathcal{L}_{h_W}^2 =&
W_t^2 + (W_t^\dagger)^2 + (W_b^T)^2 + (W_b^*)^2 - 2 W_t W_b^T - 2 W_t^\dagger W_b^* + \\
& + W_t^\dagger W_t + W_t W_t^\dagger + W_b^T W_b^* + W_b^* W_b^T - 2 W_t W_b^* - 2 W_t^\dagger W_b^T ~. 
\end{aligned} ~.
\end{equation}
The move $W$ is often of the form
\begin{equation}
W = \ketbra{\alpha}{\beta}_R ~,
\label{eq:W_single_ketbra}
\end{equation}
where $\ket{\alpha}_R$ and $\ket{\beta}_R$ are product states in the computational basis for degrees of freedom in some region $R$, i.e., $\ket{\alpha}_R = \otimes_{j \in R} \ket{m^\alpha_j}_j$, and similarly for $\ket{\beta}_R$.
Here $\{\ket{m}_j \}$ is the fixed local basis in which the operator-to-state mapping---the double space formalism---is carried out. 
Then, assuming $\alpha \neq \beta$ (treating $\alpha$ and $\beta$ as basis labels, hence for states $\bra{\beta}_R \ket{\alpha}_R = 0$), we have
\begin{equation}
\begin{aligned}
\mathcal{L}_{h_W}^2 =& -2 \ketbra{\alpha\beta}{\beta\alpha}_{R,t;R,b} - 2 \ketbra{\beta\alpha}{\alpha\beta}_{R,t;R,b} + \\
& + \ketbra{\alpha}{\alpha}_{R,t} + \ketbra{\beta}{\beta}_{R,t} + \ketbra{\alpha}{\alpha}_{R,b} + \ketbra{\beta}{\beta}_{R,b} - 2 \ketbra{\alpha,\alpha}{\beta,\beta}_{R,t;R,b} - 2 \ketbra{\beta,\beta}{\alpha,\alpha}_{R,t;R,b} ~.
\end{aligned}
\end{equation}
This manifestly preserves the ``composite spin'' subspace of the double Hilbert space defined by local basis states $\ket{m,m}_{j,t;j,b} \equiv \ket{\widetilde{m}}_j$, where it acts as
\begin{equation} 
\label{eq:P_comp_gen}
\mathcal{L}_{h_W}^2|_{\text{comp.}} = 2 \left(|\widetilde{\alpha} \rangle - |\widetilde{\beta} \rangle \right) \left(\langle \widetilde{\alpha}| - \langle \widetilde{\beta}| \right)_R ~.
\end{equation}
This is the Rokhsar-Kivelson-type form in super-Hamiltonians noted in Ref.~\cite{moudgalya2024symmetries}, App.~B, which we are generalizing and adopting to the case of impurities below.
For example, for the choice of dipole-conserving dynamics with $W_{ijkl} = S_i^- S_j^+ S_k^+ S_l^-$ satisfying $i+l=j+k$ as in the main text, and with a local spin-1/2 Hilbert space, Eq.~\eqref{eq:P_comp_gen} reads
\begin{equation}
    \delta \hat{\mathcal P}|_{\text{dip.}|\text{comp.}} = 2\left(\ket{\widetilde{\uparrow}\widetilde{\downarrow}\widetilde{\downarrow}\widetilde{\uparrow}} - \bra{\widetilde{\downarrow}\widetilde{\uparrow}\widetilde{\uparrow}\widetilde{\downarrow}}\right) \left(\ket{\widetilde{\uparrow}\widetilde{\downarrow}\widetilde{\downarrow}\widetilde{\uparrow}} - \bra{\widetilde{\downarrow}\widetilde{\uparrow}\widetilde{\uparrow}\widetilde{\downarrow}} \right)_{ijkl} .
\end{equation}
After using the basis transformation in Eq.~\eqref{eq:X_spins} to $|\widetilde{\rightarrow}\rangle$ and $|\widetilde{\leftarrow}\rangle$, this leads to the super-Hamiltonian in Eq.~\eqref{eq:P_dip}, involving one and three spin flips processes with respect to the new basis.

One may wonder why we have chosen $h_W = W + W^\dagger$ and not $h_W^\prime = i(W - W^\dagger)$ (or even the algebra generated by $h_W$ and $h_W^\prime$ independently 
A simple calculation shows that for $W$ of the form in Eq.~(\ref{eq:W_single_ketbra}), $\mathcal{L}_{h_W^\prime}^2$ also preserves the composite spin sector and $\mathcal{L}_{h_W^\prime}^2|_{\text{comp.}} = \mathcal{L}_{h_W}^2|_{\text{comp.}}$.
Hence in this case, and when studying the super-Hamiltonian in the composite spin sector, it does not matter whether we use $h_W$ or $h_W^\prime$ or both in the bond algebra (up to an overall amplitude if combining contributions to the super-Hamiltonian).
The intuition is that we are also implicitly assuming the presence of other terms [e.g., onsite in Eq.~\eqref{eq:A_dip}] in the bond algebra that select the composite spin space as the right low-energy subspace of the full super-Hamiltonian; in this case it is natural that the terms $h_W$ and $h_W^\prime$ could produce equivalent long-time dynamics in our Brownian circuits.
From now on, we will just consider the $h_W$ terms.

A more general case is when $W$ has the form
\begin{equation}
W = \ketbra{\alpha}{\beta}_R + \ketbra{\gamma}{\delta}_R + \dots ~,
\label{eq:W_multiple_ketbras}
\end{equation}
where $\ket{\alpha}_R, \ket{\beta}_R, \ket{\gamma}_R, \ket{\delta}_R, \dots$ are again product states in the computational basis.
If we require that all $\alpha, \beta, \gamma, \delta, \dots$ are distinct, then $\mathcal{L}_{h_W}^2$ preserves the composite spin space and
\begin{equation} 
\label{eq:P_comp_gen_1}
\mathcal{L}_{h_W}^2|_{\text{comp.}} = 2 \left(|\widetilde{\alpha} \rangle - |\widetilde{\beta} \rangle \right) \left(\langle \widetilde{\alpha}| - \langle \widetilde{\beta}| \right)_R + 
2 \left(|\widetilde{\gamma} \rangle - |\widetilde{\delta} \rangle \right) \left(\langle \widetilde{\gamma}| - \langle \widetilde{\delta}| \right)_R + \dots ~.
\end{equation}
This situation occurs frequently, e.g., in our calculations of the unperturbed super-Hamiltonians.
On the other hand, when $\alpha,\beta,\gamma,\delta,\dots$ are not all distinct, the contribution $\mathcal{L}_{h_W}^2$ in general does not preserve the composite spin sector and the above formula does not apply directly.
However, when $h_W$ is a perturbation (like one of our impurities) while the unperturbed super-Hamiltonian preserves the composite spin sector, we can justify considering the projection of $\mathcal{L}_{h_W}^2$ into the composite spin sector by appealing to first-order perturbation theory.
In this case, if we add a requirement that for the entering ``pairs'' $\{\alpha, \beta\}, \{\gamma,\delta\}, \dots$ in the above $W$ they are distinct as unordered 2-sets (which is natural requirement since $\gamma=\alpha,\delta=\beta$ would be a duplication in $W$, while $\gamma=\beta,\delta=\alpha$ would be a duplication in $W^\dagger$, and we are implicitly also requiring $\alpha \neq \beta$, $\gamma \neq \delta$), we have
\begin{equation}
\Pi_{\text{comp.}} \mathcal{L}_{h_W}^2 \Pi_{\text{comp.}} = 2 \left(|\widetilde{\alpha} \rangle - |\widetilde{\beta} \rangle \right) \left(\langle \widetilde{\alpha}| - \langle \widetilde{\beta}| \right)_R + 
2 \left(|\widetilde{\gamma} \rangle - |\widetilde{\delta} \rangle \right) \left(\langle \widetilde{\gamma}| - \langle \widetilde{\delta}| \right)_R + \dots ~,
\end{equation}
where $\Pi_{\text{comp.}}$ is the projector to the composite spin space.
The status of such treatment of the impurity is then (approximate) perturbative.
This is, e.g., the case for the impurities considered in Secs.~\ref{sec:tJz}~and~\ref{sec:H3}.
On the other hand, we could also treat impurity perturbations as producing independent terms $|\alpha \rangle \langle \beta|_R + \text{H.c.}$, $|\gamma \rangle \langle \delta|_R + \text{H.c.}$, etc., in the bond algebra, which is a microscopically different model of impurities but expected to have the same qualitative physics.
In this treatment and per earlier discussion of the case Eq.~(\ref{eq:W_single_ketbra}), using the r.h.s.\ of the above formula in the composite spin sector is not making any perturbative approximation.

\section{Details of the hydro-mode single-particle Hamiltonian $\mathbb{H}_\text{dip}$ on a finite chain} \label{app:dip_bound}
\subsection{Single-particle Hamiltonian}
In this appendix, we provide some details of $\mathbb{H}_\text{dip}$ on a finite chain with open boundaries.
Eq.~\eqref{eq:H_ijkl} is the full expression in any geometry as long as $i,j,k,l$ are physical sites inside the sample.
Eq.~\eqref{eq:H_dipole_s} collected the action of $\mathbb{H}_{\text{dip}}$ on $|j)$ for $j$ in the bulk, and the $J_4$ term is valid for $3 < j < L-2$, while the $J_5$ term is valid for $4 < j < L-3$.
The actions on the sites close to the boundaries are modified because fewer terms in Eq.~\eqref{eq:H_ijkl} touch those sites.
Focusing on the left boundary, the modified $J_4$ action on $|1),|2),|3)$ and the modified $J_5$ action on $|1),|2),|3),|4)$ is given by
\begin{equation} \label{eq:H_45_bound}
\begin{aligned}
\mathbb{H}_{\text{dip}}^{\text{left edge}} = J_4 \Big\{
& \big[|1) - |2) - |3) + |4) \big](1|
\;+\; \big[-|1) + 2|2) - 2|4) + |5) \big](2|
\;+\; \big[-|1) + 3|3) - |4) - 2|5) + |6) \big](3| \Big\} \\
+ J_5 \Big\{
& \big[|1) - |2) - |4) + |5) \big](1|
\;+\; \big[-|1) + 2|2) - |3) + |4) - 2|5) + |6) \big](2| \\
& \;+\; \big[-|2) + 2|3) - |4) + |5) - 2|6) + |7) \big](3| 
\;+\; \big[-|1) + |2) - |3) + 3|4) - 2|5) + |6)-2|7) + |8) \big](4| \Big\}~.
\end{aligned}
\end{equation}
Similar modifications can be spelled out near the right edge.
Diagonalizing such $\mathbb{H}_{\text{dip}}$ including all bulk and edge parts gives us the lattice modes in the dipole-preserving model in an OBC segment.
Numerical results presented in Fig.~\ref{fig:Dipole_eigenmodes_no_imp}, and panel (a) in Figs.~\ref{fig:Dipole_Ct_breakP} and ~\ref{fig:Dipole_Ct_breakPQ} were obtained by numerically diagonalizing the Hamiltonian in Eq.~\eqref{eq:H_45_bound} for $J_4=1$ and $J_5=1/\sqrt{2}$.
\subsection{Boundary conditions of the continuum equations from eigenstates of single-particle Hamiltonian} \label{subapp:QP_break}

\begin{figure*}[tb]
\includegraphics[width=0.75\linewidth]{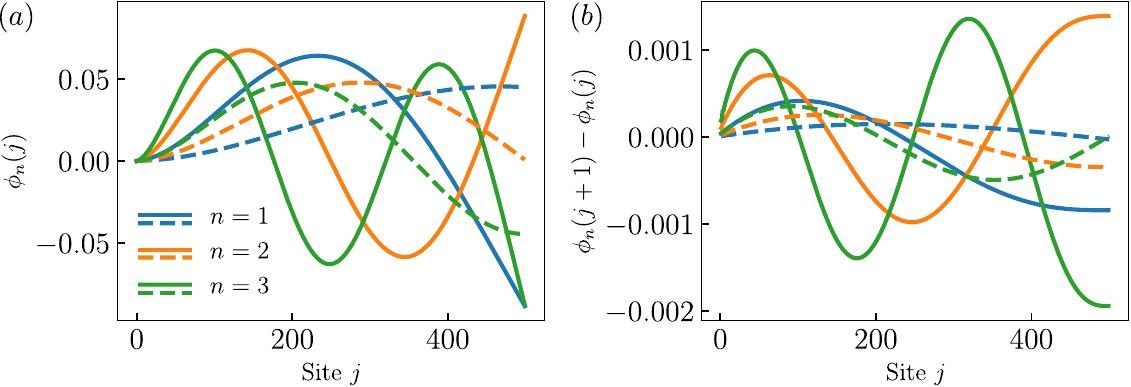}
\caption{\label{fig:H45_eigenstates_dpsi}\textbf{Eigenstates of the single-particle Hamiltonian of dipole-conserving systems with fully symmetry-breaking impurity on three consecutive sites $j_s=1, 2, 3$.}
(a) Eigenstates $\phi_n(j)$ of the single-particle Hamiltonian Eq.~\eqref{eq:H_breakPQ}, with boundary dynamics given by Eq.~\eqref{eq:H_45_bound}, for $L=500$ (solid lines) and $L=1000$ (dashed lines) and $g=1$.
(b) Finite difference $\Delta_x \phi_n(j) =\phi_n(j+1)-\phi_n(j)$, which approaches zero with increasing system size.
Note that we are showing normalized orbitals, whose overal amplitude hence decreases with $L$ as $\sim L^{-1/2}$;
however, the approach to zero that we observe in both $\phi_n(j)$ and $\nabla_x \phi_n(j)$ at the boundary is faster than this.
This suggests that the left boundary conditions $\phi_k(x)|_{x=0} = \partial_x\phi_k(x)|_{x=0}=0$ are valid emergent boundary conditions at long wavelengths for the continuum equation, as given in Sec.~\ref{subsec:dipole_breakPQ}.}
\end{figure*}

We analyze the eigenstates of the single-particle (hydro-mode) Hamiltonian $\mathbb{H}_{\mathrm{dip}}$ Eq.~\eqref{eq:H_breakPQ} with dynamics terms near the boundary given in Eq.~\eqref{eq:H_45_bound} to support the choice of the boundary conditions Eq.~\eqref{eq:Dipole_boundary_breakPQ} for the continuum equation discussed in Sec.~\ref{sec:dipole}. This analysis extends that discussed in the main text regarding the content of Fig.~\ref{fig:Dipole_eigenmodes_no_imp} in panels (c) and (f).
Figure~\ref{fig:H45_eigenstates_dpsi} shows the profiles of the first $3$ low-lying excited eigenstates (labeled by $n$) $\phi_n(j)$ and their first order finite difference $\Delta_x \phi_n(j) = \phi_n(j+1)-\phi_n(j)$. 
We find that both the eigenstates $\phi_n(j)$ and their first order finite difference $\Delta_x \phi_n(j)$ have an amplitude that tends to zero at the left boundary with increasing system size (and fixed value of $g=1$), as shown in panels (a) and (b) respectively.
These numerical results suggest that the boundary conditions $\phi_{k_n}(x)|_{x=0}=\partial_x \phi_{k_n}(x) = 0$ are valid for the continuum equation at long distances for the system with charge and dipole symmetry-breaking impurity.
We have carried out similar analysis in the absence of an impurity and for a dipole-symmetry-breaking and charge preserving impurity.

\section{Stochastic dynamics study of the spin-$1$ $3$-local and $4$-local dipole-conserving model with weak fragmentation} \label{app:H34_CA_Cz}
\begin{figure*}[bt]
\includegraphics[width=1\linewidth]{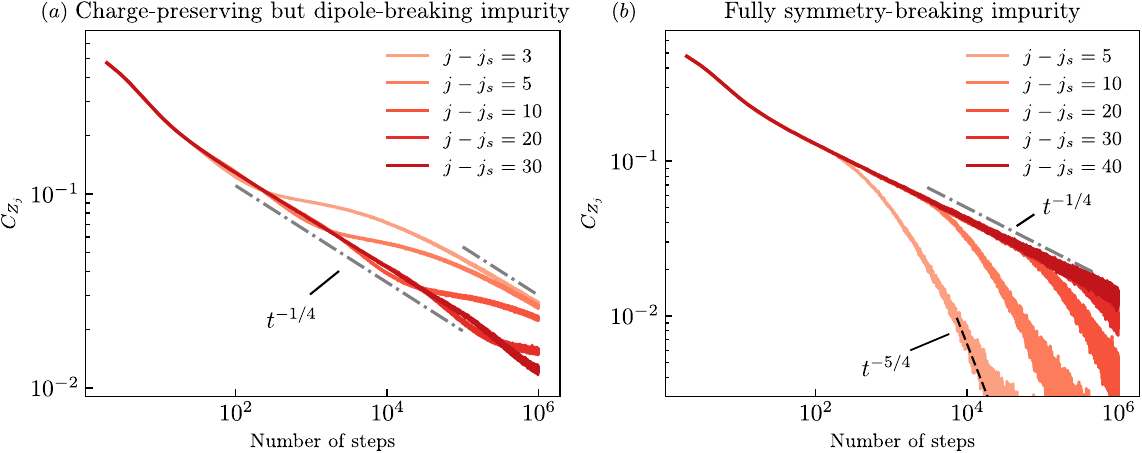}
\caption{\label{figure:H34_Cz}\textbf{Autocorrelation functions of the spin-$1$ $3$-local and $4$-local dipole-conserving model under impurity from stochastic dynamics simulations.}
The autocorrelation functions of the spin-$1$ $3$- and $4$-local dipole-conserving model under different types of impurities show similar behaviors as for the spin-$1/2$ $4$- and $5$-local dipole-conserving model discussed in Sec.~\ref{sec:dipole}, Fig.~\ref{fig:Dipole_Ct_breakPQ}(b) (note that unlike Sec.~\ref{sec:fragmentation}, we have placed the impurity at the left boundary to facilitate comparison with the figures in Sec.~\ref{sec:dipole}.
Specifically, for charge-preserving but dipole-breaking impurity in panel (a), the correlation functions scale as $C_{Z_j} \sim t^{-1/4}$ before exponential decay. 
We also see a front due to the reflective boundary with charge conservation.
On the other hand, with fully symmetry-breaking impurity in panel (b), the correlation functions transit from $t^{-1/4}$ scaling and are approaching $t^{-5/4}$ scaling when the impurity effects reach the observation location.
We expect the scaling to reach the predicted $t^{-5/4}$ more accurately at longer times and to persist until time that scales polynomially in the system size.
}
\end{figure*}
We numerically study the stochastic dynamics of the spin-$1$ $3$- and $4$-local dipole-conserving model with weak fragmentation under different types of impurity, with results in Fig.~\ref{figure:H34_Cz}. 
They exhibit similar behaviors as the spin-$1/2$ $4$- and $5$-local dipole-conserving model as discussed in Sec.~\ref{sec:dipole}, Fig.~\ref{fig:Dipole_Ct_breakPQ}(b).
Under the charge-preserving but dipole-breaking impurity, the correlation functions decay as $t^{-1/4}$, while with the fully symmetry-breaking impurity, the correlation functions transit from $t^{-1/4}$ to approximately $t^{-5/4}$. 
This indicates that our results for the spin-1/2 dipole-conserving model generalize to higher spins and different microscopic details, indicative of their broader universality.

\section{Details on the $t-J_z$ super-Hamiltonian}
In this appendix, we collect some details on the super-Hamiltonian of the $t-J_z$ model.
\subsection{Relation to the graph theory} \label{app:rel_graph}
\begin{figure*}[tb]
\includegraphics[width=15cm, scale=1]{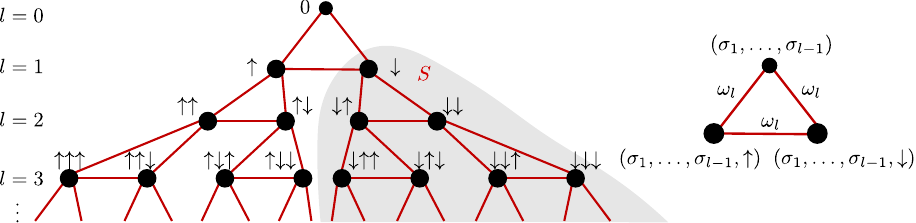}
\caption{\label{figure:graph_tJz}\textbf{Graph representation of the effective Hamiltonian of the $t-J_z$ model with boundary impurity at site $L$.} Each node corresponds to an unperturbed ground state $|G^{\{\sigma\}})$ with spin pattern $\{\sigma\}$, with weight $D_{\{\sigma\}}$ as the dimension of the Krylov subspace. The edges connecting the nodes indicate the impurity connecting the Krylov subspaces, with weight $\omega_{\{\sigma\},\{\sigma^\prime\}}$. The weight of the edges connecting nodes with $(\sigma_1, \ldots \sigma_{l-1})$, $(\sigma_1, \ldots, \sigma_l, \uparrow)$ and  $(\sigma_1, \ldots, \sigma_l, \downarrow)$ is given by $\omega_l$. 
The Cheeger's constant is given by the partition of the graph $G$ into $S$ and $G/S$, where $S$ is the shaded region including all the nodes labeled by spin patterns with $\downarrow$ as the first spin.}
\end{figure*}
First we show that the super-Hamiltonian formalism of exponentially slow relaxation of the $t-J_z$ model,  under a boundary impurity is related to the graph theory with a strong bottleneck, similar to the case introduced in Ref.~\cite{han2024exponentially, wang2025exponentiallyslowthermalization1d}.
These works studied the symmetric random unitary dynamics with a depolarizing noise at the boundary, which can be mapped to a graph with strong bottlenecks and leads to exponentially slow thermalization time.
Here, using the superoperator language and first-order degenerate perturbation theory, we show that the effective super-Hamiltonian of $t-J_z$ is exactly a normalized Laplacian $L(G)$ associated with a graph $G$ with a strong bottleneck. 
Then the analysis of the spectrum of this super-Hamiltonian can directly be related to the graph theory results of Ref.~\cite{han2024exponentially, wang2025exponentiallyslowthermalization1d}.
As discussed in the main text, for the unperturbed $t-J_z$ model, the Hilbert space separates into exponentially many Krylov subspaces, labeled by physical spin patterns $\{\sigma\} = (\sigma_1, \sigma_2, \ldots, \sigma_l)$. 
With a state-flip impurity on the right boundary, the rightmost physical spins can be flipped, such that the Krylov subspaces with spin patterns $(\sigma_1, \ldots, \sigma_{l-1}, \uparrow)$, $(\sigma_1, \ldots\sigma_{l-1}, \downarrow)$ and $(\sigma_1, \ldots\sigma_{l-1})$ are connected.
To the first-order in perturbation theory, the effective super-Hamiltonian with impurity is given by $\mathbb{H}_{\mathrm{eff}}$, with matrix 
elements $(\mathbb{H}_{\mathrm{eff}})_{\{\sigma\}, \{\sigma^\prime\}} =  (G^{\{\sigma\}}| \hat{\mathcal{V}}|G^{\{\sigma^\prime\}})$, $\hat{\mathcal{V}} = (1-\tilde{F}_{L})$, where $|G^{\sigma})$ are normalized.
Here use that the unperturbed super-Hamiltonian satisfies $\hat{\mathcal{P}}_{t-J_z|\mathrm{comp}} |G^{\{\sigma\}}) = 0$.
The effective super-Hamiltonian can then be written as a generalized Rokhsar-Kivelson (RK) Hamiltonian~\cite{castelnovo2005rk},
\begin{equation}\label{eq:tJz_RK}
    \mathbb{H}_{\mathrm{eff}} = \sum_{\{\sigma\}, \{\sigma^\prime\}} \omega_{\{\sigma\}, \{\sigma^\prime\}} \left(\frac{1}{\sqrt{D_{\{\sigma\} } }} |G^{\{\sigma\} }) - \frac{1}{\sqrt{D_{\{\sigma^\prime \} } }} |G^{\{\sigma^\prime \}}) \right)\left(\frac{1}{\sqrt{D_{\{\sigma\} } }} (G^{\{\sigma\} }| - \frac{1}{\sqrt{D_{\{\sigma^\prime \} } }} (G^{\{\sigma^\prime \}}| \right),
\end{equation}
which indicates that the unperturbed ground states with spin patterns $\{\sigma\}$ and $\{\sigma^\prime\}$ are connected by the impurity $\hat{\mathcal{V}}$, with weight $\omega_{\{\sigma\},\{\sigma^\prime\}} \geq 0$. 
And $D_{\{\sigma\}} = \binom{L}{l}$ is the dimension of the Krylov subspace labeled by a spin pattern $\{\sigma\}$ of length $l$.
The weight of the edge is given by $\omega_{\{\sigma\}, \{\sigma^\prime\}} \equiv \omega_l = \frac{1}{2} \binom{L-1}{l-1}$, between spin patterns $\{\sigma\}$ and $\{\sigma^\prime\}$ with $(\sigma_1, \ldots, \sigma_{l-1}, \uparrow)$, $(\sigma_1, \ldots\sigma_{l-1}, \downarrow)$ or $(\sigma_1, \ldots\sigma_{l-1})$. 
The ground state of the generalized RK Hamiltonian is known to be given by~\cite{castelnovo2005rk} the weighted superposition of $|G^{\{\sigma\}})$, 
\begin{equation}
    |\mathrm{GS}) = \sum_{\{\sigma\}} \sqrt{D_{\sigma}} |G^{\{ \sigma\}}) = |\mathbb{1}),
\end{equation}
which exactly corresponds to the identity matrix, the only conserved quantity of the $t-J_z$ model with state-flip impurity.

In fact, the generalized RK Hamiltonian Eq.~\eqref{eq:tJz_RK} exactly corresponds to a normalized Laplacian matrix $\hat{L}(\mathcal{G})$ that is associated with an weighted undirected graph $\mathcal{G}$~\cite{chung1997spectral, aldous-fill-2014}. 
This means that the dynamics of the quantum Hamiltonian $\mathbb{H}_{\mathrm{eff}}$ can be interpreted by the classical stochastic dynamics of the graph given by $\hat{L}$.
The graph is shown in Fig.~\ref{figure:graph_tJz}, where each node corresponds to a spin pattern $\{\sigma\}$ with weight $D_{\{\sigma\}}$. 
The nodes with spin patterns ${\{\sigma\}}$ and $\{\sigma^\prime\}$ are connected by the action of $\mathbb{H}_{\mathrm{eff}}$ with the weights $\omega_{\{\sigma\}, \{\sigma^\prime\}}$, which are denoted with the edges.
Explicitly, the unnormalized Laplacian matrix is given by $\hat{L}_0 = \hat{D} - \hat{A}$, where $(\hat{D})_{\{\sigma\}, \{\sigma\}} = D_{\{\sigma\}}$ is the degree matrix, $(\hat{A})_{\{\sigma^\prime \}, \{\sigma^\prime \}} = \omega_{\{\sigma \}, \{\sigma^\prime\}}$ is the adjacency matrix, and the normalized Laplacian matrix is given by $\hat{L} = \hat{D}^{-1/2} \hat{L}_0 \hat{D}^{-1/2}$.
The gap of the generalized RK Hamiltonian is associated with the connectivity of the graph, which can be quantified by the edge conductance. 
The edge conductance of the graph $\mathcal{G}$ is given by~\cite{chung1997spectral}
\begin{equation}
    \phi(\mathcal{G}) := \min_{\mathrm{vol}(S) \leq \mathrm{vol}(\mathcal{G})/2} \frac{|\partial S|}{\mathrm{vol} (S)},
\end{equation}
where $S$ is a partition of the graph $\mathcal{G}$, $\mathrm{vol}(S) = \sum_{\{\sigma\} \in S} D_{\{\sigma\}}$ is the total weight of nodes inside $S$, and $|\partial S| = \sum_{\{\sigma\} \in S, \{\sigma^\prime\} \notin S} \omega_{\{\sigma\}, \{\sigma^\prime\}}$ is the total weight of the edges across the boundary of $S$.
The edge conductance is determined by the ratio of the weight of edges across the boundary to the total weight of the nodes. 
When the edge conductance is small, it indicates a strong bottleneck of the graph, i.e., the dynamics flowing out of partition $S$ of $\mathcal{G}$ is slow.
The Cheeger's inequality states that the gap $\Delta$ of the normalized Laplacian is bounded by the edge conductance as
\begin{equation}
    \frac{1}{2} \phi^2(\mathcal{G}) \leq \Delta \leq 2\phi(\mathcal{G}).
\end{equation}
We can choose the partition $S$ including all the nodes that are associated with a spin pattern whose first spin is $\downarrow$, as shown in the shaded region of Fig.~\ref{figure:graph_tJz}. 
With $\mathrm{vol}(S) = \frac{1}{2}(3^L-1) \leq \mathrm{vol}(\mathcal{G})/2 = 3^L/2$, and $|\partial S| = 2 \omega_1 = 1$, the gap is upper bounded as $\Delta \leq 4/3^L$. 
This upper bound of the gap coincides with what we obtain from the SLIOM in the main text.
Therefore, the thermalization time $t_{\mathrm{th}} \sim 1/\Delta \sim \mathcal{O}(3^L)$, which is exponentially large with system size.

\subsection{First excited state of $\hat{\mathcal{P}}_{t-J_z|\mathrm{imp}}$} \label{app:num_check}
\begin{figure*}[bt]
\includegraphics[width=0.5\linewidth]{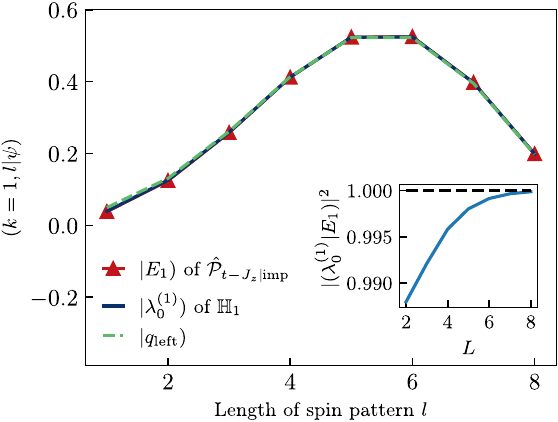}
\caption{\label{fig:tJz_superH_E1}\textbf{The first excited state $|E_1)$ of $\hat{\mathcal{P}}_{t-J_z|\mathrm{imp}}$.} 
The profile of the first excited state $|E_1)$ projected onto the $|k=1,l)$ basis, given by $(1,l|E_1)$,  looks very similar to the profile of the lowest energy eigenstate $|\lambda_0^{(1)})$ of $\mathbb{H}_1$ for $L=8$.
The leftmost SLIOM $|q_{\mathrm{left}})$ is also included for comparison.
The inset shows that $|E_1)$ can be well approximated by $|\lambda_0^{(1)})$, with $|(\lambda_0^{(1)}|E_1)|^2 \sim 1$, and the approximation improves with the increase of system sizes.
In the main text in Fig.~\ref{fig:tJz_Hk_spectrum}(a) we show that in turn $|\lambda_0^{(1)})$ is well approximated by the first SLIOM $|q_{\mathrm{left}})$.
}
\end{figure*}

We analyze the low-energy spectrum of the effective Hamiltonian $\mathbb{H}_{\mathrm{eff}} = \oplus_k \mathbb{H}_k$ in Eq.~\eqref{eq:tJz_single_H_eff} in Sec.~\ref{subsubsec:superhamlow}, which is derived from the degenerate perturbation theory for $\hat{\mathcal{P}}_{t-J_z|\mathrm{imp}}$ in Eq.~\eqref{eq:Pimp_tJz}.
Recall that the effective Hamiltonian $\mathbb{H}_{\mathrm{eff}} = \oplus_k \mathbb{H}_{k=1,\ldots,L}$ is restricted to the ``single-flip'' subspace spanned by basis $|k,l)$.
The ground state of the perturbed super-Hamiltonian, $|\mathbb{1})$, is outside of the single-flip subspace.
Therefore, the lowest-energy eigenstate of the effective Hamiltonian should approximate the first excited state of the super-Hamiltonian.
In Fig.~\ref{fig:tJz_superH_E1}, we show that the lowest energy eigenstate $|\lambda_0^{(1)})$ of $\mathbb{H}_1$ is close to $|q_{\mathrm{left}})$ (as also shown in Fig.~\ref{fig:tJz_Hk_spectrum} in the main text). 
Here we compare the lowest-energy eigenstate $|\lambda_0^{(1)})$ of $\mathbb{H}_1$ with the first excited state $|E_1)$ of $\hat{\mathcal{P}}_{t-J_z|\mathrm{comp}}$ projected onto the $|k,l)$ basis [the weight of $|E_1)$ outside of this space is found to be exponentially small].
Figure~\ref{fig:tJz_superH_E1} shows that $|\lambda_0^{(1)})$ can indeed approximate the true first excited state $|E_1)$ of the super-Hamiltonian, with the approximation improving with the increase of the system size.
This exponentially slow mode $|E_1)$ leads to the prethermal plateau of the boundary correlation function of the perturbed $t-J_z$ model.

\subsection{Effective parent Hamiltonian }
\label{subsec:parent_H0}
Here we discuss a different aspect of the $t-J_z$ super-Hamiltonian, in particular the fact that the spectra of all the $\mathbb{H}_k$ in Eq.~(\ref{eq:tJz_single_H_eff}) can be understood in terms of an effective parent super-Hamiltonian. 
Let us consider the Hamiltonian
\begin{equation}
\begin{aligned} \label{eq:H_0}
        \mathbb{H}_0 &= \sum_{l=0}^L \mu_l^{-} |0,l)(0,l| - \sum_{l=0}^{L-1} t_l \big(|0,l+1)(0, l| + \mathrm{h.c.} \big),
\end{aligned}
\end{equation}
where the onsite potentials on the $l=0$ and $l=1$ sites are given by the same function $\mu_l^{-}$ as for any other site.
Hamiltonian $\mathbb{H}_0$ in fact appears when projecting the effect of the impurity within the sector of $|G^{\sigma_1\sigma_2\dots \sigma_l})$ with no spin flips, i.e., on the subspace spanned by (their normalized forms) $\{|0,l)\}$, and admits an exact zero-energy ground state
\begin{equation}
  |q_0) \equiv \sum_{l=0}^L \sqrt{2^l \binom{L}{l}} |0,l) = |\mathbb{1}) 
\end{equation}
%
Hence, the Hamiltonian $\mathbb{H}_0$ can be thought of as a ``parent zero mode system", with all single-flip Hamiltonians $\mathbb{H}_k$ corresponding to specific ``terminations'' of the left boundary, i.e., they are obtained by removing the leftmost $k$ sites $l=0,\dots,k-1$ and then changing the chemical potential on the leftmost remaining site from $\mu_k^{-}$ to $\mu_k^{+}$.
The advantage of this ``parent'' formulation is that $\mathbb{H}_0$ can be exactly diagonalized by mapping it to a spin $\frac{L}{2}$ in the presence of a magnetic field, and the low-energy properties of $\mathbb{H}_k$ can be understood from this solution. 
Indeed, we notice that under the change of variables $l \to m = l - \frac{L}{2} \in \{-L/2, -L/2+1, \dots, L/2-1, L/2\} $, the hopping amplitude $t_l$ becomes $t_l=\frac{1}{\sqrt{2}L}\sqrt{(\frac{L}{2}-m)(\frac{L}{2}+m+1)}$, which we recognize as proportional to the matrix element connecting spin states $|j=L/2,m\rangle$ and $|j=L/2,m+1 \rangle$ under the action of
the raising operator $\hat{J}^+$ for a spin $\frac{L}{2}$ (and the lowering operator $\hat{J}^-$ connecting in the opposite direction).
Similarly, the diagonal chemical potential contribution maps to the total magnetization $\hat{J}^z$. 
Overall, we obtain the exact rewriting 
\begin{equation}\label{eq:H1_approx}
        \mathbb{H}_0 = \frac{3}{4} - \frac{3}{2L}\left( \frac{1}{3}\hat{J}_z+\frac{2\sqrt{2}}{3}\hat{J}_x\right),
\end{equation}
namely a spin $\frac{L}{2}$ with an applied field of strength $\frac{3}{2L}$ along the direction $\hat{z}'=\frac{1}{3}\hat{z}+\frac{2\sqrt{2}}{3}\hat{x}$. 
Hence, its spectrum is given by $E_{m'}=\frac{3}{4}-\frac{3}{2L}m'$ with $m'=-\frac{L}{2},\dots,+\frac{L}{2}$
The ground state energy is $0$ as expected from the identity operator $|q_0) \equiv |\mathbb{1})$ residing in this sector, and the lowest energy excitation is $\frac{3}{2L}$ scaling with $L$ as anticipated from the discussion of the excitation energy findings in $\mathbb{H}_1$ and the picture of a localized excitation on top of $|q_1)$ there.~\footnote{
Interestingly, the solution of $\mathbb{H}_0$ predicts a uniform spacing of $\frac{3}{2L}$ for the low-energy states.
In the picture of localized states on top of the ground state, this suggests that for the excitation with number $m'=1,2,\dots$, which looks like a standing wave with wavevector $\sim m'/\Delta l^{\text{loc}}_{m'}$ over localization region $\Delta l^{\text{loc}}_{m'}$, the size of the localization region itself grows with the excitation number $m'$ as $\Delta l^{\text{loc}}_{m'} \sim \sqrt{m'L}$.}
Since as we argued, the relevant low-energy physics is controlled by $l$'s far from the left boundary, we expect the same conclusions to apply to other $\mathbb{H}_k$ for general $k < \frac{2}{3}L$.

\section{Strongly fragmented $3$-local dipole-conserving model with one-site impurity} \label{app:frag_1site}
In the main text in Fig.~\ref{fig:H3_H34_gap}(a), we numerically showed that one-site impurity does not fully eliminate fragmentation in both the $H_3$ and $H_3 + H_4$ spin-1 models.
Here we demonstrate this analytically, first by appealing to simple blockade/frozen state arguments, and then by providing a detailed understanding of boundary-impurity-generated connections between original Krylov subspaces of the $H_3$ model.

In the case of $H_3$ and the impurity at the boundary, we can understand the persisting fragmentation by considering, e.g., blockade configurations of the form $|\dots + 0 \dots 0 ~\fbox{+ +}~ * \rangle$, where ``$*$'' denotes possible arbitrary spin state at the impurity location $j_s = L$; the $H_3$ term acting on the last three sites annihilates any such configuration, while all the other $H_3$ terms preserve the blockade form of the configurations (in particular, the spins at $L-2$ and $L-1$ remain frozen as indicated with the surrounding box).
For $H_3 + H_4$ and the impurity at the boundary, we can see at least some fragmentation persisting, since configurations of the form $|~\fbox{\dots - - - + + + - - - + + +}~ * \rangle$ are annihilated under all $H_3$ and $H_4$ terms (the domains with the same charge need not be regular as long as they are long enough); hence, the impurity can only change the * spin and not anything else on such states.
For a single impurity in the bulk, we can find similar blockade configurations for $H_3$ ($|\dots + 0 \dots 0 ~\fbox{+ +}~ * ~\fbox{- -}~ 0 \dots 0 - \dots \rangle$ and frozen states for $H_3 + H_4$ ($|~\fbox{\dots - - - + + + - - - + + +}~ * ~\fbox{- - - + + + - - - + + + \dots}~ \rangle$) showing that strong and at least some fragmentation respectively persist in these cases as well.
We now turn to a more detailed argument for the boundary impurity in the $H_3$ model.
Here we show that for a one-site state-flip impurity at the boundary, the spin-$1$ $3$-local dipole-conserving model remains fragmented due to the partial conservation of dipole moments and defects. 
It is sufficient to work at the level of classical transitions, e.g., considering random circuits with $3$-local dipole-conserving terms and a local state-flip impurity $G_i$.
The Krylov subspaces of the $3$-local dipole-conserving model can be uniquely labeled by (1) the first physical spin and (2) the dipole moments separated by the defects, as explained in the main text. 
Thus, we label the Krylov subspaces by $\{(\sigma_{\mathrm{left}}), (P_0, P_1, \ldots, P_{N_d})\}$
.
Note that the last physical spin $\sigma_{\mathrm{right}}$, which is also conserved, is determined by the sign of $P_{N_d}$, i.e., $\sigma_{\mathrm{right}} = + (-)$ for $P_{N_d}>0 (<0)$.
The conserved defect patterns are also determined by the $\sigma_{\mathrm{left}}$ and the sign of dipole moments $P_k$. 
For example, the first defect is $+$ ($-$) for $P_0 < 0 (>0)$ and $\sigma_{\mathrm{left}} = +$.

\begin{figure*}[bt]
\includegraphics[width=0.6\linewidth]{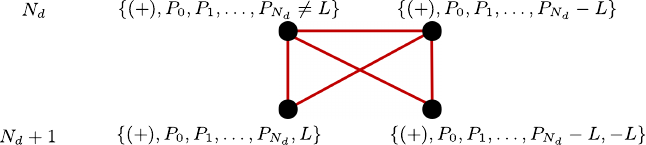}
\caption{\label{figure:graph_H3}\textbf{Graphic illustration of the $3$-local dipole-conserving model with one-site state-flip impurity at the right boundary.} Each node denotes an unperturbed Krylov subspace of the $3$-local dipole-conserving model with the number of defects $N_d \geq 2$, and the red lines indicating how the Krylov subspaces are connected by the impurity. For all original Krylov subspaces with $N_d \geq 2$, at most four Krylov subspaces are connected, with the dipole moments $P_l$ for $l<N_d$ unchanged.}
\end{figure*}

Consider an impurity on site $L$. The subspace labeled by $\{(+), P_0, P_1, \ldots P_{N_d} (\neq L)\}$ and $N_d \geq 2$, it is connected to at most three other subspaces labeled by $\{(+), P_0, P_1, \ldots, P_{N_d}-L\}$, $\{(+), P_0, P_1, \ldots, P_{N_d}, P_{N_d+1} = L\}$, and $\{(+), P_0, P_1, \ldots, P_{N_d}-L, P_{N_d+1} = -L\}$, as illustrated in Fig.~\ref{figure:graph_H3}.
Note that in some cases, a subspace is only connected to two others by the impurity. For example, for $L=3$, the one-dimensional subspace spanned by $\ket{++0}$ with $\{(+), P_0 = 1, P_1 =2\}$ is only connected to $\ket{++-}$ labeled by $\{(+), P_0, P_1 - 3\}$ and $\ket{+++}$ labeled by $\{(+)\}, P_0, P_1, 3$, when the subspace $\{(+), P_0-3, -3\}$ does not exist.
Note that for example, for the subspace $\{(+), P_0, P_1, \ldots, P_{N_d}, P_{N_d+1} = L\}$, the dipole moment $P_{N_d+1} = L$ means that the $(N_d+1)$th defect is at site $L$ and is completely frozen under the action of $H_3$ terms, thus the subspace cannot be further connected to any other subspaces than $\{(+), P_0, P_1, \ldots P_{N_d} (\neq L)\}$ and $\{(+), P_0, P_1, \ldots, P_{N_d}-L\}$ by the impurity at site $L$.

\begin{figure*}[bt]
\includegraphics[width=0.5\linewidth]{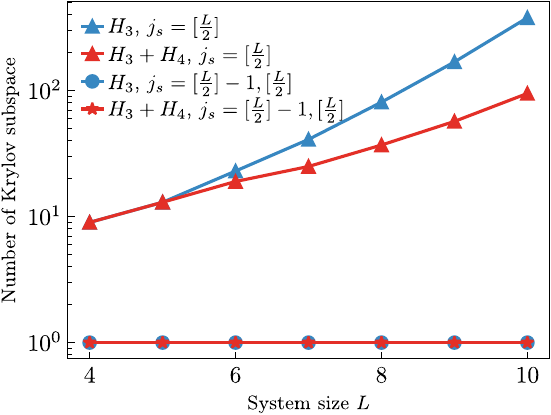}
\caption{\label{figure:H3_H34_gap_bulkimp}\textbf{Number of Krylov subspaces of spin-$1$ dipole-conserving models with bulk impurities.} For one local impurity in the bulk (with $[\frac{x}{2}]=\frac{x}{2}$ for even $x$ and $[\frac{x}{2}]=\frac{x-1}{2}$ for odd $x$), 
both the $3$-local dipole-conserving model and the $3$-local and $4$-local dipole-conserving models have exponentially many Krylov subspaces, indicating fragmentation. While with impurities acting on two sites in the bulk, all symmetries of both models are broken, resulting in one Krylov subspace.
}
\end{figure*}

Therefore, under a one-site local impurity at the boundary, the $3$-local dipole-conserving system remains fragmented due to the defect patterns that are almost conserved.
In fact, the number of Krylov subspaces under one-site local impurity at the boundary $K_{\mathrm{pert}}$ is reduced to $K_{\mathrm{pert}} \approx K/4 \gtrsim \mathcal{O}(2.11^{L})$\cite{2020_sala_ergodicity-breaking}, with $K$ as the number of Krylov subspaces, based on previous analysis. The scaling was shown in Fig.~\ref{fig:H3_deg_gap_corre}.
We also numerically evaluate the number of Krylov subspaces for a state-flip impurity located in the bulk, which also appears to show exponential scaling with system size and indicates fragmentation (Fig.~\ref{figure:H3_H34_gap_bulkimp}).

\section{$t-J_z$ model with impurities at both boundaries}\label{app:imp_both_boundaries}


Here, we consider the $t-J_z$ model with impurities at both boundaries, and show that: (i) the system thermalizes in polynomial time (instead of exponential) with system size; and (ii) the system does not have prethermal plateaus of correlation functions.

\begin{figure*}
\includegraphics[width=1\linewidth]{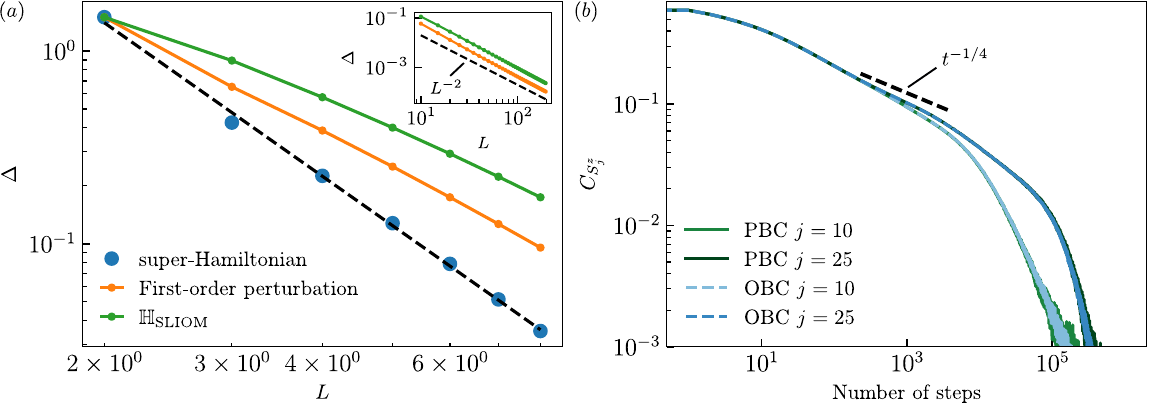}
\caption{\label{fig:tJz_two_imp}\textbf{$t-J_z$ model with impurities at both boundaries.} 
(a) The energy gap of the perturbed super-Hamiltonian $\hat{\mathcal{P}}_{t-J_z|\mathrm{two\ imp}}$ [Eq.~\eqref{eq:H_tjz_twoimp}], with the fit showing scaling as $\sim L^{-2.6(1)}$.
The energy gap is upper bounded by the lowest eigenvalue of the first-order perturbation theory given by $\mathbb{H}_{\mathrm{eff|two\ imp}}$, which in turn is upper-bounded by the left SLIOM effective hopping Hamiltonian $\mathbb{H}_{\mathrm{SLIOM}}$; these bounds scale approximately as $L^{-2}$ as shown by the offset dashed line in the inset (note that these calculations allow much larger sizes than for the full $\hat{\mathcal{P}}_{t-J_z|\mathrm{two\ imp}}$).
(b) Bulk correlation functions $C_{S_j^z}$ for OBC and PBC chains with $L=50$ and impurities at $j_s = 1, L$, calculated using corresponding stochastic cellular automaton simulations and averaging over $10^8$ random realizations.
The correlation functions of the PBC and OBC systems essentially coincide, and decay first polynomially and then exponentially without showing any prethermal plateaus.
}
\end{figure*}

First, we study the gap of the perturbed super-Hamiltonian $\hat{\mathcal{P}}_{t-J_z|\mathrm{two \ imp}} = \hat{\mathcal{P}}_{t-J_z|\mathrm{comp}} + \hat{\mathcal{V}}_1 + \hat{\mathcal{V}}_L$, with $\hat{\mathcal{V}}_{j_s} = 1-\tilde{F}_{j_s}$ [cf.~Eq.~\eqref{eq:Pimp_tJz}], within the composite spin subspace as defined in the main text. 
Numerically, we find that the gap scales as $L^{-2.6(1)}$ as shown in Fig.~\ref{fig:tJz_two_imp}(a), which is distinct from the exponentially small gap for a single impurity at one boundary only.
Therefore, the $t-J_z$ model perturbed by impurities at both boundaries thermalizes in polynomial time, which is qualitatively faster than the exponential thermalization time in the case of a single impurity at one boundary only.
The key physics is that the left SLIOMs, which are exponentially localized away from the right boundary and can thus avoid effects of the right boundary impurity, are extended (in terms of non-trivial operator strings) near the left boundary and cannot avoid effects of the left boundary impurity in the same way.
This can be seen in the energy of the trial state $|q_k)$ under the perturbations.
As discussed in the main text, with a single impurity at the right boundary, the left SLIOMs $q_k$ with $k < 2L/3$ are only weakly perturbed, which become approximate conserved quantities with super-Hamiltonian energy $\Delta_{k} = (q_k|\hat{\mathcal{V}}_L|q_k)/(q_k|q_k) \sim e^{-c L}$. 
This leads to the exponentially slow thermalization rate.
However, with impurities at both boundaries, these $|q_k)$ obtain a finite trial energy with $(q_k|\hat{\mathcal{V}}_1+\hat{\mathcal{V}}_L|q_k)/(q_k|q_k) \sim \mathcal{O}(1)$.
We will show detailed calculations and discuss how the SLIOMs can adjust to lower the trial energy later.
To gain more understanding of the polynomial in $L$ thermalization time, we can also study the effective Hamiltonian $\mathbb{H}_{\mathrm{eff|\mathrm{two\ imp}}}$ using the first-order perturbation theory in the space spanned by the basis states $\{|k,l)\}$, similar to the study in Sec.~\ref{sec:tJz}:
\begin{equation}\label{eq:H_tjz_twoimp}
\begin{aligned}
    \mathbb{H}_{\mathrm{eff|two\ imp}} &= \mathbb{H}_{\mathrm{eff}}[\hat{\mathcal{V}}_L] + \mathbb{H}_{\mathrm{eff}}[\hat{\mathcal{V}}_1]~, \\
\mathbb{H}_{\mathrm{eff}}[\hat{\mathcal{V}}_L] &= 
\sum_{k=1}^L \mu_k^{+} |k,k)(k,k| + \sum_{k=1}^L \sum_{l=k+1}^L \mu_l^{-} |k,l)(k,l| - \sum_{k=1}^{L-1} \sum_{l=k}^{L-1} t_l \left[|k,l+1)(k, l| +\mathrm{h.c.}\right], \\
\mathbb{H}_{\mathrm{eff}}[\hat{\mathcal{V}}_1] &=  \sum_{l=1}^L \mu_l^+ |1,l)(1,l| + \sum_{k=2}^L \sum_{l=k}^{L} \mu_l^- |k,l)(k,l| - \sum_{k=1}^{L-1} \sum_{l=k}^{L-1} t_l \left[ |k+1, l+1)(k,l|+\mathrm{h.c.} \right],
\end{aligned}
\end{equation}
where $\mathbb{H}_{\mathrm{eff}}[\hat{\mathcal{V}}_{j}]$ is given by the impurity at site $j$. Therefore, $\mathbb{H}_{\mathrm{eff}}[\hat{\mathcal{V}}_{L}]$ is exactly given by $\sum_{k=1}^L \mathbb{H}_k$ in Eq.~\eqref{eq:tJz_single_H_eff}.
Recall that the basis state $|k,l)$ corresponds to the unperturbed ground state $|G^{\{\sigma\}})$ with the spin pattern $(\widetilde{\rightarrow}_1, \widetilde{\rightarrow}_2, \ldots \widetilde{\rightarrow}_{k-1}, \widetilde{\leftarrow}_k, \widetilde{\rightarrow}_{k+1}, \ldots, \widetilde{\rightarrow}_l)$. 
As discussed in the main text, the impurity at the right boundary $\hat{\mathcal{V}}_L$ can add or remove a $\widetilde{\rightarrow}$ from the right, thus connecting $|k,l)$ to $|k, l\pm1)$.
On the other hand, the impurity at the left boundary $\hat{\mathcal{V}}_1$ can add or remove a $\widetilde{\rightarrow}$ from the left, connecting $|k,l)$ to $|k\pm1,l\pm1)$.
The $\mathbb{H}_{\mathrm{eff}}[\hat{\mathcal{V}}_1]$ can also be directly obtained from $\mathbb{H}_{\mathrm{eff}}[\hat{\mathcal{V}}_L]$, by counting the spin patterns from the right, where now $|k,l) \rightarrow |l-k+1,l)_{\mathrm{right}}$. 
This accounts for the rearrangements in the sums for the “onsite terms” and the “hopping” being now between $|k, l)$ and
$|k \pm 1, l \pm 1)$, manifestly mixing the sectors with different $k$.
For simplicity, we assume the impurity strength to be the same at the two boundaries, which explains why the same ``onsite potentials'' and ``hopping amplitudes'' appear in the two lines, 
with $\mu_l^-$ and $t_l$ defined in Eq.~\eqref{eq:tJz_single_H_eff} and $\mu_l^+ = (1+\frac{l}{2L})$. 
The lowest eigenvalue of $\mathbb{H}_{\mathrm{eff|two\ imp}}$ scales approximately as $L^{-2.1(1)}$ from fitting the data in the inset of Fig.~\ref{fig:tJz_two_imp}(a). 
%
This provides an upper bound on the energy gap of $\hat{\mathcal{P}}_{t-J_z|\mathrm{two\ imp}}$.
We note that, unlike the full super-Hamiltonian, we can study the first-order effective Hamiltonian for large sizes where the scaling can be accurately determined. We emphasize, however, that it is only a variational bound and not the full answer, but this is sufficient to assert the polynomial rather than exponential scaling.

This approximate $L^{-2}$ scaling of the gap of $\mathbb{H}_{\mathrm{eff|two\ imp}}$ can be understood by further restricting the super-Hamiltonian to the space of the left SLIOMs, obtaining an effective hopping problem of the SLIOMs.
Using the expression of the SLIOMs $\{|q_k) \}$ in the $\{|k,l) \}$ basis, Eq.~\eqref{eq:qk_kl}, we obtain
\begin{equation}
\mathbb{H}_{\mathrm{SLIOM}} = \sum_{k=1}^L v_k |\tilde{q}_k)(\tilde{q}_k| + \sum_{k=1}^{L-1} \bigg[w_k
 |\tilde{q}_{k+1})(\tilde{q}_{k}| + \mathrm{h.c.} \bigg], \quad 
 v_k = (\tilde{q}_k| \hat{\mathcal{V}}_1 + \hat{\mathcal{V}}_L |\tilde{q}_k), \quad 
 w_k = (\tilde{q}_{k+1}| \hat{\mathcal{V}}_1 |\tilde{q}_k) ~.
\end{equation}
For ease of notation, here we use normalized $|\tilde{q}_k)$, and we only show non-zero matrix elements.
Thus, $\hat{\mathcal{V}}_L$ does not have any off-diagonal matrix elements in the left SLIOM basis, while the diagonal matrix elements $(\tilde{q}_k|\hat{\mathcal{V}}_L |\tilde{q}_k)$ were calculated in the main text, Eqs.~\eqref{eq:exact_Delta_k} and \eqref{eq:Deltak} (direct calculation without using $\mathbb{H}_k$ is easier in this case).
We next obtain the needed pieces for the calculations of $(\tilde{q}_k|\hat{\mathcal{V}}_1 |\tilde{q}_k)$ and $(\tilde{q}_{k+1}| \hat{\mathcal{V}}_1 |\tilde{q}_k)$:
\begin{equation}\label{eq:H_tjz_twoimp_sliom}
\begin{aligned}
    &(q_k|q_k) = \sum_{l=k}^L 2^l \binom{L}{l} = \sum_{l=0}^L 2^l \binom{L}{l} - \sum_{l=0}^{k-1}2^l \binom{L}{l} = 3^L - \sum_{l=0}^{k-1} 2^l \binom{L}{l} ~; \\
    &(q_k| \hat{\mathcal{V}}_1 |q_k) = 
    \sum_{l=k}^L 2^l \binom{L}{l} \mu_l^- = \sum_{l=0}^L 2^l \binom{L}{l} \mu_l^- - \sum_{l=0}^{k-1} 2^l \binom{L}{l} \mu_l^- = 2 \cdot 3^{L-1} - \sum_{l=0}^{k-1} 2^l \binom{L}{l} \mu_l^-, \quad k \neq 1 ~; \\
    &(q_{k=1}| \hat{\mathcal{V}}_1 |q_{k=1}) = 
    \sum_{l=1}^L 2^l \binom{L}{l} \mu_l^+ = 4 \cdot 3^{L-1} -1, \quad k = 1; \\
    &(q_{k+1}|\hat{\mathcal{V}}_1|q_k)= - \sum_{l=k}^{L-1} t_l \sqrt{2^{l+1} \binom{L}{l+1}} \sqrt{2^l \binom{L}{l}} = -\sum_{l=k}^{L-1} \frac{L-l}{L} 2^l \binom{L}{l} 
    = -3^{L-1} + \sum_{l=0}^{k-1} \frac{L-l}{L} 2^l \binom{L}{l}.
\end{aligned}
\end{equation}
In each line, we also show a form that will be convenient for the subsequent semi-analytical treatment.
Putting everything together, we then evaluated the lowest eigenvalues of $\mathbb{H}_{\mathrm{SLIOM}}$ for varying $L$ and compared with those of $\mathbb{H}_{\mathrm{eff|two \ imp}}$, as shown in the inset of Fig.~\ref{fig:tJz_two_imp}(a).
In both cases we see scaling $\mathcal{O}(L^{-2})$, suggesting that $\mathbb{H}_{\mathrm{SLIOM}}$ captures such qualitative property of $\mathbb{H}_{\mathrm{eff|two \ imp}}$.
Note that the estimation with the left SLIOMs here is not expected to be accurate compared to $\mathbb{H}_{\mathrm{eff|two \ imp}}$, as the right SLIOMs, a set of approximately conserved quantities under $\hat{\mathcal{V}}_1$, are completely ignored here.
To gain semi-analytical understanding, we note that the following common factor appearing in the above expressions, $P(l,L) \equiv \frac{1}{3^L} 2^l \binom{L}{l} = \left(\frac{2}{3}\right)^l \left(\frac{1}{3}\right)^{L-l} \binom{L}{l}$, is a binomial distribution with probabilities $p_A = 2/3$ and $p_B = 1 - p_A = 1/3$.
For very large $L$, this distribution is sharply peaked around $l=2L/3$; near this value, it is a Gaussian with variance $2L/9$ for deviations $O(\sqrt{L})$, while if we fix $l = fL$ with $f<2/3$, it is exponentially small $\sim e^{-c_f L}$ with calculable $c_f$.
Hence, we estimate $\sum_{l=0}^{k-1} g(l) 2^l \binom{L}{l} \lesssim 3^L e^{-c_f L} G(L)$ for any polynomial function $g(l)$---e.g., such as appearing in Eq.~(\ref{eq:H_tjz_twoimp_sliom})---with some polynomial $G(L)$. 
Therefore, for $k \lesssim 2/3$, the normalization factor $(q_k|q_k) \approx 3^L$. 
Similarly, the diagonal matrix elements are $(q_1|\hat{\mathcal{V}}_1|q_1) \approx 4\cdot 3^{L-1}$ and $(q_k|\hat{\mathcal{V}}_1|q_k) \approx 2\cdot 3^{L-1}$ for $k\neq1$.
The off-diagonal matrix elements are $(q_{k+1}|\hat{\mathcal{V}}_1|q_k) \approx -3^{L-1}$.
On the other hand, for $k \lesssim 2L/3$, the contribution from $\hat{\mathcal{V}}_L$ is negligible, 
see Eq.~\eqref{eq:Deltak}.
Therefore, the effective left SLIOM Hamiltonian can be approximated by 
\begin{equation} \label{eq:H_sliom}
\mathbb{H}_{\mathrm{SLIOM}} \approx \frac{4}{3}|\tilde{q}_1)(\tilde{q}_1| +\sum_{k=2}^{2L/3} \frac{2}{3} |\tilde{q}_k)(\tilde{q}_k| - \sum_{k=1}^{2L/3} \frac{1}{3} \bigg[|\tilde{q}_{k+1})(\tilde{q}_k| + \mathrm{h.c.} \bigg] + \mathbb{V}_{[2L/3, L]} ~.
\end{equation}
$\mathbb{V}_{[2L/3, L]}$ includes the rest of the terms with $k > 2L/3$.
We see that this is a hopping problem with an impurity (extra repulsive potential) at $k=1$, uniform hopping and potential at $1 < k\lesssim 2L/3$, and a large potential wall due to $\mathbb{V}_{[2L/3, L]}$ at $k\gtrsim 2L/3$.
The profile of the lowest energy eigenstate is approximately $|\psi_\kappa\rangle \sim \sum_{k=1}^{2L/3} \sin(\kappa k) |\tilde{q}_k)$ with $\kappa \sim 1/L$ and energy $\sim \kappa^2 \sim 1/L^2$, and the coefficients of $|\tilde{q}_k)$ with $k\gtrsim 2L/3$ are close to zero [e.g., $\kappa \approx \pi/(2L/3)$ will achieve smooth vanishing of the profile near $k=0$ and $k=2L/3$].
Therefore, we expect that the scaling of the energy gap is approximately $L^{-2}$, based on the analysis of a similar problem that arose in the U($1$) case in Sec.~\ref{sec:U1_break}.
We thus see that this variational left SLIOM standing wave already explains the $L^{-2}$ scaling observed in the first-order perturbation theory as shown in the inset of Fig.~\ref{fig:tJz_two_imp}(a). 

Finally, we study the stochastic dynamics of the bulk correlation functions, for the OBC and PBC chains with $L=50$ and impurities at two sites $j_s=1,L$; the results are shown in Fig.~\ref{fig:tJz_two_imp}(b). 
The correlation functions are very close in the OBC and PBC cases, indicating the essential equivalence of OBC and PBC chains under this setting as argued in Sec.~\ref{subsec:tJz_twoimp}.
The correlation functions first decay polynomially, and then decay exponentially, without showing the prethermal plateaus, which is distinct behavior from the case with an impurity at one boundary.

\section{Evolution of the magnetization from a fully polarized state} \label{sec:magnetization}
Let us prepare an initial state $\ket{\psi_0}$, and compute the evolution of an operator $\hat{O}(t)$ (rather than of its correlation function) under the Brownian circuit dynamics.
Then, after averaging over various Brownian circuit realizations and using Eq.~\eqref{eq:O_t}, we find
\begin{equation}
    \overline{\langle \psi(t)|\hat{O}|\psi(t)\rangle} = \textrm{tr}\left[ \ketbra{\psi_0}{\psi_0} \, \overline{\hat{O}(t)}\right] = 
    \big( \ketbra{\psi_0} {\psi_0} \big| e^{-\hat{\mathcal{P}}t} \big|\hat{O} \big), 
\end{equation}
where in the last expression, we have used the operator to state mapping and the super-Hamiltonian formulation.
In particular, we are interested in the evolution of the magnetization $Z_{\textrm{tot}}= \sum_j Z_j$, upon preparing the system in the initial product state $|\psi_0\rangle = \ket{\uparrow }^{\otimes L}$.
The corresponding projector onto the maximal magnetization sector, $\ketbra{\psi_0}{\psi_0}$, in the super-Hamiltonian formulation maps to the fully polarized state of the composite spins: $\big| \ketbra{\psi_0}{\psi_0} \big) = |\tilde{\uparrow} \rangle^{\otimes L}$.
In the absence of impurities, this is an element of the commutant for all the systems considered in the main text, and hence $\hat{\mathcal{P}}|\tilde{\uparrow} \rangle^{\otimes L}=0$ and the total magnetization is constant and equals to $L$.
On the other hand, the magnetization dynamics becomes non-trivial in the presence of a symmetry-breaking impurity.

\subsection{Evolution of the magnetization in the U($1$)-symmetric systems perturbed by impurities}
First, we consider a spin-$1/2$ chain with U$(1)$-symmetric dynamics perturbed by impurities, as in Sec.~\ref{sec:U1} in the main text.
Recall that the magnetization operator lies within the single spin-flip sector spanned by normalized $\{|j)\}$ defined after Eq.~(\ref{eq:H_U1}),  and is given by $\big|Z_{\textrm{tot}} \big)
= 2^{L/2} \sum_j |j)$, cf.\ Eq.~(\ref{eq:Zj}).
In our models with impurities, the super-Hamiltonian preserves the single spin-flip sector, with action given by, e.g., Eq.~(\ref{eq:U1_Heff_imp}).
Denoting by $|\alpha) = \sum_j \phi_\alpha(j)|j)$ the corresponding normalized single-particle eigenstates with eigenenergies $E_\alpha$, we find that 
\begin{equation} \label{eq:M_t}
    \overline{\langle \psi(t)|Z_{\textrm{tot}}|\psi(t)\rangle} = \sum_\alpha e^{-E_\alpha t} \bigg|\sum_j \phi_\alpha(j)\bigg|^2 = L \sum_\alpha e^{-E_\alpha t} \big|( k=0|\alpha)\big|^2,
\end{equation}
where the zero momentum wave $|k=0) = \sum_{j=1}^L |j) / \sqrt{L}$ is in fact proportional to $\big|Z_{\textrm{tot}}
\big)$.
Hence, if there are no impurities and $|k=0)$ is an exact zero-energy eigenstate [and orthogonal to all the other eigenstates $|\alpha)$], we conclude that the magnetization remains invariant as expected. 
More generally, with impurities that break the U$(1)$ symmetry, we observe that the magnetization evolution from the specified initial state is not sensitive to the spatial structure of the single-particle eigenstates $\phi_\alpha(j)$, as the final expression only depends on the overlap with the zero-momentum state.

For an illustration, let us consider a pair of symmetry-breaking impurities, each of them lying on one boundary of the system. 
(The case with an impurity only at one boundary can be analyzed similarly with appropriate orbitals as in the main text.)
Then, the total magnetization is not conserved and showcases non-trivial dynamics.
Following the discussion in Sec.~\ref{sec:U1} and working in the continuum limit, we expect that low-energy eigenstates of the perturbed super-Hamiltonian correspond to solutions of the free-particle problem with absorbing boundary conditions (on both boundaries), namely $\phi_n(x)=\sqrt{\frac{2}{L}}\sin(\frac{\pi n x}{L})$ with $n=1,2,3,\dots$ and the continuum coordinate $x \in [0, L]$, but with the same energy dispersion relation $E_n ¨=c \left(\frac{\pi n}{L}\right)^2$, for some constant $c$ set by the unperturbed super-Hamiltonian in the bulk.
Inserting these expressions into Eq.~\eqref{eq:M_t}, one then finds
\begin{equation} 
    \overline{\langle \psi(t)|Z_{\textrm{tot}}|\psi(t)\rangle} = \sum_{n\,\text{odd}} e^{- c \left(\frac{ \pi n}{L}\right)^2 t} \frac{8L}{\pi^2n^2},
\end{equation}
since only odd $n$ have a non-zero overlap with the state $|k=0)$.
What is important here is that few lowest-energy eigenstates like $n=1$ or $n=3$ have $O(1)$ overlap with the $|k=0)$ state, which one can see immediately by considering the shapes of these orbitals.
Hence, only a fraction of the magnetization can decay until the time $t^* \sim E_1^{-1} \sim L^2$.
For later times $t > E_1^{-1}$, $\overline{\langle \psi(t)|Z_{\textrm{tot}}|\psi(t)\rangle}$ decays to zero exponentially fast (as also expected from the decay of two-point correlation functions).

\subsection{Evolution of the magnetization in the $t-J_z$ model with impurities at both boundaries}

While this analysis cannot be fully analytically carried over to the evolution of the magnetization in the $t-J_z$ model in the presence of symmetry-breaking impurities at both boundaries, we can exploit the approximate SLIOM Hamiltonian formulation in Eq.~\eqref{eq:H_sliom} of the previous section, to lower bound the time scale beyond which the magnetization decays exponentially fast.
In particular, approximating low-lying excitations of $\mathbb{H}_{\text{SLIOM}}$ via $|\psi_\kappa) \sim \sum_{k=1}^{2L/3} \sin(\kappa k) |\tilde{q}_k)$ [which we recall is a rather crude approximation which neglects the contributions from the other type (i.e., right) SLIOMs], we argue that the time scale beyond which the global magnetization starts to decay exponentially fast, is lower bounded by $L^2/g$, where $g$ is the impurity strength [cf.\ Eq.~(\ref{eq:Pimp_tJz})].
The evolution of magnetization is given by
\begin{equation}
    \big(\ketbra{\psi_0}{\psi_0} \big| e^{-\hat{\mathcal{P}}t} \big|Z_{\text{tot}} \big)
    = \sum_\mu e^{-E_\mu t} \big(\ketbra{\psi_0}{\psi_0} \big| E_\mu\big) \big(E_\mu \big|Z_{\text{tot}} \big)
    \overset{t\gg 1}{\gtrsim} \sum_\kappa e^{-c'g\kappa^2 t} \big(\ketbra{\psi_0}{\psi_0} \big|\psi_\kappa \big) \big(\psi_\kappa \big|Z_{\text{tot}} \big).
\end{equation}
The first sum is over all orthonormal eigenstates $|E_\mu)$ of the super-Hamiltonian.
We then made an approximation keeping only the trial low-energy states constructed in the SLIOM subspace of the commutant as in Appendix~\ref{app:imp_both_boundaries}.
The trial states are ``standing waves'' in the SLIOM label $k$ with ``wavevectors''
$\kappa_n \approx \frac{n\pi}{2L/3}$, $n=1,2,3,\ldots$, such that the orbitals smoothly vanish when $k$ approaches $0$ or $2L/3$; $c'$ is an $O(1)$ number determined from the dimensionless $\mathbb{H}_{\mathrm{SLIOM}}$ in Eq.~(\ref{eq:H_sliom}), while we have also restored the impurity strength $g$.
First, using that $|Z_{\text{tot}}) = \sum_{k=1}^L |q_k)$ we find 
\begin{equation}
(Z_{\text{tot}}|\psi_\kappa) = \sum_{k'=1}^L \frac{1}{\sqrt{L}} \sum_{k=1}^{2L/3}  \sin(\kappa k) \frac{(q_{k'}|q_k)}{\||q_k)\|} \approx \frac{3^{L/2}}{\sqrt{L}}\sum_{k=1}^{2L/3}\sin(\kappa k),
\end{equation}
where we used the fact that $(q_{k'}|q_k) = \delta_{k',k} \||q_k)\|^2 \approx \delta_{k',k} 3^L$.
Now, using expression Eq.~\eqref{eq:qk_super} we find
\begin{equation}
\big(\ketbra{\psi_0}{\psi_0} \big|\psi_\kappa\big) = \frac{1}{\sqrt{L}} \sum_{k=1}^{2L/3} \sin(\kappa k) \frac{(\tilde{\uparrow} \ldots \tilde{\uparrow} |q_{k})}{\||q_k)\|} \approx \frac{1}{3^{L/2}\sqrt{L}} \sum_{k=1}^{2L/3} \sin(\kappa k)(\tilde{\uparrow}\ldots\tilde{\uparrow} |q_{k}) = \frac{1}{3^{L/2}\sqrt{L}} \sum_{k=1}^{2L/3} \sin(\kappa k).
\end{equation}
Putting everything together we find
\begin{equation}
    \big(\ketbra{\psi_0}{\psi_0} \big| e^{-\hat{\mathcal{P}}t} \big|Z_{\text{tot}} \big) \gtrsim \sum_\kappa e^{-c'g \kappa^2 t}  \frac{1}{L} \left|\sum_{k=1}^{2L/3} \sin(\kappa k) \right|^2 
    \geq A_1 e^{-c'g\kappa_1^2 t} L,
\end{equation}
where $A_1$ is an $O(1)$ number, and in the last equation we have kept only the contribution from the lowest-energy trial state which is also expected to have the largest $A_1$.

Using an approximate analysis based on the left SLIOMs, we thus have shown that with impurities at both boundaries, the time scale $t^*$ before exponential decay of magnetization is lower bounded as $t^* > \mathcal{O}(L^2/g)$.
While this is a loose bound, it indicates the presence of polynomial in $L$ time scale for the relaxational behavior of magnetization.
We note that the numerical analysis in Ref.~\cite{wang2025exponentiallyslowthermalization1d} found that $t^* \sim \mathcal{O}(L^{3.5})$, which we cannot derive yet and which remains an open question.

To conclude, we want to point out that while there is some similarity between the solution of the U$(1)$ case in the previous subsection and the discussion of the $t-J_z$ model here, there are also qualitative differences.
Thus, in the U$(1)$ case the characteristic decay time $t^* \sim E_1^{-1} \sim L^2$ is in fact independent of the impurity strength for large enough $L$, since the new lowest-energy mode is obtained by the adjustments of the low-energy excitations of the original model.
[Note that in principle we could have obtained a similar bound in the $t-J_z$ model by simply appealing to its U$(1)$ global spin conservation.]
Instead, in the $t-J_z$ model we have constructed low-energy trial states by working with only the original exact zero modes $\{|q_k)\}$; the ``hopping'' in the effective Hamiltonian for $\{|q_k)\}$ in Appendix~\ref{app:imp_both_boundaries} is set by the impurity strength $g$, and hence the upper-bound on the relaxation rate $\sim g/L^2$ is also parametrically small in the impurity strength $g$, while the $1/L^2$ comes from the ``adjustments'' in the superpositions of $|q_k)$'s.
We suspect that by bringing also the original low-energy excitations of the $\mathcal{P}_{t-J_z}$ into play we can lower the trial energy further, but we have not been able to come up with an actual construction; this remains an interesting open problem.

\end{document}